\documentclass[prx,aps,twocolumn,amssymb,floatfix,eqsecnum,superscriptaddress]{revtex4-2}
\usepackage{ulem}

\usepackage{graphicx}
\usepackage{amsmath}
\usepackage{amstext}
\usepackage{amssymb}
\usepackage[colorlinks,citecolor=blue]{hyperref}
\usepackage{graphicx}
\usepackage{amsmath}
\usepackage{amstext}
\usepackage{longtable}
\usepackage{amssymb}
\usepackage{amsfonts}
\usepackage{longtable,booktabs}
\usepackage{hyperref}\usepackage{url}%Turn on when output to pdf file

\usepackage{amsbsy}
\usepackage{dcolumn}
\usepackage{amsthm}
\usepackage{bm}
\usepackage{multirow}
\usepackage{hyperref}
\hypersetup{
colorlinks, bookmarks=true,breaklinks=true,linkcolor=magenta, citecolor=cyan, linktocpage=true, urlcolor=blue
}

\usepackage{mathrsfs}
\usepackage{amsfonts}
\usepackage{amsbsy}
\usepackage{dcolumn}
\usepackage{bm}
\usepackage{multirow}
\usepackage{color}
\def\Z{\mathbb{Z}}

\newcommand{\g}[1]{{\mathbf{g}}}

 \usepackage{graphicx}
    \usepackage{amssymb}
    \usepackage{amsthm}
    \usepackage{mathtools}
\usepackage{float}

      \theoremstyle{remark}
      
\begin{document}

  \title{{\fontfamily{ptm}\fontseries{b}\selectfont Edge theories of  2D fermionic symmetry protected topological phases  protected by unitary Abelian symmetries}}

\author{Shang-Qiang Ning}
%\email{sqning91@hku.hk}
   \affiliation{Department of Physics and HKU-UCAS Joint Institute for Theoretical and Computational Physics, The University of Hong Kong, Pokfulam Road, Hong Kong, China}
  
  \author{Chenjie Wang}
%\email{cjwang@hku.hk}
   \affiliation{Department of Physics and HKU-UCAS Joint Institute for Theoretical and Computational Physics, The University of Hong Kong, Pokfulam Road, Hong Kong, China}
 
   \author{Qing-Rui Wang}
   \affiliation{Department of Physics, Yale University, New Haven, CT 06511-8499, USA}

\author{Zheng-Cheng Gu}
\email{zcgu@phy.cuhk.edu.hk}
   \affiliation{Department of Physics, The Chinese University of Hong Kong, Shatin, Hong Kong, China}

   \begin{abstract}  
Abelian Chern-Simons theory, characterized by the so-called $K$ matrix, has been quite successful in characterizing and classifying Abelian fractional quantum hall effect (FQHE) as well as symmetry protected topological (SPT) phases, especially for bosonic SPT phases. However, there are still some puzzles in dealing with fermionic SPT(fSPT) phases. In this paper, we utilize the Abelian Chern-Simons theory to study the fSPT phases protected by arbitrary Abelian total symmetry $G_f$. 
Comparing to the bosonic SPT phases, fSPT phases with Abelian total symmetry $G_f$ has three new features: (1) it may support gapless majorana fermion edge modes, (2) some nontrivial bosonic SPT phases may be trivialized if $G_f$ is a nontrivial extention of bosonic symmetry $G_b$ over $\mathbb{Z}_2^f$, (3) certain intrinsic fSPT phases can only be realized in interacting fermionic system. We obtain edge theories for various fSPT phases, which can also be regarded as conformal field theories (CFT)  with proper symmetry anomaly.
In particular, we discover the construction of Luttinger liquid edge theories with central charge $n-1$ for Type-III bosonic SPT phases protected by $(\mathbb{Z}_n)^3$ symmetry and the Luttinger liquid edge theories for intrinsically interacting fSPT protected by unitary Abelian symmetry. The ideas and methods used in these examples could be generalized to derive the edge theories of fSPT phases with arbitrary unitary Abelian total symmetry $G_f$. 
    \end{abstract}
%\date{{\currenttime, \small\today} }
\date{{ \small\today}}
  \maketitle
\tableofcontents  

 \section{Introduction}
 
Recently, tremendous progress has been made towards understanding gapped phases of quantum matter. It has been pointed out that the entanglement pattern is a unique feature to characterize gapped quantum phases. A state only has ``short-range entanglement" if and only if it can be connected to an un-entangled state (i.e., a direct product state or an atomic insulator state) via a local unitary transformation; otherwise, it has "long-range entanglement”. In the presence of global symmetry, even the "short-range entangled" phases can have many different classes. Among them one class is the conventional symmetry breaking phase described by the Landau theory. However, to our surprise, there exists a new class of topological phases - the symmetry protected topological (SPT) phases\cite{gu09,chenScience2012,chen13} associated with any global symmetry in any dimension.  So far, the SPT phases have been quite well-understood in many aspects for both interacting bosonic and fermionic systems, including the classification\cite{chenScience2012,chen13,GuWen2014,kapustin14a,kapustin14,freed14, wen15, wangc-science,chong14,kapustin14,cheng15,freed16,WangGu2017,WangGu2020,Kapustin2017,Juven2018}, characterization\cite{levin_gu_model, Xie12chiralsymm, cheng2014,threeloop,ran14,wangcj15,wangj15,Juven2015,lin15,wanggu16,juven16,Juven2018,WCWG2018}, boundary-bulk correspondence\cite{vishwanath13,wangc13,chen14, bonderson13,wangc13b,fidkowski13,chen14a,metlitski14,metlitski15,witten15,wangPRX2016,Fidkowski2018}, construction of exactly solvable models\cite{chen11b,fidkowski11,chenScience2012,chen13, levin_gu_model, GuWen2014,Gaiotto2016, Fidkowski1604, Chen17}, field theories\cite{freed14,gaiotto16,freed16,morgan16,Kapustin2017,Morgan2018,Kapustin2018},  model realization\cite{kanemele2005, Ning14, bosonic_TI_zheng_xin_liu_2014_prl, yin_chen_he_bqhe_2015_prl, bilayer_graphen_bosonic_SPT_model_cenke_2014_prb, Juven14edge} and experimental discovery\cite{KaneRMP} as well. One of the most striking phenomena of SPT phase is that even though the bulk is short-range entangled without  any fractionalized excitation, its boundary can not be short-range entangled symmetric gapped state. It must be gapless, breaking symmetry(spontaneously or explicitly) or topological ordered state(for the boundary of 3D SPT phases) with fractionalized excitations, due to the anomalous(non-onsite) symmetry action on the boundary. 
%in 3 and higher dimensions, which is due to the symmetry anomaly hosted by the boundary. 

In two dimension, Abelian  Chern-Simons(ACS) theory is a powerful and simple tool to characterize and classify gapped phases such as Abelian fractional quantum Hall effect(FQHE) and bosonic SPT protected by Abelian symmetry. Especially that ACS theory admits a quite elegant boundary-bulk correspondence, which benefits those being interested in the edge theories.  For example, it is quite straightforward to get the chiral Luttinger liquid edge theory description for Abelian FQHE and the (non-chiral) Luttinger liquid theory description with proper anomalous symmetry action for bosonic Abelian SPT phases.  
%n some cases of  symmetry-enriched topological order. 
ACS theory has also been used in studying fermionic SPT(fSPT) phases in Ref.\cite{Lu2012} to obtain a minimal subset classification, however, there is still lacking of systematical and complete understanding. 
%where an important and also more interesting case is missing, that is the gapless majorana fermion edge fields and sometimes does not obtain the right trivialization of bosonic SPT.  
Very recently, the K-matrix formulation of some interesting gapless edge theories of fSPT phases is discussed, e.g., Ref.\cite{Levin2018} provides a valuable example with $\Z_2\times \Z_2^f$ symmetry.

In this paper, we utilize the ACS theory to obtain the edge theories for fSPT phases with Abelian unitary total symmetry $G_f$ in a systematical way.  We derive and identify the gapless edge theories with proper anomalous symmetry realization for all root phases, and also obtain the relations between root phases and phases with other symmetry realization on the edge.  
In general $G_f$ could be a central extension of bosonic symmetry $G_b$ over fermion parity symmetry $\Z_2^f$, %which can be a trivial extension, or nontrivial extension, 
characterized by a second group cohomology class $\omega_2\in H^2(G_b, \Z_2^f)$. For the trivial extension, $G_f=G_b\times \Z_2^f$, while for nontrival extension, the precise way to express $G_f$ is described by a short exact sequence. For simplifying notations, we just denote them as  $G_b \times_{\omega_2} \Z_2^f$. 
%Throughout the whole paper we will restrict all our discussions to those cases with Abelian total symmetry $G_f$. 

%consider both the trivial and nontrivial $\Z_2^f$-extension of $G_b$ and $G_b$ is restricted to  the unitary Abelian group. 
In particular, we construct the Luttinger liquid edge theories of Type-III root states protected by $G_b=(Z_n)^3$ for the first time. 
%It can be realized by non-chiral Luttinger liquid with central charge equal to $n-1$. 
It is natural to ask what is the lowest bound of central charge for a conformal field theory(CFT) that can realize such kind of symmetry anomaly. %For the Type-III anomaly of bosonic SPT of $(Z_n)^3$, 
Our construction suggests that it should be $n-1$. Moreover, we also construct the Luttinger liquid edge theories for the intrinsic interacting fSPT phases protected by total symmetry $G^f=\Z_4^f\times \Z_4\times \Z_4$. From the view point of CFT, our results reveal new types of symmetry anomalies in these multi-component bosonic CFT or spin CFT.

The rest of the paper is organized as following: in Sec.\ref{review}, we review some useful knowledge about the ACS for study fSPT, especially, we show in Sec.\ref{anomaly_detection} two ways to detect the symmetry anomaly of the edge theory, one is the so-called null vector criterion in Sec.\ref{null_vector} and the other is to check the projective representation potentially carried by symmetry flux in Sec.\ref{proj_rep}.
% In Sec.\ref{main_result}, we summarize our main results.
  We carefully study the examples with trivial central extension of $G_b$ by $\Z_2^f$ in Sec.\ref{trivial_extension}, and then nontrivial extension cases in Sec.\ref{nontrivial_extension}. In Sec.\ref{type-III}, we discuss the construction of edge theory with Type-III anomaly of bosonic SPT with $(\Z_n)^3$. 
  In Sec.\ref{sec_intrinsic}, we discuss the edge theory of very interesting case: intrinsically interacting fSPT phases.
  Conclusion and discussion is in Sec.\ref{discussion}. Some other examples and other solutions of the examples in the main text will  be discussed in appendix.

\section{Overview}
\label{review}

In this section, we first review the main knowledge that we will use for examples studied in Sec.\ref{review1}-\ref{anomaly_detection}. Then we will summarize our results in Sec.\ref{main_result}. 

\subsection{K matrix formulism for fSPT}
\label{review1}

%Abelian Chern-Simons field theory has been extensively applied to describe the Abelian topological order as well as SPT phases. 
Generally, a $(U(1))^n$ Chern-Simons theory can take the following form
\begin{align}
\mathcal{L}=\frac{K_{IJ}}{4\pi} \epsilon^{\mu\nu\lambda} a_{\mu}^I\partial_\nu a_\lambda^J+  a^I_\mu j_{I}^\mu+...
\label{CS}
\end{align}
where $K$ is a symmetric integral matrix, $\{a^I\}$ is a set of one-form gauge fields and $\{j_I\}$ are the corresponding currents that couple to the $a^I$ gauge fields. The theory has an emergent symmetry related to the relabeling of the gauge fields $a^I$. Namely, two theories $\mathcal{L}[a^I]$ and $\mathcal{L}[\tilde a^I]$ related by $a^I=W_{IJ} \tilde a^J$, where $W$ is a $n\times n$ integral unimodular matrix,  actually describe the same gapped phase.  As a result, not every $K$  labels a distinct phase, but only  $K$ up to the $SL(n,\mathbb{Z})$ transformation. 

\begin{figure}[t]
   \centering
   \includegraphics[width=9cm]{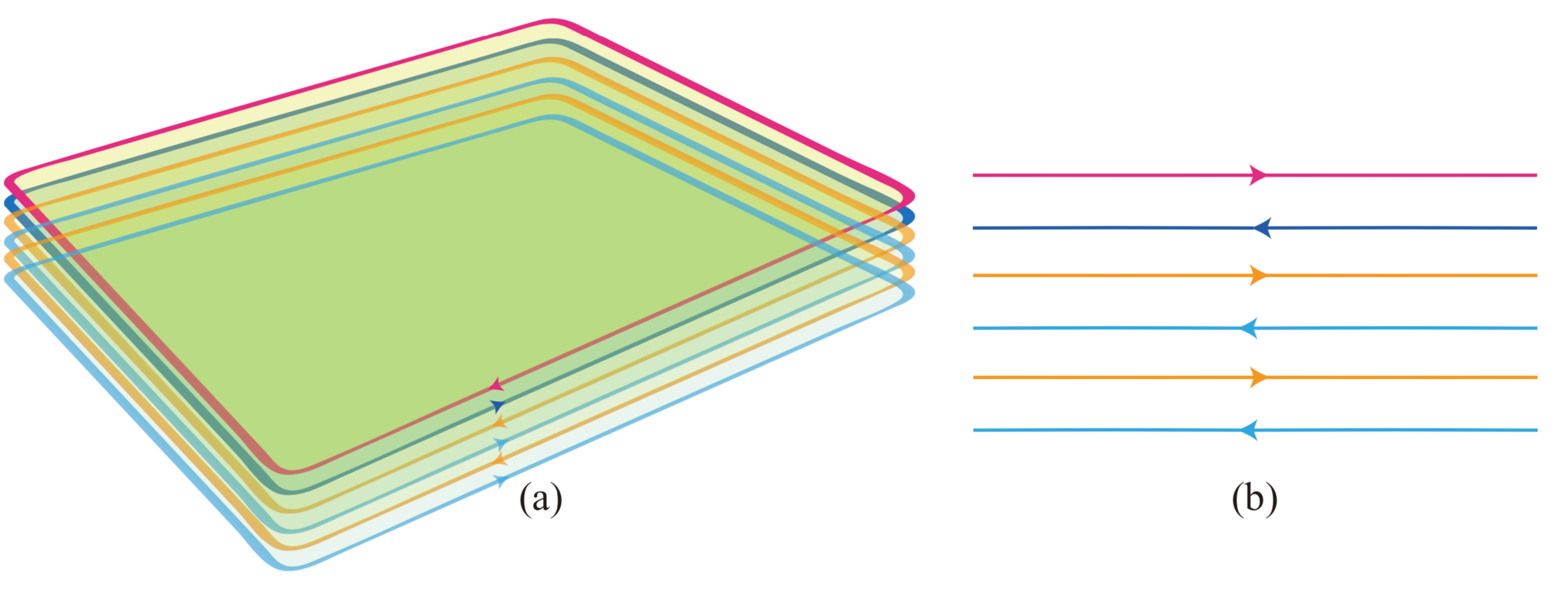} % requires the graphicx package
   \caption{ The bulk described by ACS action has a natural Luttinger liquid(s) boundary. (a) The bulk and boundary of one example with K matrix $K=\sigma_z\oplus \sigma_x\oplus \sigma_x$;  (b) the non-chiral Luttinger liquid boundary has six edge fields, three propogating in clockwise and the other three anticlockwise.}
   \label{fig_bb_correspondence}
\end{figure}

The topological order described by Abelian Chern-Simons theory hosts Abelian anyon excitations. An anyon can be labeled by an integer vector $l=(l_1,l_2,...,l_n)$. The self statistics of an anyon $l$ is given by 
\begin{align}
\theta_l=\pi l^T\cdot K^{-1} \cdot l
\end{align}
and the mutual statistics of two anyons $l$ and $l'$ is given by
\begin{align}
\theta_{l,l'}=2\pi l^T\cdot K^{-1} \cdot l'.
\label{mutual_statistics}
\end{align}
A bosonic excitation means that the self statistics is multiple of $2\pi$ while a fermion means that the self statistics is $\pi$ modular $2\pi$. The total number of anyons and the ground state degeneracy on torus are both given by $|\det{K}|$. In SPT phases, there is no anyons and the ground state is non-degenerate on any closed manifold, so we should require that $|\det(K)|=1$. In our later discussions, we will consider the presence of additional global symmetries. An external global $U(1)$ symmetry can be described by a charge vector by $q$. Then, the charge carried by an anyon excitation $l$ is 
\begin{align}
Q=q^T\cdot K^{-1} \cdot l
\label{charge_of_excitation}
\end{align}
We note that readers shall not be confused with the $U(1)$ gauge symmetries of $a^I$ and the global $U(1)$ symmetry.

The K matrix Chern-Simons theory admits a well-known edge-bulk correspondence (see Fig.\ref{fig_bb_correspondence}) In a system with open boundary, the edge theory  corresponding to (\ref{CS}) can take the form
\begin{align}
\mathcal{L}_{edge}=\frac{K_{IJ}}{2\pi} \partial_x \phi_I \partial_t \phi_J+v_{IJ} \partial_x \phi_I \partial_x \phi_J+...
\label{edge}
\end{align} 
where $\phi_I$ is the chiral bosonic field on the edge and related to the bulk dynamical gauge field by $a_\mu^I=\partial_\mu\phi_I$. An anyon labeled by $l$ on the edge can be created by $e^{i l^T\cdot \phi}$ where $\phi=(\phi_1, \phi_2,...,\phi_n)$.

\subsection{Symmetry implementation} 
\subsubsection{Definition of  symmetry in fSPT}

Any fermionic system has the fermionic parity invariance. Namely,  if we denote the total symmetry of an fSPT by $G_f$, then the fermion parity symmetry $\Z_2^f=\{1, P_f \}$ has to be a normal subgroup of $G_f$. The quotient group $G_b=G_f/\Z_2^f$ is the bosonic part of symmetry in the fSPT. Therefore, $G_f$ is a central extension of $G_b$ by $\Z_2^f$ which is labeled by the second group cohomology $\omega_2\in H^2(G_b,\Z_2^f)$. More precisely, $G_f$ defines as
\begin{align}
0\longrightarrow \Z_2^f \longrightarrow G_f \longrightarrow G_b \longrightarrow 0
\end{align}
If the extension is trivial, $G_f=G_b\times \Z_2^f$, otherwise we denote $G_f=G_b\times_{\omega_2} \Z_2^f$.
 Note that the trivial element in $G_b$, which we denote as $e$, can be represented by $1$ or $P_f$.  For example, since $H^{2}(\Z_2, \Z_2^f)=\Z_2$,  there are two extensions of $G_b$. The trivial one is just $G_b=\Z_2\times \Z_2^f$ with two order-2 generators while the nontrivial one is $G_f=\Z_2\times_{\omega_2} \Z_2^f=\Z_4^f$ whose  generator  $g$ is  order-4 and squares to $P_f$.
Through out this paper, we focus on the Abelian $G_f$.

\subsubsection{Implementing symmetry on the edge}
We now consider how a symmetry is implemented in the edge theory. Under a symmetry operation $g$, the edge field $\phi_I$ transform as
\begin{align}
g: \phi_I(x)\rightarrow W_{IJ}^g \phi_J(x)+\delta \phi^g_I
\label{symmetrytransf1}
\end{align}
where the repeated $J$ is summed over and $W^g$ is an integral unimodular matrix such that 
\begin{align}
K=\eta_g (W^g)^TKW^g
\label{symmetrytransf2}
\end{align}
where $\eta_g=\pm $ corresponding to   unitary $g$ or  anti-unitary $g$. $\delta \phi^g$ is a constant vector up to $2\pi$. Accordingly, the excitation on the edge created by $e^{il^T\cdot \phi(x)}$ transform as 
\begin{align}
g: e^{il^T\cdot \phi(x)}\rightarrow e^{il^T\cdot W\cdot \phi(x)} e^{il^T\cdot \delta \phi^g}
\end{align}
%An excitation is invariant under the symmetry is defined as $l^T\cdot \delta\phi^g=0$ mod $2\pi$ for all group elements. 

The quantities $W^g$ and $\delta\phi^g$ in the transformation (\ref{symmetrytransf1}) can not be arbitrary. Besides (\ref{symmetrytransf2}), another natural constraint is that they should be compatible with the group structure of $G_f$. Specially, let us take $\{g_1,g_2,..,g_k\}$ to be the generators of a finite symmetry group. They satisfy a set of group relations in the following form 
\begin{align}
\prod_{i}^kg_i^{n_i}=1
\label{group_relation}
\end{align} 
where $n_i$ are integer.
Then, acting both sides of (\ref{group_relation}) on the edge fields according to (\ref{symmetrytransf1}), we will require that
\begin{align}
\prod_{i}^kg_i^{n_i}:\phi\rightarrow \phi \quad \text{ mod} \quad 2\pi
\end{align}
This set of conditions constrain possible values that $W^g$ and $\phi^g$ can take. 

However, among the solutions $\{W^{g_i},\delta \phi^{g_i}\}$ of the above conditions, there is some redundancy.  Two solutions related by the following gauge transformation are treated equivalently:
\begin{subequations}
\label{gauge_transformation}
\begin{align}
&\tilde W^{g_i}=U^{-1} W^{g_i} U, \\
 &\delta \tilde\phi^{g_i}=U^{-1}[\delta\phi^{g_i}-(1-\eta_{g_i} W^{g_i}) \Delta \phi]
\end{align}
\end{subequations}
where $U$ is a $n\times n$ integral unimodular matrix  such that $U^TKU=K$ and $\Delta \phi$ is a constant vector up to $2\pi$ and $\eta_g=\pm$ denotes the group element $g$ to be unitary/anti-unitary.

 Thoughout this paper, we use the notation of $[K, \{W^{g_i}, \delta \phi^{g_i}\}]$ to denote a consistent  realization of a certain symmetry in the bulk and also on the edge whose low energy physics are described by ACS theory with K matrix $K$ and its canonical Luttinger liquid edge theory. The consistency is guaranteed  by that it is a solution for constraint equations enforced by symmetry. Below we  call $[K, \{W^{g_i}, \delta \phi^{g_i}\}]$ as state, phase or solution interchangeably.  Without causing confusion, we sometimes  omit $K$ matrix and only list the realization of generator(s) of symmetry group  in the notation $[K, \{W^{g_i}, \delta \phi^{g_i}\}]$.

\subsubsection{Fermion parity operator}
Since every fSPT is invariant under the fermion parity, here we pay special attention to the fermion parity operator. It is well-known that any $2\times 2$ K matrix for fSPT can transform into $\sigma_z$ via proper modular transformation. Then, for $K=\sigma_z$, the fermion parity should realize as\cite{Lu2012}
%
%Under symmetry action $g$, 
%\begin{align}
%a_I \rightarrow s(g) W^g_{IJ} a_J,
%\end{align}
%and
%\begin{equation}
%(W^g)^T K W^g=K, \quad W^g\in GL(Z,2)
%\end{equation}
%so that the theory is invariant under symmetry action.  On the boundary, due to the relation $a_I^\mu=\partial_\mu \phi_I$, the symmetry action on $\phi_I$ may be
%\begin{align}
%\phi_I(x) \rightarrow s(g) W_{IJ}^g \phi_J(x)+\delta\phi_I^g. 
%\label{symmetry on phi}
%\end{align}
%Note that $s(g)$ is +1 if g is unitary and -1 if g is antiunitary.
%
%In the fermion system, we always have the fermion parity symmetry, which is implemented as 
\begin{align}
W^{P_f}=1_{2\times 2}, \quad \delta \phi^{P_f}=\pi\begin{pmatrix} 1 \\1\end{pmatrix}.
\label{fermionparity}
\end{align}
To justify it, we consider two basic facts: (1) two basic fermionic excitations $e^{i\phi_1}, e^{i\phi_2}$ aquire a minus sign under the fermion parity, and any bosonic excitation $e^{i (l_1\phi_1+l_2\phi_2)}$ with $l_1+l_2=0$ mod 2 is invariant under fermion parity, (2) with only $\Z_2^f$ symmetry, any bosonic excitation can condense to gap out the edge modes without breaking any symmetry.

In the following, most of our examples are with $K=\sigma_z$, and we always assume the fermion parity is realized as (\ref{fermionparity}). For those with larger dimensional $K$ matrix in this paper, $K$ would take the form as $\sigma_z\oplus \sigma_x\oplus ...\oplus \sigma_x$, then the fermion parity is simply generalized into the form 
\begin{align}
W^{P_f}=1_{2\times 2}\oplus1_{2\times 2} \oplus ... \oplus 1_{2\times 2} , \\\quad \delta \phi^{P_f}=\pi\begin{pmatrix} 1 \\1\end{pmatrix}\oplus \begin{pmatrix} 0 \\0\end{pmatrix}\oplus ... \oplus \begin{pmatrix} 0 \\0\end{pmatrix}.
\label{fermionparity2}
\end{align}
%Note that $K = \sigma_x$ describe bosonic SPT phases.

\subsection{Stacking of SPT phases}
\label{review2}
For two fSPT phases described by two ACS theories $[K, \{W^{g_i}, \delta \phi^{g_i}\}]$ and $[\tilde K, \{\tilde W^{g_i}, \delta \tilde \phi^{g_i}\}]$, the stacking of these two phases forms a new phase, which is described by a new ACS theory
 $[K_s=K\oplus \tilde K, \{W_s^{g_i}=W^{g_i}\oplus \tilde W^{g_i}, \delta\phi^{g_i}_s=\delta \phi^{g_i}\oplus \delta \tilde \phi^{g_i}\}]$. If the classificaion of fSPT is an Abelian group, then the stacking operation is the group operation of the classification group.   We also use the notation $[K, \{W^{g_i}, \delta \phi^{g_i}\}]^{-1}$ as the inverse phase of $[K, \{W^{g_i}, \delta \phi^{g_i}\}]$ in the classification group.  The stacking operation can also be well-defined for many  phases. In particular, we might use the notation  $[K, \{W^{g_i}, \delta \phi^{g_i}\}]^{\oplus n}$ as   phase  that  is a stacking of $n$-copy phases of $[K, \{W^{g_i}, \delta \phi^{g_i}\}]$.

 For example, the classification $\Z_2^f\times \Z_2$ fSPT is $\Z_8$ whose generator is denoted by $w$, if the root phase labeled by $w$ can be realized by  an ACS theory with  $[K_1, \{W_1^{g_i}, \delta \phi_1^{g_i}\}]$, then two copies of $w$ stack to a new phase labeled by $w^2$ and similarly eight copies stack to $w^8=1$ which is trivial and hence whose edge can be symmetrically gapped out. The root state can also be realized by another ACS theory with $[K'_1, \{{W'}_1^{g_i}, \delta {\phi'}_1^{g_i}\}]$, then the stacking theory 
 $[K_2=K_1\oplus K_1', \{W_2^{g_i}=W_1^{g_i}\oplus {W'}_1^{g_i}, \delta\phi_2^{g_i}=\delta {\phi}_1^{g_i}\oplus \delta {\phi'}_1^{g_i}\}]$ also describes a phase labeled by $w^2$.

%When an ACS theory, no matter whether from stacking operation or not, describes the trivial phase, in principle, we can symmetrically gap out the edge fields. However, sometimes it is not easy to find the proper interaction terms to induce the symmetric gap, but it becomes easier to do so if we stack an additional ACS theory that realizes the trivial phase.

\subsection{Detecting the symmetry anomaly on the boundary}
\label{anomaly_detection}
\begin{figure}[t]
   \centering
   \includegraphics[width=9cm]{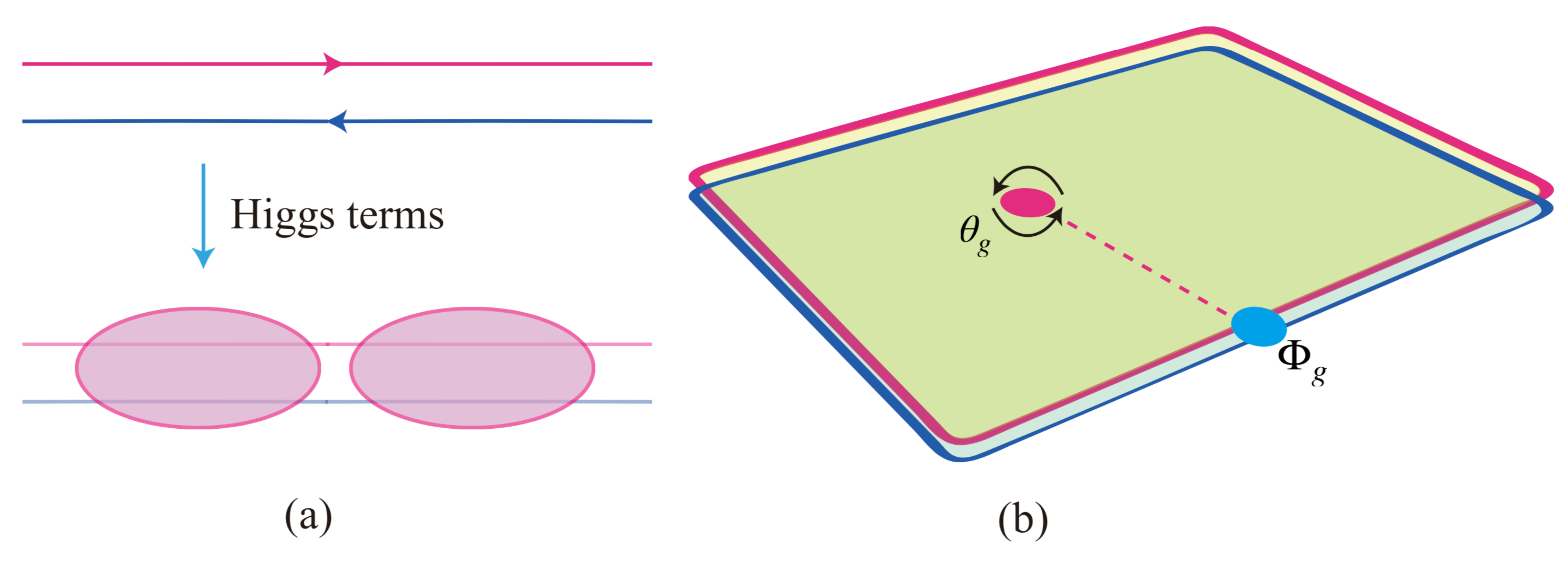} % requires the graphicx package
   \caption{Two approaches to detect the symmetry anomaly. (a) If there exist symmetric Higgs term(s) that can drive the edge Luttinger liquid to a symmetric short range entangled state, the bulk is a trivial SPT; otherwise, it is nontrivial SPT. (b) The  symmetric flux  are inserted, whose topological spin or (projective) representation of other symmetry can be used to detect whether a SPT is nontrivial or not. If the symmetry $g$ does not permute edge fields, but only shift them by a phase shift $\delta\phi^g$, the symmetry defect on the boundary can be created by applying the ``frationalized" operator $\Phi_{\text{g}}=e^{iK^{-1}\cdot \delta\phi^g\cdot \phi}$. }
   \label{fig_symmetry_flux}
\end{figure}

Here we discuss two different ways to dectect whether the symmetry on boundary is anomalous or not. The first one, as discuss in  Sec.\ref{null_vector}, directly studies the stability of the edge fields, while the second one, in Sec.\ref{proj_rep}, turns to study the the topological  properties of symmetry flux that can characterize  SPT phase (see Fig.\ref{fig_symmetry_flux}). In practice, the first one is convenient for proving a phase  is trivial since the symmetric Higgs terms for symmetrically and fully gapping out the edge fields may be easily constructed, while the latter may be especially useful to assert that a phase is nontrivial and also assert which nontrivial phase it belongs to.

\subsubsection{Ingappability without breaking symmetry}
\label{null_vector}

The nontrivial topology of  nontrivial SPT phases can be manifest on their boundaries where the symmetry is anomalous. The existence of symmetry anomaly on the boundary is in fact one  defining property of SPT phases and has direct physical consequence  that the boundary can not be adiabatically connected to symmetric short range entangled states. In general, the boundary of nontrivial SPT can only be gapless, spontaneously symmetry breaking or develop topological order if not breaking symmetry. For 2D nontrivial SPT, as there is no nontrivial topological order in 1D, their 1D edges can only be  gapless  without breaking symmetry or gapped with breaking symmetry.  

Based on the properties, we can detect whether a SPT phase is nontrivial or not by studying the stability of its edge modes. If its edge modes can be symmetrically fully gapped out, then it must be a trivial SPT, otherwise, it is a nontrivial one. 
We can call this criterion as ingappability criterion. 
In particular, for the SPT phases realized by $K$ matrix ACS theories, we have the following two pratical ways to see whether they are nontrivial or not based on the ingappibility criterion.

The first one is the so-called null-vector criterion that is directly related to the bosonic edge fields.  To symmetrically gap out the bosonic edge fields (\ref{edge}), we can condense some bosons,  namely by adding the Higgs terms 
\begin{align}
\mathcal{L}_{hig}=\sum_{i} C_i \cos(l_i^T\cdot \phi+\alpha_i) 
\end{align}
The perturbative terms should satisfy the following conditions. First, it should be symmetric under every symmetry,  
$g:\mathcal{L}_{hig}\rightarrow \mathcal{L}_{hig}$.  Second, for a $2n\times 2n$ K matrix, it needs $n$ different Higgs terms with vectors $l_{1,2,...,n}$ to fully gap out the edge. The $n$ vectors $l_1, l_2, \dots, l_n$ should be linearly independent.  Third, the Higgs terms should satisfy the so-called null-vector condition. To illustrate it, we construct $n$ different integer vectors $\Lambda_{i}:=K^{-1}\cdot l_i$. The null vector condition then states that the corresponding edge theory can be fully gapped out if and only if the following conditions are satisfied
\begin{align}
& \Lambda_i^TK\Lambda_i=0\, \text{ for all\text{ }} i \\
& \Lambda_{i}^TK\Lambda_j=0\, \text{ for\, all\, pairs\, of }\,  i,j
\end{align}
Fourth,  it is required that no  spontaneously symmetry breaking occurs, once the edge fields are  fully gapped by $\mathcal{L}_{hig}$. For this, it is required that the greatest common divisor of all the $n\times n$ minors of the matrix $(\Lambda_1, \Lambda_2, ..., \Lambda_n)$ is $\pm1$.

If the Higgs terms satisfying the above conditions exist, the edge can be fully gapped without breaking any symmetry. Then, the symmetry realization in the edge theory is anomaly-free. This means the bulk fSPT is trivial.  Otherwise, if the edge modes can not be symmetrically gapped out, it means the symmetry realization  is anomalous, which indicates the bulk fSPT is non-trivial.

The second one utilizes the so-called refermionization of the scalar bosonic edge fields (\ref{edge}). 
For many cases, the K matrix we study takes the form as $K=(\sigma_z)^{\oplus n}$, whose edge fields are denoted $\phi_{i}, i=1,2,...,2n$. We can define $2n$ fermions by $\psi_{i}\sim e^{i\phi_i}$ and transform the edge theory (\ref{edge}) in terms of bosonic edge fields with certain radius into an edge theory in terms of fermionic fields. Then the staibility of this fermionic edge theory can be studied by checking whether there is symmetric mass (interaction) term to gap out the edge fields. 
%If the   components of fermions is large enough, it is also necessary to consider the symmetric interactions that may also symmetrically gap out the edge fields. 
% We , this way is quite convenient for those trivial edge theories. 
 As  in many examples as follows, transforming the bosonic edge field theories  into a fermionic one may be more simpler to find the symmetric mass terms to fully gap out the edge fields.

We stress that the above two ways are not totally different from each other. In fact, they are equivalent in certain cases.  However, we illustrate them here explicitly just for pratical purpose.

\subsubsection{Symmetry flux}
\label{proj_rep}
\label{projective_rep}

Whether the symmetry realization is anomalous or not can also be checked by the  properties of symmetry fluxes. Here, we consider two examples for illustration. First let us consider $\Z_N$ symmetry group for examples. One possible realization of the generator $g_1$ of $\Z_N$ group is
\begin{align}
W^{g_1}=1, \quad \delta \phi^{g_1}=\frac{2\pi}{N}K^{-1}\cdot \chi_{g_1}
\label{symmetrycub1}
\end{align}
where $\chi_{g_1}$ is an integer vector.  $K$ is the K matrix for SPT phase. The phase shift $\delta \phi^{g_1}$ has a physical meaning, that is the symmetry charge carried by the excitation created by $e^{i\phi}$.  More precisly, we can denote the  charge vector of the $n$ different fundamental excitations $e^{i\phi_i}$ by $q=\chi_{g_1}$.  Therefore, via (\ref{charge_of_excitation}), the symmetry charge of the excitation labeled by $l$  is $Q=q^T\cdot K^{-1} \cdot l$, if we view $\Z_N$ as a subgroup of $U(1)$.  Now we consider inserting the elementary symmetry flux $\frac{2\pi}{N}$ in the system,  and the Berry phase accumulated when braiding the excitation $l$ around this symmetry flux is simply given by
\begin{align}
\theta_{g,l}=\frac{2\pi}{N}Q=2\pi \frac{\chi_{g_1}^T}{N}\cdot K^{-1} \cdot l
\end{align}
Compared with (\ref{mutual_statistics}), the effective label of the symmetry flux in this K matrix framework is $l=\frac{\chi_{g_1}}{N}$
 which means that the ``fractionalized" vertex operator $e^{i\frac{1}{N}  \chi^T_{g_1}\cdot \phi}$ can be treated as the symmetry flux of this $Z_N$ symmetry with the same $K$ matrix (see.Fig.\ref{fig_symmetry_flux}).  There actually is no such kind of fractionalized dynamical excitations in the SPT bulk. Since by inserting  symmetry flux in the system, we have to create an branch cut which is singular, we understand that this ``fractionalized excitation" is actually a static defect which lies at the end of the branch cut. Nevertheless, we can continue to calculate the ``topological spin" of the symmetry flux as 
 \begin{align}\theta_{g_1}= \frac{\pi}{N^2} \chi_{g_1}^T\cdot K^{-1} \cdot \chi_{g_1}
 \end{align} which would become the true value of topological spin of gauge flux once we gauge this symmetry. If the edge can be gapped out without breaking the symmetry, it is required that $N\theta_{g_1}=0$ modulo $2\pi$. If this condition is satisfied, it means that there exists a bosonic symmetry flux and then we can condense it without breaking the symmetry. This physical intuitive understanding can be refered to more rigorous derivation in Ref.\cite{cheng2014}.
%  To justify this, we can compare the mutual statistics between the anyon created by this  "fractionalized" vertex operator and the excitation $e^{i\phi_i}$ and the Berry phase accumulated braiding the same amount symmetry charge around the symmetry flux.

Let us consider another example. For the non-Abelian root state protected by $(\Z_N)^3$  symmetry, a sufficient property is  that the symmetry flux corresponding to one $\Z_N$ subgroup carries the fundamental projective representation of the rest $\Z_N\times \Z_N$ subgroup. Supposed one of the $\Z_N$ symmetry subgroup realizes as (\ref{symmetrycub1}), ad the other two are realized as, generally speaking, $\{ W^{g_2}, \delta \phi^{g_2}\}$ and $\{ W^{g_3}, \delta \phi^{g_3}\}$ where $g_2$ and $g_3$ are the two generators of the rest two $\Z_{N}$ subgroups. Then acting on the symmetry flux related to the first $\Z_N$ subgroup, the representative matrices of $g_2$ and $g_3$ should form the fundamental projective representation of $\Z_N\times \Z_N$ group\cite{Juven14edge}.  

%\subsection{Group structure of phase}

\subsection{Main Results} 
\label{main_result}

In this section, we will present our main results: we aim to find out the edge theories for various fSPT in the K matrix formulism. The root phase results are summarized in Table.\ref{mainresult}. 
% Here  we consider the cases with Abelian $G_f$.  Depending on whether there is $\Z_2^f$ extension by the same $G_b$, there can be different $G_f$.   We consider both the trivial and nontrivial $\Z_2^f$-extended $G_f$. 
%For trivial $\Z_2^f$ extension, we discuss carefully the $\Z_2\times \Z_2^f$, $\Z_4\times \Z_2^f$ and $\Z_2\times \Z_2 \times \Z_2^f$ symmetries. Besides, for the nontrivial $\Z_2^f$ extension, here we discuss carefully   $\Z_8^f$, $\Z_2\times \Z_4^f$ and $\Z_4\times \Z_4^f$ ($\Z_4^f$ is also discuss in the Appendix.\ref{z4f}).    We also discuss the edge theories of the so-called Type-III bosonic SPT, such as $(\Z_2)^3, (\Z_3)^3, (\Z_4)^3, (\Z_5)^3$ and intrinsically interacting fSPT, such as the one of $\Z_4^f\times \Z_4\times \Z_4$.
 Here we briefly summarized the results  we obtain. In the following, for most cases, we take the K matrix as $K=\sigma_z$. So if there is no claim on the explicit form of  K matrix, it is assumed that $K=\sigma_z$. 
 
\begin{enumerate}

\item 
%(1)
$\Z_2\times \Z_2^f:$  
%To obtain the well-known gapless majorana fermion edge theory, we start with $K=\sigma_z$ whose edge is a $c=1$ Luttiner liquid. We can add  symmetric perturbative  interactions which would drive the $c=1$  Luttinger liquid to two counter-propogating gapless majorana fermion. 
% and then after identifying the symmetry realization that acts in a anomalous way on these bosonic edge fields, we can add  symmetric perturbative  interactions which would drive the $c=1$  Luttinger liquid to the well-known Ising cricality with $c=1/2$, i.e. with two counter-propogating gapless majorana fermion.  
%The symmetry transformation of these two majorana fermions is anomalous, which prevend it being gapped out by a symmetric mass. In this case, this edge is stable, at least perturbatively.
%The symmetry action which satisfies the above consideration on the edge fields $(\phi_1, \phi_2)$ can be chosen to be $W^{g}=\sigma_z$ and $\delta \phi^{g}=0$, and the symmetric interactions terms are $U[\text{cos}(\phi_1+\phi_2)+\text{cos}(\phi_1-\phi_2)]$.  In fact, the above choice is not unique, we can also choose  $W^{g}=\sigma_z$ and $\delta \phi^{g}=(\pi, 0)^T$, then the symmetric interaction terms would become $U[\text{cos}(\phi_1+\phi_2)-\text{cos}(\phi_1-\phi_2)]$. In both cases, it can be seen that eight copies of the two majorana fermions left can be symmetrically gapped out, leading to the $\Z_8$ classification. 
The classification of fSPT protected by this symmetry is $\Z_8$. The root phase is identified as that with $W^g=\sigma_z, \delta\phi^g=0$.  The physics of this root phase is that the edge hold two gapless majorana fermions that propogate in opposite directions.
%We  find that following  important relations between the phases realized by different symmetry actions: 
%(a) the phase $[\sigma_z, (\pi, 0)]$ is the inverse of the root phase;
%(b) the phase $[-\sigma_z, (0, 0)]$ is the inverse of the root phase and  the phase $[-\sigma_z, (0, \pi)]$ is the same phase as the  root phase.
%(c) the phase $[1, (0, \pi )]$  realizes the phase of two copy of the root phase, and combine the fact four copies of the phase  $[1, (0, \pi )]$ are trivial, it can also be seen that eight copies of root phase are trivial. 
We note that the majorana fermion edge theory for SET emerged from ACS theory  is also discussed in Ref.\cite{Lu13SET} 
and for $\Z_2\times \Z_2^f$ SPT in Ref.\cite{Levin2018}. However we discuss more thoroughly here, including how other phases are related to each other, as see the phase relation (\ref{eqn_z2z2f_relation_1})-(\ref{eqn_z2z2f_relation_2}).
\item
%(2) 
$\Z_4 \times \Z_2^f:$  The classification of fSPT protected by this symmetry is $\Z_8\times \Z_2$, we find that the root phase for the $\Z_8$ classification is the phase with $W^{g}=1_{2\times 2}$, $\delta \phi^g=(0, {\pi}/{2})^T$. In this case, the edge fields when being gapless is a $c=1$ Luttinger liquid.  
The root phase for $\Z_2$ classification is identified by the phase with $K=(\sigma_z)^{\oplus 3}$ and $W^{g}=\sigma_z\oplus (1_{2\times 2})^{\oplus 2}, \delta \phi^g=(0,0)^T\oplus(0,\pi/2)^T\oplus (0,\pi/2)^T$. This root phase in fact is a stacking phase that consists of the phase with $K=\sigma_z$ and $W^{g}=\sigma_z, \delta \phi^g=0$  and  two copies of that above $\Z_8$ root phase, which  admits odd number of gapless majorana fermions on the edge.

% transformation with $W^g=\sigma_z, \delta \phi^g=0$ is also a symmetry realization of the $\Z_4\times \Z_2^f$ group. One can easily see that eight copies of them are trivial, which naively conclude it is another $\Z_8$ root phase since this case can admit odd number of gapless majorana fermions on the edge. This is contridictive to the classification, as stated above.  To solve this puzzle, we find that two copies of the phase  with $W^g=\sigma_z, \delta \phi^g=0$ can be trivialized by adding four copies of the   root phase with $W^g=1_{2\times 2}, \delta \phi^g=0$.  The  root phase with $W^g=1_{2\times 2}, \delta \phi^g=(0, {\pi}/{2})^T$ is not more fundamental than the phase with $W^g=\sigma_z, \delta \phi^g=0$ since upon gauging, the former is Abelian topological order and the latter is non-Abelian one.   Therefore, we treat the phase with $W^{g}=\sigma_z, \delta \phi^g=0$ stacking two copies of that above root phase as a new root state for the $\Z_2$ classification which upon gauging becomes a non-Abelian topological order.

Besides these two root phases, there are also many other allowed states. We also  show that the relations between other phases and the root ones, such as (\ref{eqn_z4z2f_phase_relation_i0})-(\ref{eqn_z4z2f_phase_relation_i6}).

%(3)
\item
 $\Z_2\times \Z_2 \times \Z_2^f: $ The classification of this symmetry is $\Z_8\times \Z_8 \times \Z_4$, hence there are three root phases, two of which are just the ones that protected by a single $\Z_2$  subgroup alone, which are easily obtained by choosing the other $\Z_2$ subgroup being totally trivial.  These two root states are just the ones with odd number of gapless majorana fermions.

The third root phase is identified  to the one with $W^{g_1}=W^{g_2}=-1_{2\times 2}, \delta \phi^{g_1}=(0,\pi), \delta \phi^{g_2}=0$. We note that the fermion parity is to realize as $W^{P_f}=1_{2\times 2}, \delta \phi^{P_f}=(\pi, \pi)^T$.   We find  that this root phase can be trivialized by four copies of them, but not by two copies, hence it is not a $\Z_2$ root phase. We also find that  when acting on the symmetry flux labeled by $g_1$, the $\Z_2\times \Z_2^f$ form the projective representation, which indicates that the root phase is a ferminic  non-Abelian SET upon gauging the whole symmetry.
    
    There are various other allowed states, in addition to these three root ones, whose relations to the roots ones are also obtained in Sec.\ref{sec_z2z2z2f_groupstru} and also in Appendix.\ref{append:z2z2z2d}.
    
       \item
%    (4).   
    $\Z_8^f:$ The classification of fSPT protected by this symmetry is $\Z_2$.  The root phase is identified as the one with  $W^g=1_{2\times 2}, \delta \phi^{g}=(\pi/4, 3\pi/4)$.   We find that to see that  two copies of this root state is indeed trivial; a easy way is to stack them with additional two trivial phases. Compared to the $\Z_4\times \Z_2^f$ symmetry where $\Z_4$ is trivial extension of $\Z_2^f$, we  find that due to the presence of nontrivial $\Z_2^f$ extension, the symmetry transformations as $W^g=\pm\sigma_z, -1$  are not consistent with the symmetry, which indicates that the edge with odd number of gapless majorana fermion are not consistent in this case.  We also discuss  typical phases relations, such as  (\ref{z8f_phase_relation2})-(\ref{z8f_phase_relation4}) between the root  and other allowed states.

%    (5) 
\item
    $\Z_2\times \Z_4^f: $ The classification of this symmetry $\Z_4$. The root state is identified as the one with $W^{g_1}=W^{g_2}=1_{2\times 2}, \delta \phi^{g_1}=(0, \pi), \delta \phi^{g_2}=(\pi/2, \pi/2)$.  Similarly to $\Z_8^f$, the  realizations of  symmetry $\Z_2\times \Z_2\times \Z_2^f$ with $W^{g_1}=\pm\sigma_z$ or  $W^{g_2}=\pm\sigma_z$  which give rise to the edge with odd number of gapless majorana fermion are not consistent in this case.  We also discuss  typical phases relations, such as  (\ref{z4fz2_phase_relation1})-(\ref{z4fz2_phase_relation3}), between the root and other allowed states.

%    (6)  
    \item $\Z_4\times \Z_4^f$: The classification of this symmetry $\Z_8\times \Z_2$.  The root state  for the $\Z_2$ classification is identified as the one with $W^{g_1}=W^{g_2}=1_{2\times 2}$ and $\delta \phi^{g_1}=(\pi/2, -\pi/2), \delta \phi^{g_2}=(\pi/2, \pi/2)$, while the one for the $\Z_8$ classification is with $W^{g_1}=W^{g_2}=1_{2\times 2}$ and $\delta \phi^{g_1}=(\pi/2, \pi/2), \delta \phi^{g_2}=(0, \pi/2)$. One way to see that they are indeed different root states is to see the ``topological spin" of symmetry flux $g_2$, which is $0$ for the former one and $\frac{\pi}{16}$ for the latter.  We also discuss the relations between other states and the two root ones, as (\ref{eqn_z4fz4_phase_structure_1})-(\ref{eqn_z4fz4_phase_structure_3}).
    
%    Even though we have seen in the above results that the nontrivial $Z_2^f$ extension may exclude some interesting anomalous symmetry realization on the edge fields,  there also exsits the  nontrivial bosonic SPT phases which are always nontrivial in fermionic system no matter how the bosonic symmetry $G_b$ is extended by the $Z_2^f$.  We consider this symmetry realization on the edge fields of this kinds of bosonic SPT, which  debed as Type-III anomaly. \\
    
%    (7) 
    \item $G_b=\Z_n\times \Z_n \times \Z_n: $ We find the symmetry realization with the Type-III anomaly on the edge fields of the root state. One way to detect  whether the symmetry realization on the edges is with Type-III anomaly is to check whether one of the symmetry flux carries the projective representation of the left subgroup. We come up with a construction of  the realization of the Type-III anomaly of root phases protected by $(\Z_n)^3$ with central charge $c=n-1$, and illustrating that by the examples of $n=2,3,4,5$ explicitly.
    
    \item $G_f=\Z_4^f\times \Z_4\times \Z_4:$ We construct the edge theory of the intrinsically interacting fSPT root phase that can also  be realized with central charge $c=3$. Fingerprint of this root phase is that the $\frac{\pi}{2}$ flux defect of $\Z_4^f$ carries the fundamental projective representation of the other $\Z_4\times \Z_4$ symmetry.  We also argue that phase with two these root phase stacking is equivalent to the Type-III nontrivial bSPT protected by $\Z_2\times \Z_4\times \Z_4$ symmetry. In this sense, we can call that the root state of intrinsically interacting fSPT  of $\Z_4^f\times \Z_4\times \Z_4$ is the square root of the Type-III nontrivial bSPT protected by $\Z_2\times \Z_4\times \Z_4$ symmetry. 
    \end{enumerate}
      \begin{table*}[t] 
    \centering
    \begin{tabular}{c|c|c|c|c}
    \hline\hline
  \footnotesize{Symmetry} &  \footnotesize{Classification }& \footnotesize{Generators}&\footnotesize{K matrix and  symmetry transformation}& note \\
  \hline
%  $Z_2^f$ & 0 & g & $K=\sigma_z$ &  & \\
%  $Z_ 4^f$ & 0 & g & $K=\sigma_z$ & &\\ 
   $\Z_8^f$ & $\Z_2$ & $g$ & \footnotesize{$K=\sigma_z$, $W^{g}=1_{2\times 2}, \delta \phi^{g}=\frac{\pi}{4}$ $\begin{pmatrix} 1\\-1 \end{pmatrix}$} & \\
   \cline{1-4}
   $\Z_2\times \Z_2^f$ & $\Z_8$ & $g$ & \footnotesize{$K=\sigma_z$, $W^{g}=\sigma_z, \delta \phi^{g}=0$}&\\
    \cline{1-4}
  \multirow{2}{*}{$\Z_4\times \Z_2^f$ }& \multirow{2}{*}{$\Z_8\times \Z_2$} & \multirow{2}{*}{ $(g, P_f)$} &  \footnotesize{$root$ $1$: $K=\sigma_z$,  $W^{g}=1_{2\times 2},$ $\delta \phi^{g} =\frac{\pi}{2} \begin{pmatrix} 0\\1\end{pmatrix}$} & \\
   & & & \footnotesize{$root$ $2$: $K=\sigma_z^{\oplus3}$, $W^{g}=\sigma_z\oplus 1_{4\times 4}, \delta \phi^g = \frac{\pi}{2}  \begin{pmatrix} 0\\0\end{pmatrix}\oplus  \begin{pmatrix} 0\\1\end{pmatrix} \oplus  \begin{pmatrix} 0\\1\end{pmatrix}$} &   \\
%   \hline
%   $Z_6\times Z_2^f$ & & & & \\
    \cline{1-4}
   $\Z_2\times \Z_4^f$ & $\Z_4$ & $(g_1,g_2)$ & \footnotesize{$K=\sigma_z$, $W^{g_1}=W^{g_2}=1_{2\times 2}, \delta \phi^{g_1}=\pi \begin{pmatrix} 0\\1 \end{pmatrix}, \delta \phi^{g_2} =\frac{\pi}{2} \begin{pmatrix} 1\\1 \end{pmatrix} $} &  \\
    \cline{1-4}
  \multirow{2}{*}{ $\Z_4\times \Z_4^f$} & \multirow{2}{*}{ $\Z_2\times \Z_8$ }&  \multirow{2}{*}{$(g_1,g_2)$ }& \footnotesize{ $root $ $1:$ $K=\sigma_z, W^{g_{1,2}}=1_{2\times 2}, \delta \phi^{g_1}=\frac{\pi}{2} \begin{pmatrix} 1\\-1\end{pmatrix}, \delta \phi^{g_2}=\frac{\pi}{2} \begin{pmatrix} 1\\1\end{pmatrix}$} &   \\
  & & & \footnotesize{$root$ $2: $ $K=\sigma_z$, $W^{g_{1,2}}=1_{2\times 2}, \delta \phi^{g_1}=\frac{\pi}{2}\begin{pmatrix} 1\\1\end{pmatrix}, \delta \phi^{g_2}=\frac{\pi}{2} \begin{pmatrix} 0\\1 \end{pmatrix}$} &  \\
    \cline{1-4}
  \multirow{3}{*}{ $\Z_2\times\Z_2\times \Z_2^f$} & \multirow{3}{*}{$(\Z_8)^2 \times \Z_4$} & \multirow{3}{*}{$(g_1,g_2, P_f)$} & \footnotesize{$root\, 1:$ $K=\sigma_z, W^{g_1}=\sigma_z, W^{g_2}=1_{2\times 2}, \delta \phi^{g_1}=0, \delta \phi^{g_2}=0$} &   \\
  & & & \footnotesize{ $root\, 2:$ $K=\sigma_z, W^{g_1}=1_{2\times 2}, W^{g_2}=\sigma_z, \delta \phi^{g_1}=0, \delta \phi^{g_2}=0$} & \\
  & & & \footnotesize{$root\, 3:$ $K=\sigma_z, W^{g_1}=W^{g_2}=-1_{2\times 2}, \delta \phi^{g_1}=\pi\begin{pmatrix} 0 \\1 \end{pmatrix}, \delta \phi^{g_2}=0$} & \\
%  \hline
%  \multicolumn{5}{c}{\multirow{2}{*}{Root of type-III bosonic SPT}} \\
%  \multicolumn{5}{c}{ }  \\
 \cline{1-4}
 \multirow{3}{*}{  } & & \multirow{3}{*}{} & \multirow{3}{*}{ \footnotesize{$type$-$III$} $root$ : $K=\sigma_x, W^{g_1}=-W^{g_2}=-W^{g_3}=1_{2\times 2},$} &  \\
 $\Z_2\times \Z_2\times \Z_2$ & $type$-$III:$ &$(g_1,g_2,g_3)$ &  & \\ 
 & $\Z_2$ & & $ \delta \phi^{g_1}=\pi \begin{pmatrix} 1\\0\end{pmatrix}, \delta \phi^{g_2}=0, \delta \phi^{g_3}=\pi \begin{pmatrix} 0 \\1 \end{pmatrix}$ &  \footnotesize{(below we denote $-i$ by $\bar{i}$)}\\
    \cline{1-4}
 \multirow{3}{*}{  } & & \multirow{3}{*}{} & \multirow{3}{*}{ \footnotesize{$type$-$III$} $root$ : $K=(\sigma_x)^{\oplus 2}, W^{g_1}=1_{4\times 4}, 
 W^{g_2}=A_3, $} &   \\
 $\Z_3\times \Z_3\times \Z_3$ & $type$-$III$:  &$(g_1,g_2,g_3)$ &  &\multirow{3}{*}{\footnotesize{$A_3=\begin{pmatrix} 
 0&0&\bar{1}&0\\  
                                         0 & \bar{1} & 0 & \bar{1}\\
                                         {1} & 0 & \bar{1} & 0\\
                                         0& {1} & 0 & 0  
  \end{pmatrix}$}  }   \\ 
 &$\Z_3$ & & $W^{g_3}=(A_3)^2, \delta \phi^{g_1}=\frac{2\pi}{3} \begin{pmatrix} {1}\\0 \\{-1} \\ 0\end{pmatrix},\delta \phi^{g_2}=0, \delta \phi^{g_3}=\frac{\pi}{3} \begin{pmatrix} 0 \\1 \\0 \\1 \end{pmatrix}$ & \\
  \cline{1-4}
 \multirow{3}{*}{  } & & \multirow{3}{*}{} & \multirow{3}{*}{ \footnotesize{$type$-$III$} $root$ : $K=(\sigma_x)^{\oplus 3}, W^{g_1}=1_{6\times 6}, 
 W^{g_2}=A_4, $} &  \\
 $\Z_4\times \Z_4\times \Z_4$ & $type$-$III:$ &$(g_1,g_2,g_3)$ &  &\multirow{3}{*}{\footnotesize{$A_4=\begin{pmatrix}  
                                           0& 0 &\bar{1}&0&0&0\\  
                                         0 & \bar{1} & 0 & \bar{1} & \bar{1} &0\\
                                         0 & 0 & \bar{1} & 0 & 0& {1}\\
                                         0&0& 0 & 0   & 1&0\\
                                         0& {1}& 0& 0&0 & 0\\
                                         1& 0& \bar{1}& 0&0&0
   \end{pmatrix}$}  }  \\ 
 & $\Z_4$& & $W^{g_3}=(A_4)^3,  \delta \phi^{g_1}=\frac{\pi}{2} {\begin{pmatrix} 1\\0 \\ -1 \\ 0 \\ 0\\2\end{pmatrix}}, \delta \phi^{g_2}=0,\delta \phi^{g_3}=\frac{\pi}{2} {\begin{pmatrix} 0 \\1  \\ 0 \\1 \\ 1\\ 0\end{pmatrix}} $ & \\
 & & & & \\
  \cline{1-4} 
 \multirow{3}{*}{  } & & \multirow{3}{*}{} & \multirow{3}{*}{ \footnotesize{$type$-$III$} $root$ : $K=(\sigma_x)^{\oplus 4}, W^{g_1}=1_{8\times 8}, 
 W^{g_2}=A_5,$} &  \\
 $\Z_5\times \Z_5\times \Z_5$ & $type$-$III:$ &$(g_1,g_2,g_3)$ &  &\multirow{3}{*}{\footnotesize{$A_5=\begin{pmatrix} 
                                         0& 0 &0&1&\bar{1}&0 & 0 & 0\\  
                                         0 & 0 & 1 & 0 & 0 &0 & 0& 0\\
                                         0& 0& 0 & 0&0&0&1&0\\
                                         0 & 0 & 0 & 0 & \bar{1}& 0&0&1\\
                                         1& 0 & 0 & 0   & \bar{1} &0& 0&0\\
                                         0& 1& 0& 0&0 & 0 & 0&0\\
                                         0& \bar{1}& \bar{1}& 0&0&\bar{1} & \bar{1} &0\\
                                         0&0&0&0&\bar{1} & 0 & 0 &0
   \end{pmatrix}$}  }  \\ 
 &$\Z_5$ & & $ W^{g_3}=(A_5)^4, \delta \phi^{g_1}=\frac{2\pi}{5} \begin{pmatrix} 3\\0\\0\\2\\-1\\0\\0\\1\end{pmatrix}, \delta \phi^{g_2}=0, \delta \phi^{g_3}=\frac{2\pi}{5} \begin{pmatrix} 0 \\1 \\1 \\0\\0\\1\\1\\0\end{pmatrix}$ & \\
\cline{1-4}
  \multirow{3}{*}{  } & & \multirow{3}{*}{} & \multirow{3}{*}{ \footnotesize{$intrinsic$} $root$ : $K=\sigma_z\oplus (\sigma_x)^{\oplus 2}, W^{g_1}=1_{6\times 6}, 
 W^{g_2}=\tilde A_4, $} &  \\
 $\Z_4^f\times \Z_4\times \Z_4$ & $intrinsic$ &$(g_1,g_2,g_3)$ &  &\multirow{3}{*}{\footnotesize{$\tilde A_4=\left(
\begin{array}{cccccc}
 1 & 0 & 0 & \bar 2 & 0 & \bar 1 \\
 0 & 1 & 0 & 2 & 0 & \bar 1 \\
 0 & 0 & 0 & 0 & 0 & 1 \\
 1 & \bar 1 & 0 & \bar 2 & 1 & 0 \\
 0 & 0 & 0 & 1 & 0 & 0 \\
 2 & 2 & 1 & 0 & 0 & \bar 2 \\
\end{array}
\right)$}  }  \\ 
 & $\Z_4$& & $W^{g_3}=(\tilde A_4)^3,  \delta \phi^{g_1}=\frac{\pi}{2} {\begin{pmatrix} 1\\1 \\ 0 \\ 0 \\ 0\\0\end{pmatrix}}, \delta \phi^{g_2}=0,\delta \phi^{g_3}=\frac{\pi}{2} {\begin{pmatrix} 2 \\0  \\ 2 \\1 \\ 1\\ 2\end{pmatrix}} $ & \\
 & & & & \\

   \hline\hline
    \end{tabular}
    \caption{Symmetry transformation on edge fields of fSPT root state. For some $G_f$ symmetry, we list all their root states while for $G_b=(\Z_n)^3$, we only list the root state of type-III bosonic SPT which can not be trivialized in fermionic system.  For $G_f=\Z_4^f\times \Z_4\times \Z_4$, we only list the root state for the intrinsically interacting fSPT.}
    \label{mainresult}
\end{table*}

%      \begin{table*}[t] 
%    \centering
%    \begin{tabular}{c|c|c|c|c}
%    \hline\hline
%  {$G_f$} &  classification & generator & K matrix and  symmetry transformation & note \\
% \hline
%   $Z_2\times Z_2\times Z_2$ & & & &  \\
%   \hline
%    $Z_3\times Z_3\times Z_3$ & & & &  \\
%    \hline
%     $Z_4\times Z_4\times Z_4 $ & & & &  \\
%     \hline
%      $Z_5\times Z_5\times Z_5$ & & & &  \\
%   \hline\hline
%    \end{tabular}
%    \caption{Symmetry transformation on edge fields of Type-III bosonic SPT root state.}
%    \label{mainresult}
%\end{table*}

\section{$G_b\times \Z_2^f$ type of symmetry group}
\label{trivial_extension}

Here we consider the examples with $G_f=G_b\times \Z_2^f$. More specifically, we consider carefully three examples: $\Z_2\times \Z_2^f$,  $\Z_4\times \Z_2^f$ and $\Z_2\times \Z_2 \times \Z_2^f$ whose classifiations of SPT  are $\Z_8$, $\Z_8\times \Z_2$ and $(\Z_8)^2\times \Z_2$ respectively.

\subsection{$\Z_2\times \Z_2^f$ symmetry}
\label{sec_trivial_exten_1}

\subsubsection{Symmetry realization}
Here we figure out all  possible symmetry realization with the simplest K matrix i.e., $K=\sigma_z$. 

The generators of this symmetry group are denoted as $g$ and $P_f$  with the group relation as $g^2=1$ and  $P_f^2=1$.   As mentioned above, the parity realizes as (\ref{fermionparity}) and the group relation of $g$ indicates 
\begin{align}
(W^{g})^2=1_{2\times 2}.
\label{group relation}
\end{align} 
Therefore, taking into acount the constraint (\ref{symmetrytransf2}), $W^g$ can take $\pm 1_{2\times 2}$, $\pm \sigma_z$ and we have 
\begin{align}
W_{IJ}^g\delta \phi_J^g+\delta\phi_I^g=0. 
\label{eqn on phi}
\end{align}
where the repeated $J$ is summed. Note that $W^g$ and $\delta \phi^g$  consist the full implementation of symmetry action $g$ in this system. Below we will solve (\ref{eqn on phi})  for $\delta\phi^g$ with different $W^g$.
\begin{enumerate}
\item 
For $W^g=1_{2\times 2}$, 
via
solving  (\ref{eqn on phi}),  we get
$\delta \phi^g=\pi(t_1, t_2)^T$ with  $t_1, t_2=0, 1$ and then we can denote the SPT phases correspongding to  $t_1,t_2$ by $[1_{2\times 2}, t_1,t_2]$. 

\item
For $W^{g}=-1$, the equation 
   (\ref{eqn on phi}) does not impost any constraint on $\delta \phi^g$, therefore $\delta\phi^g=(\theta_1,\theta_2)^T$ with $\theta_{1,2}\in [0,2\pi)$. However,
via  the gauge transformation on $\phi$, we can get 
$\delta \phi^g =0$.
\item
For $W^g=\sigma_{z}$, via
solving (\ref{eqn on phi}), we get $
\delta \phi^g = (n \pi, \theta)^T$ where $\theta \in [0, 2\pi]$ and $n=0,1
$.
Using the gauge transformation, we can shift $\theta=0$. Therefore
\begin{align}
\delta \phi^g =\begin{pmatrix} n \pi\\ 0 \end{pmatrix}, \text{ }  n=0,1
\label{eqn_z2_phi_root_1}
\end{align}
We denote the phases related to $W^g=\sigma_z$ as $[\sigma_z, n]$.
%Here we first study   $[\sigma_z, 0]$. (The other one  $[\sigma_z, 1]$ is discussed in Sec.\ref{sec_z2z2f_groupstru}). 

\item For $W^g=-\sigma_z$, via solving (\ref{eqn on phi}), we get $
\delta \phi^g = ( \theta, n \pi,)^T$ with $\theta\in [0,2\pi)$ and $n=0,1$. Using the gauge transformation, we can shift $\theta=0$. Therefore
\begin{align}
\delta \phi^g =\begin{pmatrix} 0\\ n \pi \end{pmatrix}, \text{ }  n=0,1
\label{eqn_z2_phi_root_2}
\end{align}
We denote the phases related to $W^g=\sigma_z$ as $[-\sigma_z, n]$.
\end{enumerate}

%For the solution with $W^g=1_{2\times 2}$, 
%via
%solving the (\ref{eqn on phi}),  we get
%$\delta \phi^g=\pi(t_1, t_2)^T$ with  $t_1, t_2=0, 1$ and then we can denote the SPT phases correspongding to  $t_1,t_2$ by $[t_1,t_2]$.  In fact, this case was treated  in Ref.\cite{Lu2012} which found that   $[t_1,t_2]=[0,0]$, $[1,1]$ are trivial while $[t_1,t_2]=[0,1]=[1,0]^{-1}$ is topological nontrivial and furthermore only $[1,0]^{\oplus 4}$ is trivial, giving to a $Z_4$ classification. 
%
%However, it is well known that the true classification of $\Z_2\times \Z_2^f$ symmetric SPT is classified by $Z_8$ which implies the solution $[0,1]$ does not correspond to the root phase whose edge  holds two counter-propagating gapless majorana fermion fields.   One may guess that the root state may need $W^g=-1$ or $\pm \sigma$. 
%As for the solution with $W^{g}=-1$, via
%solving equation  (\ref{eqn on phi}),
%and  the gauge transformation on $\phi$, we can get 
%$\delta \phi^g =0$, which indicates that we can 
%symmetrically gap out the edge fields via the symmetric Higgs term
%$\text{cos}(\phi_1+\phi_2)$, which implies this solution is trivial. Then the root state might need $W^g=\pm \sigma_z$. It is a correct conjecture, shown as follows.

% The symmetry realizations corresponding to various $[W^g, \delta\phi^g]$ are related to  solutions for the constaints (\ref{group relation}) and (\ref{eqn on phi}).
  Below we will first show that symmetry realization corresponding to the root for classification and then discuss how  phases related to other symmetry realizations relate to the root one.

\subsubsection{Root phase}
\label{z2z2f_majorana}
Here we show that the root phase is identified as  the phase with $[\sigma_z,0]$, namely $W^{g}=\sigma_z$ and $\delta\phi^g=0$.

%For $W^g=\sigma_{z}$, via
%solving (\ref{eqn on phi}), we get $
%\delta \phi^g = (n \pi, \theta)^T$ where $\theta \in [0, 2\pi]$ and $n=0,1
%$.
%Using the gauge transformation, we can shift $\theta=0$. Therefore
%\begin{align}
%\delta \phi^g =\begin{pmatrix} n \pi\\ 0 \end{pmatrix}, \text{ }  n=0,1
%\label{eqn_z2_phi_root_1}
%\end{align}
%We denote the phases related to $W^g=\sigma_z$ as $[\sigma_z, n]$.
%Here we first study   $[\sigma_z, 0]$. (The other one  $[\sigma_z, 1]$ is discussed in Sec.\ref{sec_z2z2f_groupstru}). 
The physics of the root phase is that its edge  holds two robust counter-propagating gapless majorana fermion fields. 
For $[\sigma_z,0]$, to obtain the  gapless Majorana fermion edge from the Luttinger liquid edge theory (\ref{edge}), we consider the following
 symmetric Higgs terms 
\begin{align}
S_{edge}^1=\sum_l g_l \int dxdt \text{ cos}[l(\phi_1+\phi_2)+\alpha_l] \nonumber \\+\text{cos}[l(\phi_1-\phi_2)+\alpha_l]
\label{sigmaz:symmetrymass}
\end{align}
The two Higgs terms with the same coupling constant is guaranteed by the symmetry action. Since $\phi_1+\phi_2$ and $\phi_1-\phi_2$ do not commute, they can not condense simutaneously. Naively, one might conclude that the edge modes $\phi_1$ and $\phi_2$ remain gapless Luttinger liquid state and propagate in opposite directions. However, things are not so disppointed.  Due to the fact that the coupling constant are always the same for $\text{ cos}[l(\phi_1+\phi_2)+\alpha_l]$ and $\text{cos}[l(\phi_1-\phi_2)+\alpha_l]$, the nonzero $g_l$ can drive the Luttinger liquid to some other nontrivial universality class. For simplicity, we first consider the most relevant case $l=1$. Then the edge theory is
\begin{align}
S_{edge}=&\frac{1}{4\pi}\int dx dt  (-1)^{i-1}\partial_t \phi_i \partial_x \phi_i+ \partial_x \phi_i v_{ij} \partial_x \phi_j \nonumber \\
&+ g_1 \int dxdt \text{ cos}(\phi_1+\phi_2)+\text{cos}(\phi_1-\phi_2)
\label{l=1}
\end{align}
where the repeated $i,j$ are summed and $\alpha_1$ is  absorbed.
To be invariant under symmetry, $v_{12}=v_{21}=0$. Without affecting the symmetry anomaly, we tune  $v_{11}=v_{22}=v$ for convenience.   Under basis transformation $\phi_1=\phi+\theta$, $\phi_2=\phi-\theta$, the edge theory can be quantized to be  
\begin{align}
H=\frac{v}{2\pi}\int dx  (\partial_x \phi)^2+(\partial_x \theta)^2\nonumber \\+ g_1\int dx \text{ cos}(2\phi)+\text{cos}(2\theta).
\label{l=1:2}
\end{align}

It is well-known that these Higgs terms $g_1$ leads the system to lie at the Ising cricality.  To see this,  define  majorana fermion  $\eta_{R,L}^{1,2}$ by
\begin{subequations}
 \label{refermionization}
\begin{align}
&\eta_{R}^1+i\eta_R^2=\frac{1}{\sqrt{\pi}} e^{i(\phi-\theta)} = \frac{1}{\sqrt{\pi}} e^{i\phi_2} \\
&\eta_{L}^1+i\eta_L^2=\frac{1}{\sqrt{\pi}} e^{-i(\phi+\theta)} =\frac{1}{\sqrt{\pi}} e^{-i \phi_1}
\end{align}
\end{subequations}
Recalling that the symmetry $g$ transforms $\phi_{1,2} \rightarrow \pm \phi_{1,2}$,  under $g$ the majorana fermions  transform as
\begin{align}
\eta_L^{1,2}\rightarrow \eta_L^{1,2}, 
 \eta_R^1 \rightarrow \eta_R^1, 
\eta_R^2 \rightarrow -\eta_R^2.
\end{align} 
Under this refermionization,  
\begin{align}
H=\frac{v}{2\pi}\int dx \eta_R^a \partial_x \eta_R^a- \eta_L^a\partial_x \eta_L^a
+ im \eta_R^1 \eta_L^2
\end{align}
where the repeated   $a$ $(=1,2)$ is summed and $m\propto g_1$. The mass term $m$ is symmetric, which would symmetrically gap out  the majorana fermions $\eta_R^1$ and $\eta_L^2$, leaving the effective edge Hamiltonian is
\begin{align}
H_{eff}= \int dx \text{ } \eta_R^2 \partial_x \eta_R^2- \eta_L^1\partial_x \eta_L^1.
\end{align}
As under $\Z_2$ symmetry, they transform as 
\begin{align}\eta_R^2 \rightarrow - \eta_R^2\,,
\eta_L^1 \rightarrow \eta_L^1,
\label{two_gapless_majorana_symmetry}
\end{align} 
the mass term $i \eta_R^2 \eta_L^1$  is not allowed by symmetry, and the gapless majorana edge is robust.   Therefore, this edge belong to a nontrivial state. In fact, this solution is also obtained in Ref.\cite{Levin2018}.

If we put eight copies of this majorana edge theory, we can symmetrically gap out the edge by adding a symmetric interaction (See Ref.\cite{Fidkowski10, ryu_zhang_2012, Gu_levin_2014}).
%\begin{align}
%H_{int}=A(\sum_{a=1}^7 \eta_L^{1, a} \eta_R^{2,a})^2+B(\sum_{a=1}^7 \eta_L^{1, a} \eta_R^{2,a}) \eta_L^{1,8}\eta_R^{2,8}
%\end{align}
Therefore,  eight copies of $[\sigma_z, 0]$ is equivalent to be trivial, namely,
\begin{align}
[\sigma_z, 0]^{\oplus 8}=1.
\end{align}
where we have denoted the trivial phase as $1$.

\subsubsection{Group structure of phases}
\label{sec_z2z2f_groupstru}

Here we will show  how other phases (realized by $K=\sigma_z$ and different $W^g,\delta\phi^g$) relate to the root one. We first discuss the case with $W^g=1_{2\times 2}$ and then discuss another case with $W^g=\sigma_z$. Finally, we will show that the case with $W^g=-1_{2\times 2}$ is always trivial and the case with $W^g=-\sigma_z$ can always be related to those with $W^g=\sigma_z$.

For $W^g=1_{2\times 2}$, 
%via
%solving the (\ref{eqn on phi}),  we get
%$\delta \phi^g=\pi(t_1, t_2)^T$ with  $t_1, t_2=0, 1$ and then we can denote the SPT phases correspongding to  $t_1,t_2$ by $[1_{2\times 2}, t_1,t_2]$. 
 in fact, this case was treated  in Ref.\cite{Lu2012} which found that   $[1_{2\times 2},0,0]$ and $[1_{2\times 2},1,1]$ are trivial while $[1_{2\times 2}, 0,1]=[1_{2\times 2},1,0]^{-1}$ is topological nontrivial and furthermore only $[1_{2\times 2},1,0]^{\oplus 4}$ is trivial, giving to a $\Z_4$ classification. 

%However, it is well known that the true classification of $\Z_2\times \Z_2^f$ symmetric SPT is classified by $Z_8$ which implies the solution $[0,1]$ does not correspond to the root phase whose edge  holds two counter-propagating gapless majorana fermion fields.   One may guess that the root state may need $W^g=-1$ or $\pm \sigma$. 
%As for the solution with $W^{g}=-1$, via
%solving equation  (\ref{eqn on phi}),
%and  the gauge transformation on $\phi$, we can get 
%$\delta \phi^g =0$, which indicates that we can 
%symmetrically gap out the edge fields via the symmetric Higgs term
%$\text{cos}(\phi_1+\phi_2)$, which implies this solution is trivial. Then the root state might need $W^g=\pm \sigma_z$. It is a correct conjecture, shown as follows.

In particular, we are going to show the relation between phase $[1_{2\times 2}, 0,1]$ and phase $[\sigma_z, 0]$ that is missed in Ref.\cite{Lu2012} and \cite{Levin2018}. The relation is 
\begin{align}
[\sigma_z, 0]\oplus [\sigma_z,0]=[1_{2\times 2}, 0,1],
\label{eqn_z2z2f_relation_1}
\end{align}
which is equivalent to  that the stacking system 
\begin{align}
[\sigma_z, 0]\oplus [\sigma_z,0]\oplus [1_{2\times 2}, 1,0]
\label{eqn_z2z2f_relation_1_1}
\end{align}
is trivial.
The edge theory of $[1_{2\times 2}, 1, 0]$ is two-component Luttinger liquid, and can be translated into majorana fermion basis as
\begin{align}
H_{eff}= \int dx \xi_R^a \partial_x \xi_R^a- \xi_L^a\partial_x \xi_L^a
\end{align}
where the repeated $a$ is summed and we have denoted the bosonic edge fields of $[1_{2\times 2}, 1, 0]$ as $\tilde \phi_{1,2}$ and defined $\xi_{R,L}^1+i \xi_{R,L}^2=\frac{1}{\sqrt{\pi}}  e^{i\pm\tilde\phi_{2,1}}$.
For  $[1_{2\times 2}, 1, 0]$,  under symmetry transformation, $\tilde \phi_1 \rightarrow  \tilde \phi_1+\pi$, $\tilde \phi_2 \rightarrow \tilde\phi_2$, which indicates that under symmetry transformation,
\begin{align}
\xi_R^a\rightarrow \xi_R^a\,,
\xi_L^a\rightarrow -\xi_L^a.
\end{align} 
Now we consider stacking system (\ref{eqn_z2z2f_relation_1_1}). We note that the edge fields  of the former two root phases are denoted as $\eta_R^2, \eta_L^1$ and $\chi_R^2, \chi_L^1$ and those of the latter one are denoted as $\xi_R^a, \xi_L^a$, $a=1,2$. In fact, we can symmetrically gap out the edge by the following symmetric mass terms
\begin{align}
i m_1\eta_R^2 \xi_L^1+
im_2\chi_R^2 \xi_L^2+
im_3\eta_L^1 \xi_R^1+ 
im_4\chi_L^2 \xi_R^2.
\end{align}
Therefore, the stacking system (\ref{eqn_z2z2f_relation_1_1}) is trivial and then (\ref{eqn_z2z2f_relation_1}) is proved.

%So far, the relations among phases we have are
%\begin{align}
%&[\sigma_z, 0]^{\oplus 8}=1 \nonumber \\
%&[1_{2\times 2}, 0, 1]^4=1 \nonumber \\
%&[\sigma_z, 0]\oplus [\sigma_z, 0]=[1_{2\times 2}, 0, 1]
%\end{align}
%Therefore, we conclude that $[\sigma_z, 0]$ is the generator and all the phases form a classification by $Z_8$.

Next, we consider the phase $[\sigma_z, 1]$, which is related to the root by 
\begin{align}
[\sigma_z,1]=[\sigma_z,0]^{-1}.
\label{eqn_z2z2f_relation_2}
\end{align}
Following the similar discussion, 
we can get the edge Halmitonian  of $[\sigma_z,1]$ 
\begin{align}
H=\frac{v}{2\pi}\int dx  (\partial_x \phi)^2+(\partial_x \theta)^2\nonumber \\+ g_1\int dx \text{ cos}(2\phi)-\text{cos}(2\theta).
\end{align}
Compare to (\ref{l=1:2}), the minus sign of $\text{cos}(\theta)$ comes from the fact that $\phi_1 \rightarrow -\phi_1 +\pi$ in case $[\sigma_z, 1]$. Using the same refermionization as (\ref{refermionization}), we get (the tilde label is added on the hat for this case to differ from  the case above)
\begin{align}
H_{eff}= \int dx \text{ } \tilde{\eta}_R^1 \partial_x \tilde{\eta}_R^1- \tilde{\eta}_L^2\partial_x \tilde{\eta}_L^2
\end{align}
and 
under symmetry, 
\begin{align}
\tilde{\eta}_R^1  \rightarrow \tilde{\eta}_R^1\,,
 \tilde{\eta}_L^2 \rightarrow -\tilde{\eta}_L^2
 \end{align} 
Therefore, if we stack a $[\sigma_z,0]$ and $[\sigma_z, 1]$, we can add two symmetric mass terms $i \eta_R^2 \tilde{\eta}_L^2$ and $i \eta_L^1 \tilde{\eta}_R^1$ to symmetrically gap out the edge. Therefore, (\ref{eqn_z2z2f_relation_2}) is proved.

As for $W^{g}=-1$,  since
$\delta \phi^g =0$, we can 
symmetrically gap out the edge fields via the symmetric Higgs term
$\text{cos}(\phi_1+\phi_2)$, which implies the phase with $W^g=-1$ is trivial. 

 Finally, the phases with $W^g=-\sigma_{z}$ is related to the root one via the relation
\begin{align}
[-\sigma_z,n]=[\sigma_z,n]^{-1}
\end{align}
 where $n=0,1$. To show this relation, we consider the stacking system $[\sigma_z,n]\oplus [-\sigma_z,n]$ whose bosonic edge fields are denoted by $\phi_{1,2}$ and $\tilde \phi_{1,2}$.  Under symmetry, these bosonic fields transform as 
 \begin{align}
 g: \phi_i\rightarrow \epsilon_{ij}\phi_j+\delta_{1,i}n\pi,\,\tilde\phi_i\rightarrow \epsilon_{ji}\tilde\phi_j+\delta_{2,i}n\pi
 \end{align}
  where the repreated $j$ is summed.
 We can symmetrically fully gap out the edge fields by adding the Higgs terms $\text{cos}(\phi_1+\tilde \phi_2)$ and $\text{cos}(\phi_2+\tilde \phi_1)$. On the other hand,  similar to (\ref{refermionization}), we define majorana fermions $\eta_{R,L}^{1,2}$ and $\tilde\eta_{R,L}^{1,2}$, which transform under symmetry as 
 \begin{subequations}
 \begin{align}
 g:&\,\eta_R^i\rightarrow \epsilon_{ij}\eta_R^j,\,\eta_L^i\rightarrow (-1)^n\eta_L^j,\\
 &\eta_R^i\rightarrow (-1)^n\eta_R^j,\,\eta_L^i\rightarrow \epsilon_{ij}\eta_L^j.
 \end{align}
 \end{subequations}
 where the repreated $j$ is summed.
 We can fully gap out the edge fields by add the following symmetric mass terms
 \begin{align}
 im_{1i}\eta_R^i\tilde\eta_L^i+ im_{2i}\tilde\eta_R^i\eta_L^i
 \end{align}
 where the repreated $i$ is summed.

\subsection{$\Z_4\times \Z_2^f$ symmetry}

\subsubsection{Symmetry realization}
Here we figure out all  possible symmetry realization with the simplest K matrix i.e., $K=\sigma_z$. 

For this symmetry, we have a simple group relation $g^4=1$ where $g$ is the generator of $\Z_4$ subgroup,
 which inidcates that $(W^g)^4=1$.
Besides, $W^g$ also has to satisfy
$(W^g)^TKW^g=K.$ Therefore, $W^g$ can take $\pm 1_{2\times 2}$, $\pm \sigma_z$.
For $\delta \phi^g$, it has to satisfy the  relation
\begin{align}
(W^g)^3\delta \phi^g +(W^g)^2\delta \phi^g +W^g\delta \phi^g +\delta \phi^g =0 \label{z4delphi}
\end{align}
Below we will solve (\ref{eqn on phi})  for $\delta\phi^g$ with different $W^g$.
\begin{enumerate}
\item For
$W^g=1_{2\times 2}$, 
from (\ref{z4delphi}), we have $
4 \delta \phi^g=0 \text{ mod } 2\pi$, which indicates 
\begin{align}
\delta \phi^g= \frac{\pi}{2} \begin{pmatrix}  t_1 \\ t_2\end{pmatrix}, \text{ } t_{1,2}=0,1,2, 3
\end{align}
We denote the phases related to the solution with $W^g=1_{2\times 2}$ and $\delta \phi^g=\pi/2(t_1,t_2)^T$ as $[1_{2\times 2}, t_1,t_2]$.

\item For $W^g=-1_{2\times 2}$, the equation (\ref{z4delphi}) does not have constraint on $\delta\phi^g$, hence $\delta\phi^g=(\theta_1,\theta_2)$ with $\theta_{1,2}\in[0,2\pi)$. However, via gauging transformation, we can shift $\delta\phi^g=0$.

\item  For $W^g=\sigma_z$,
 from (\ref{z4delphi}) and via gauge transformation,  we have
\begin{align}
\delta \phi^g=\frac{2\pi}{4} \begin{pmatrix} t \\0\end{pmatrix}, t=0,1,2,3.
\end{align}
 We denote the phases related to these solutions as $[\sigma_z, t]$.
 
 \item  For $W^g=-\sigma_z$,
 from (\ref{z4delphi}) and via gauge transformation,  we have
\begin{align}
\delta \phi^g=\frac{2\pi}{4} \begin{pmatrix} 0\\t\end{pmatrix}, t=0,1,2,3.
\end{align}
 We denote the phases  related to these solutions as $[-\sigma_z, t]$.

\end{enumerate}

\subsubsection{Root phase for $\Z_8$ classification}
\label{z4z2f_1}
Here we will show that the root phase for the $\Z_8$ classification is $[1_{2\times 2}, 0, 1]$, which in fact hold a  Luttinger liquid  with anomalous symmetry on the edge.

The physics of this root phase is that  in this phase, the topological spin of symmetry flux related to $g$ is $ {\pi}/{16}$ or $-\pi/16$ modulo ${\pi}/{2}$\cite{Chenjie2016}.  The period $\pi/2$ comes from the attaching charge-1 particle to the symmetry flux, which does not affect the symmetry flux content. As in Sec.\ref{proj_rep}, the symmetry flux for $[1_{2\times 2}, 0, 1]$ is represented by $l_g=(0, -1/4)^T$ where $T$ denote the transposition operation. Its topological spin can be computed, that is $\theta_{l_g}=\pi l_g^TK^{-1}l_g=-{\pi}/{16}$.   Furthermore,  stacking eight copies of $[1_{2\times 2}, 0, 1]$, the topological spin becomes $\pi/2$ which is trivial.  Therefore, we indeed can treat the phase $[1_{2\times 2}, 0, 1]$ as the root for $\Z_8$ classification.

To further justify the statement, we study the structure of edge fields straightforwardly. Instead, we use the ingappability  criterion (see Sec.\ref{null_vector}) for assert whether a (stacking) phase is trivial or not. First, we claim the following relation between phases
\begin{align}
&[1_{2\times 2}, 0, 2]^{\oplus2}=1\label{z4102}\\
&[1_{2\times 2}, 0, 2]=[1_{2\times 2}, 0, 1]\oplus [1_{2\times 2}, 1, 2] \label{z4101b}\\
& [1_{2\times 2}, 1, 2]=[1_{2\times 2}, 0, 1]^{\oplus 3}. \label{z4101a}
%&[1_{2\times 2}, 0, 1]^{\oplus 3}\oplus [1_{2\times 2}, 2, 3] =1\label{z4101c}
\end{align}
These phase relations are proved in Sec.\ref{sec_z4z2f_groupstru} where the ingappability  criterion of edge fields are mainly used.
Plugging (\ref{z4101a}) into (\ref{z4101b}), we obtain 
\begin{align}
[1_{2\times 2}, 0, 2]=[1_{2\times 2}, 0, 1]^{\oplus 4},\label{eqn_z2z2f_relation_3}
\end{align} which plugs in (\ref{z4102}) to give that
\begin{align}
[1_{2\times 2}, 0, 1]^{\oplus 8}=1\label{z4101}.
\end{align}
Therefore, we see that indeed eight copies of $[1_{2\times 2}, 0, 1]$ is trivial. 

One question is whether the phase $[1_{2\times 2},0,2]$ is  nontrivial or not.  This can be answered by computing the topological spin of symmetry flux, which turns out to be $\frac{\pi}{4}$ mod $\frac{\pi}{2}$.  Therefore, it is a nontrivial phase.  On the other hand, we can also justify it by checking the symmetric Higgs terms.  Under symmetry, its bosonic edge fields  $\phi_{1,2}$ transform as 
\begin{align}
g: \phi_1 \rightarrow \phi_1,\, \phi_2\rightarrow \phi_2+\pi.
\label{eqn_symmetry_z2z2f}
\end{align}
The lowest order Higgs terms $\text{cos}(\phi_1\pm \phi_2+\alpha_\pm)$ explicitly break the symmetry  transformation (\ref{eqn_symmetry_z2z2f}). The next order Higgs terms  $\text{cos}(2\phi_1\pm 2\phi_2+\alpha_\pm)$ is symmetric under (\ref{eqn_symmetry_z2z2f}) but their condensation both spontaneously break symmetry. This observation implies that the phase $[1_{2\times 2},0,2]$ is indeed nontrivial.  So from (\ref{eqn_z2z2f_relation_3}), the four copies of the root is not trivial, further justify it is indeed a $\Z_8$ root.
%One may wonder whether there exists other interactions beyond the above Higgs terms that can symmetrically gap out the edge fields, like more symmetric Higgs terms that involve other fields of trivial phases that can be stacked to $[1_{2\times 2},0,2]$. However, it is always impossible since we can check the nontrivial topological property in the bulk, that is 

 \subsubsection{Root phase for $\Z_2$ classification}

% To achieve this root state, we consider the solution with $W^g=\sigma_z$.
% From (\ref{z4delphi}) and via gauge transformation,  we have
%\begin{align}
%\delta \phi^g=\frac{2\pi}{4} \begin{pmatrix} t \\0\end{pmatrix}, t=0,1,2,3.
%\end{align}
% We denote the phases(solutions) as $[\sigma_z, t]$.
 
  Here we show that $[\sigma_z, 0]\oplus [1_{2\times 2},0,1]^{\oplus 2}$ can be identified as  the root phase for $\Z_2$ class whose edge can hold odd number of majorana fermions

%First of all, we show that  $[\sigma_z, 0]\oplus [1_{2\times 2},0,1]^{\oplus 2}$ can be treated as the root state for $Z_2$ class whose edge can hold odd number of majorana fermions.
First of all, we study the phase $[\sigma_z,0]$.
For this case, under symmetry, 
$\phi_1 \rightarrow \phi_1 $ and $\phi_2 \rightarrow -\phi_2$, then similar to Sec.\ref{z2z2f_majorana},
 we can add symmetric Higgs terms:
 \begin{align}
g[ \text{cos}(\phi_1+\phi_2)+\text{cos}(\phi_1-\phi_2)] \label{z4mass}
 \end{align}
which together with the free part, will lead to an Ising cricality. To see this conveniently, we use the refermionization trick (\ref{refermionization}) to define the four majorana fermions
%\begin{align}
$\eta_R^1+i \eta_R^2= \frac{1}{\sqrt{\pi}} e^{i\phi_2}$,
$\eta_L^1+i \eta_L^2= \frac{1}{\sqrt{\pi}} e^{-i\phi_1}$.
%\end{align}
 Then (\ref{z4mass}) will becomes 
%\begin{align}
$im \eta_R^1 \eta_L^2$
%\end{align}
 which will gap out the two majorana fermion $\eta_R^1$ and $\eta_L^2$. Therefore, only  $\eta_R^2$ and $\eta_L^1$ remain gapless.  Under symmetry, they transform in the same way as (\ref{two_gapless_majorana_symmetry}, i.e.,
 \begin{align}
 g: \eta_R^2 \rightarrow -\eta_R^2,\, \eta_L^1\rightarrow  \eta_L^1,
 \label{eqn_z2z2_mf_sym}
 \end{align}
 so they are stable against symmetric perturbations, indicating this edge theory is nontrivial. As indicated in  Sec.\ref{z2z2f_majorana}, eight copies of (\ref{eqn_z2z2_mf_sym}) is trivial since we can symmetrica gap out all the edge fields by four-fermion interactions. However, here for the $\Z_4\times \Z_2^f$ fSPT, we have more choice of phases to stack to this phases with majorana  fermion edge fields, it may reduce the number of copies that is necessary to obtain a trivial phase.
 Below we can show that two copies of $ [\sigma_z, 0]$ stacking with some other phases become trivial, that is 
 \begin{align}
 [\sigma_z, 0]^{\oplus2}\oplus [1_{2\times 2}, 2, 0] =1. 
 \end{align}
To show this relation, we denote the majorana fermion for another $[\sigma_z, 0]$ as $\tilde \eta_R^2$ and $\tilde \eta_L^1$ and the two edge boson fields for $[1_{2\times 2}, 2, 0]$ as $\tilde \phi_1$ and $\tilde \phi_2 $ which transform under symmetry 
 \begin{align}
g: \tilde  \phi_1  \rightarrow  \tilde \phi_1+\pi,\,  \tilde \phi_2\rightarrow  \tilde \phi_2  
\end{align}
 Similarly, we define four majorana  fermions from these two boson fields
% \begin{align}
$ \xi_R^1+i \xi_R^2= \frac{1}{\sqrt{\pi}} e^{i\tilde \phi_2},$ 
 $\xi_L^1+i \xi_L^2= \frac{1}{\sqrt{\pi}} e^{-i\tilde\phi_1}$,
%\end{align}
 which transform under symmetry as
  \begin{align}
g:  \xi_R^i\rightarrow \xi_R^i, \,   \xi_L^{i} \rightarrow -\xi_L^i 
\end{align}
where $i=1,2$.
% \begin{align}
%g: \begin{pmatrix} \xi_R^1 \\ \xi_R^2 \\ \xi_L^{1} \\ \xi_L^2 \end{pmatrix} \rightarrow  \begin{pmatrix} \xi_R^1 \\\xi_R^2 \\ -\xi_L^1\\ -\xi_L^2 \end{pmatrix}   
%\end{align}
 Therefore, we can add the symmetric mass terms 
 \begin{align}
 i m_1 \eta_ R^2 \xi_L^1+ im_2 \tilde \eta_R^2 \xi_L^2+ im_3 \xi_R^1 \eta_L^1 +im_4 \xi_R^2\tilde \eta_L^2
 \end{align}
 to full gap out all the edge mode without breaking symmetry.
 Recall that $[1_{2\times 2}, 2, 0]=[1_{2\times 2}, 0, 2]^{-1}=[1_{2\times 2}, 0, 2]=[1_{2\times 2}, 0, 1]^{\oplus 4}$. Therefore, the combination 
 \begin{align}
 [\sigma_z, 0]\oplus [1_{2\times 2}, 0, 1]^{\oplus 2}
 \end{align}
 is the  root state for the $\Z_2$ classification with odd number of majorana fermions at the edge.

\subsubsection{Group structure of phases}
\label{sec_z4z2f_groupstru}

Here we show that the relations between other phases  realized by $K=\sigma_z$ and the two root ones.  
Since the two root phases generate $Z_8\times Z_2$ classification, we use a two component vector $r=(r_1,r_2)$  with $r_1=0,1,2...,7$ and $r_2=0,1$ to denote a certain phase. We coin this  vector of a phases as \textit{structure factor} of phase.  In particular, the  fundamental phases correspond to the basic structure factors
\begin{subequations}
\begin{align}
&r([1_{2\times 2},0,1])=(1,0)\\
&r([\sigma_z, 0]\oplus [1_{2\times 2},0,1]^{\oplus 2})=(0,1)\\
&r([\sigma_z, 0])=(6,1).\label{eqn_z4z2f_phase_relation_i0}
\end{align}
\end{subequations}
Using the strucure factors, the stacking operation becomes the (modular) additive of the three component vector.
%Then the first root state is labeled by $r([1_{2\times 2},0,1])=(1,0)$ and the second root state is $r([\sigma_z, 0]\oplus [1_{2\times 2},0,1]^{\oplus 2})=(0,1)$, and also $r([\sigma_z, 0])=(6,1)$.
Here we illustrate  the following nontrivial relations between some phases and the root ones,
\begin{subequations}
\begin{align}
&r([1_{2\times 2}, 0, 3])=(1,0)\label{eqn_z4z2f_phase_relation_i1}\\
&r([1_{2\times 2}, 0, 2])=(4,0)\label{eqn_z4z2f_phase_relation_i2}\\
&r([1_{2\times 2}, 1, 2])=(3,0)\label{eqn_z4z2f_phase_relation_i3}\\
%&r([1_{2\times 2}, 1, 3])=(0,0)\\
&r([1_{2\times 2}, 2, 3])=(5,0)\label{eqn_z4z2f_phase_relation_i4}\\
&  r([\sigma_z,1])=  r([\sigma_z,3])=(5,1)\label{eqn_z4z2f_phase_relation_i5}\\
&  r([\sigma_z,2])=(2,1)\label{eqn_z4z2f_phase_relation_i6}
%&  r([-\sigma_z,0])=(2,1)\\
%&  r([-\sigma_z,1])= r([-\sigma_z,3])=(3,1)\\
%&  r([-\sigma_z,2])=(6,1)
\end{align}
\end{subequations}
We only illustrate the phase $[1_{2\times 2}, t_1, t_2]$ with $t_1<t_2$ simply due to the relation (\ref{eqn_z4z2f_phase_relation_1}). The phases $[1_{2\times 2}, t, t]$ are trivial since the  Higgs term $\text{cos}(\phi_1-\phi_2)$ can symmetrically gap out their edge fields. We do not illustrate the phases  with $W^{g}=-\sigma_z$  due to (\ref{eqn_z4z2f_phase_relation_2}), namely they can be straightforwardly  related to those with  $W^g=\sigma$.
We also note that for  the case with $W^g=-1_{2\times 2}$, similar to the discussion of the phase with $W^g=-1$ in Sec.\ref{sec_trivial_exten_1},  the phase here with $W^{g}=-1_{2\times 2}$  is also trivial.  

Now we first consider the phases with $W^g=1_{2\times 2}$. 
The first relation  
\begin{align}
[1_{2\times 2}, t_1, t_2]=[1_{2\times 2}, t_2, t_1]^{-1}
\label{eqn_z4z2f_phase_relation_1}
\end{align}
is correct since the edge fields of  stacking system 
$[1_{2\times 2}, t_1, t_2]\oplus [1_{2\times 2}, t_2, t_1]$  can be symmetrically gapped out by adding symmetric Higgs terms $\text{cos}(\phi_1-\phi_4)$ and $\text{cos}(\phi_2-\phi_3)$.
  So we only need to consider the cases with $t_1<t_2$.

  Before proceeding,  we can  show the following simpler relations:
  \begin{subequations}
  \begin{align}
  &[1_{2\times 2}, 0, 3]=[1_{2\times 2}, 0, 1]\label{eqn_z2z2f_relation_4}\\
  &[1_{2\times 2}, 2, 3]=[1_{2\times 2}, 1, 2]^{-1}\label{eqn_z2z2f_relation_5}\\
  &[1_{2\times 2}, 1, 3]=1\label{eqn_z2z2f_relation_6}
  \end{align}
  \end{subequations}
  We note that using these relations together with  (\ref{z4102})-(\ref{eqn_z2z2f_relation_3}) can directly lead to (\ref{eqn_z4z2f_phase_relation_i1})-(\ref{eqn_z4z2f_phase_relation_i4}). In particular, the relation (\ref{eqn_z2z2f_relation_4}) and (\ref{z4101a})  directly leads to (\ref{eqn_z4z2f_phase_relation_i1}) and (\ref{eqn_z4z2f_phase_relation_i3})  respectively while  (\ref{eqn_z2z2f_relation_3}) based on  (\ref{z4102})-(\ref{z4101a})   leads to (\ref{eqn_z4z2f_phase_relation_i2}).

  To show (\ref{eqn_z2z2f_relation_4}) is equivalent to show that  the stacking system $[1_{2\times 2}, 0, 1]\oplus [1_{2\times 2}, 3, 0]$ is trivial, which is correct since  its edge fields  can symmetrically gapped out by Higgs terms $A_1=\text{cos}(\phi_1+\phi_4)$ and $A_2=\text{cos}(\phi_2+\phi_3)$. Similarly, the two Higgs terms $A_1$ and $A_2$ can also symmetrically gap out all the edge fields of the stacking system $[1_{2\times 2}, 1, 2]\oplus [1_{2\times 2}, 2, 3]$, so that the relation  (\ref{eqn_z2z2f_relation_5}) is correct.
%  To show (\ref{eqn_z2z2f_relation_5}),  the edge fields of the stacking system $[1_{2\times 2}, 1, 2]\oplus [1_{2\times 2}, 2, 3]$ can symmetrically gapped out by the same Higgs terms as (1).
 Moreover, (\ref{eqn_z2z2f_relation_6}) is correct since the Higgs term $\text{cos}(\phi_1+\phi_2)$ can symmetrically gap out the edge fields.

%  Therefore, we still only need to consider the two cases: $[1_{2\times 2}, 0, 2]$, $[1_{2\times 2}, 1, 2]$. 
   We now are going to show the relations (\ref{z4102})-(\ref{z4101a}). To show (\ref{z4102}),  we can equivalently consider a stacking system
\begin{align}
[1_{2\times 2}, 0, 2]^{\oplus 2}\oplus [1_{2\times 2}, 1, 3]^{\oplus2} =1 \label{z4102a}
\end{align}
since $[1_{2\times 2}, 1, 3]$ is trivial.  We assume that the two edge fields for the two $[1_{2\times 2}, 0, 2]$ and two $[1_{2\times 2}, 1, 3]$ as $\phi_R^a, \phi_L^a$  and $\tilde\phi_R^a, \tilde\phi_L^a$ $(a=1,2)$.  These fields transform under symmetry as
\begin{subequations}
\begin{align}
g: &\, \phi_R^a \rightarrow    \phi_R^a ,\, \phi_L^a \rightarrow \phi_L^a+\pi \\
&  \tilde \phi_R^a  \rightarrow  \tilde \phi_R^a +\frac{\pi}{2},\, \tilde \phi_L^a  \rightarrow  \tilde \phi_L^a-\frac{\pi}{2}. 
\end{align}
\end{subequations}
%\begin{align}
%g: \begin{pmatrix} \phi_R^a \\ \phi_L^a \end{pmatrix} \rightarrow   \begin{pmatrix} \phi_R^a \\ \phi_L^a+\pi\end{pmatrix} \\
%g: \begin{pmatrix} \tilde \phi_R^a \\ \tilde \phi_L^a \end{pmatrix} \rightarrow   \begin{pmatrix} \tilde \phi_R^a +\frac{\pi}{2}\\ \tilde \phi_L^a-\frac{\pi}{2}\end{pmatrix}
%\end{align}
We can fully gap out these edge fields by the following symmetric Higgs terms
\begin{subequations}
\begin{align}
&\text{cos}(\phi_R^1+\phi_R^2+\tilde \phi_L^1-\tilde \phi_L^2) \nonumber\\
&\text{cos}(\phi_L^1-\phi_L^2+\tilde \phi_R^1-\tilde \phi_R^2) \nonumber\\
&\text{cos}(\phi_R^1+\phi_L^2-\tilde \phi_R^1+\tilde \phi_L^1) \nonumber\\
&\text{cos}(-\phi_L^1+\phi_R^2-\tilde \phi_R^1+\tilde \phi_L^1).\nonumber
\end{align}
\end{subequations}
It can be shown that these Higgs terms  do not lead to spontaneously symmetry breaking, namely they satisfy the so-call null vector criterion in Sec.\ref{null_vector}.
Therefore, we prove (\ref{z4102a}).

%Furthermore, we consider the phase relations among $[1_{2\times 2}, 0,1]$, $[1_{2\times 2}, 0,2]$ and $[1_{2\times 2}, 1,2]$.
%In fact, we can show that 
% \begin{align}
%&[1_{2\times 2}, 0, 1]\oplus [1_{2\times 2}, 1, 2] =[1_{2\times 2}, 0, 2]\label{z4101b}\\
%&[1_{2\times 2}, 0, 1]^{\oplus 3}= [1_{2\times 2}, 1, 2] \label{z4101a}
%\end{align}

%To show (\ref{z4101b}), 

To show (\ref{z4101b}), we can equivalently to show that the stacking system
\begin{align}
[1_{2\times 2}, 0, 1]\oplus [1_{2\times 2}, 1, 2] \oplus [1_{2\times 2}, 2, 0]
\label{eqn_z2z2f_stacking_1}
\end{align}
is trivial. This  can be shown  by adding the following symmetric Higgs terms $
\text{cos}(\phi_R^1-\phi_L^3),
\text{cos}(\phi_R^2-\phi_L^1)$ and 
$\text{cos}(\phi_R^3-\phi_L^2) $
which fully gap out the edge fields withtout breaking symmetry and where $\phi^{1,2,3}_{\alpha}$ $(\alpha=R,L)$ denote the right and left moving fields of $[1_{2\times 2}, 0, 1], [1_{2\times 2}, 1, 2]$ and $[1_{2\times 2}, 2,0]$ respectively.

To prove (\ref{z4101a}), we first 
recall that $[1_{2\times 2}, 1, 2]^{-1}=[1_{2\times 2}, 2, 3]$. Then to prove (\ref{z4101a})  is equivalent to prove 
\begin{align}
[1_{2\times 2}, 0, 1]^{\oplus 3}\oplus [1_{2\times 2}, 2, 3] =1\label{z4101c}
\end{align}
We denote the edge fields for the three $[1_{2\times 2}, 0, 1]$ as $\phi_R^a, \phi_L^a$ $(a=1,2,3)$ and those for $[1_{2\times 2}, 2, 3]$ as $\phi_R^4, \phi_L^4$. Therefore, the following Higgs terms will symmetrically gap out the edge fields wiouth breaking symmetry:
\begin{subequations}
\begin{align}
&\text{cos}(\phi_R^1+\phi_R^2+ \phi_L^3+ \phi_L^4) \nonumber\\
&\text{cos}(\phi_L^1+\phi_L^2+ \phi_R^3+ \phi_R^4) \nonumber\\
&\text{cos}(\phi_R^1-\phi_L^2- \phi_R^3+ \phi_L^3) \nonumber\\
&\text{cos}(-\phi_L^1+\phi_R^2- \phi_R^3+\phi_L^3).\nonumber
\end{align}
\end{subequations}
Therefore, we prove (\ref{z4101a}).

Now we consider the case with $W^g=\sigma_z$.
  We will show the three phase relations 
  \begin{subequations}
  \begin{align}
 & [\sigma_z, 1]\oplus [1_{2\times 2}, 0,1] =[\sigma_z, 0] \label{z4z2fstructure} \\
 &  [\sigma_z, 2] =[\sigma_z, 0]^{-1} \label{z4z2fstructure2} \\
 &  [\sigma_z, 3] \oplus [1_{2\times 2}, 0, 1] =[\sigma_z, 0] \label{z4z2fstructure3}
  \end{align}
  \end{subequations}
  We note that these three relations (\ref{z4z2fstructure})-(\ref{z4z2fstructure3}) directly imply the stucture factors (\ref{eqn_z4z2f_phase_relation_i5})-(\ref{eqn_z4z2f_phase_relation_i6}).
  
To prove (\ref{z4z2fstructure}), we denote the edge fields for  $ [\sigma_z, 1]$ and $ [1_{2\times 2}, 0,1]$ as $\phi_1, \phi_2$ and $\tilde \phi_1, \tilde \phi_2$ respectively which transform under symmetry as
 \begin{subequations}
   \begin{align}
g:&\,   \phi_1\rightarrow  \phi_1+\frac{\pi}{2},\,  \phi_2  \rightarrow   -\phi_2 \\
  & \tilde  \phi_1 \rightarrow  \tilde \phi_1 ,\, \tilde \phi_2  \rightarrow   \tilde \phi_2+ \frac{\pi}{2}. 
%\end{align}
\end{align} 
\end{subequations}
%  \begin{align}
% & g: \begin{pmatrix}   \phi_1 \\  \phi_2 \end{pmatrix} \rightarrow   \begin{pmatrix}  \phi_1+\frac{\pi}{2} \\  -\phi_2\end{pmatrix}  \\
%  &g: \begin{pmatrix} \tilde  \phi_1 \\ \tilde \phi_2 \end{pmatrix} \rightarrow   \begin{pmatrix} \tilde \phi_1 \\ \tilde \phi_2+ \frac{\pi}{2}\end{pmatrix} 
%%\end{align}
%\end{align} 
We can add symmetric Higgs term 
%\begin{align}
$\text{cos}(\phi_1-\tilde \phi_2)
$%\end{align}
to gap out these two fields and then the edge theory is effectively described by the two fields $\tilde \phi_1, \phi_2$ which transform under symmetry as
  \begin{align}
   g:  \tilde  \phi_1 \rightarrow \tilde \phi_1 ,\,  \phi_2\rightarrow  -\phi_2 
%\end{align}
\end{align} 
which is the same transformation as those in $[\sigma_z, 0]$.  Therefore, we have shown the relation (\ref{z4z2fstructure}).  Taking similar steps, we can also show the relation (\ref{z4z2fstructure3}).
% \subsubsection{$[\sigma_z, 2]$}
 
   To show the relation (\ref{z4z2fstructure2}), we stack two phases $  [\sigma_z, 2] \oplus [\sigma_z, 0]$ whose edge fields are denoted as $\phi_1, \phi_2$ and $\tilde \phi_1, \tilde \phi_2$. Under symmetry, they transform as
   \begin{align}
 g:&\,   \phi_i \rightarrow  (-)^{i-1}\phi_i+\delta_{1,i}{\pi},\,
  \tilde  \phi_i\rightarrow (-)^{i-1}\tilde\phi_i
%\end{align}
\end{align} 
Through  refermionization as  (\ref{refermionization})
the Higgs terms
%\begin{align}
$g[\text{cos}(\phi_1+\phi_2)-\text{cos}(\phi_1-\phi_2)] $ and 
$h[\text{cos}(\tilde \phi_1+\tilde\phi_2)+\text{cos}(\tilde \phi_1-\tilde \phi_2)]$ can
%\end{align}
symmetrically gap out half of the majorana fields similarly as (\ref{z4mass}) and leave four majorana fermions being gapless,  which transform under symmetry as
\begin{subequations}
\begin{align}
  g:\,\, & \eta_R^1 \rightarrow \eta_R^1, \, \tilde \eta_R^2\rightarrow  -\tilde \eta_R^2 \\
&  \eta_L^2 \rightarrow -\eta_L^2,\, \tilde \eta_L^1  \rightarrow   \tilde \eta_L^1 
 \end{align} 
 \end{subequations}
%  \begin{align}
% & g: \begin{pmatrix}   \eta_R^1 \\\tilde \eta_R^2\\ \eta_L^2 \\ \tilde \eta_L^1 \end{pmatrix} \rightarrow   \begin{pmatrix}  \eta_R^1 \\-\tilde \eta_R^2\\ -\eta_L^2 \\ \tilde \eta_L^1 \end{pmatrix}  \\
% \end{align} 
Therefore, we can further gap out this four majorana fermions by adding the symmetric mass terms
%\begin{align}
$im \eta_R^1 \tilde \eta_L^1 +i\tilde m \tilde \eta_R^2\eta_L^2
$%  \end{align}
  This indicates that we can fully symmetrically gap out the edge of the phase $  [\sigma_z, 2] \oplus [\sigma_z, 0]$. Therefore, we prove (\ref{z4z2fstructure2}). 
  
%  To summarize, using the two-component vector to denote phases, from (\ref{z4z2fstructure})-(\ref{z4z2fstructure3}), we have
%  \begin{align}
%&  r([\sigma_z,1])=  r([\sigma_z,3])=(5,1)\\
%&  r([\sigma_z,2])=(2,1)
%  \end{align}

 Finally, we consider the case with $W^g=-\sigma_z$. We will show that they  are not independent and can be related to those with $W^g=\sigma_z$.
%  From (\ref{z4delphi}) and gauge transformation, we can get 
%  \begin{align}
%  \delta \phi^g= \frac{2\pi}{4}\begin{pmatrix} 0 \\ k \end{pmatrix}, k=0,1,2,3.
%  \end{align}
  More explicitly, we can show that 
  \begin{align}
   [-\sigma_z, k]\oplus [\sigma_z, k]=1.
   \label{eqn_z4z2f_phase_relation_2}
  \end{align}
Denote the edge fields of this two phase by $\phi_{1,2}$ and $\tilde \phi_{1,2}$ respectively. This can easily be shown by considering the symmetric Higgs terms
%\begin{align}
$\text{cos}(\phi_2 +\tilde \phi_1)+\text{cos}(\phi_1-\tilde \phi_2)
$%\end{align}
which can fully gap out the edge fields of $[-\sigma_z, k]\oplus [\sigma_z,k]$ without breaking symmetry.  From (\ref{eqn_z4z2f_phase_relation_2}), (\ref{eqn_z4z2f_phase_relation_i5}) and (\ref{eqn_z4z2f_phase_relation_i6}), we have the structure factors for the phases with $W^g=-\sigma_z$, that is,
\begin{subequations}
  \begin{align}
&  r([-\sigma_z,0])=(2,1)\\
&  r([-\sigma_z,1])= r([-\sigma_z,3])=(3,1)\\
&  r([-\sigma_z,2])=(6,1).
  \end{align}
  \end{subequations}
%
%To remind that $\tilde \phi_1, \tilde \phi_2$ of $[\sigma, k]$ transform under symmetry as
%
% \begin{align}
%  &g: \begin{pmatrix} \tilde  \phi_1 \\ \tilde \phi_2 \end{pmatrix} \rightarrow   \begin{pmatrix} \tilde \phi_1 +\frac{\pi}{2}k\\ -\tilde \phi_2\end{pmatrix} 
%%\end{align}
%\end{align}

\subsection{$\Z_2\times \Z_2 \times \Z_2^f$ symmetry}

\subsubsection{Symmetry realization}
Here we figure out all  possible symmetry realization with the simplest K matrix i.e., $K=\sigma_z$.

We denote $g_1$ and $g_2$ as the two generators of the two symmetry subgroups which satisfy group relations $g_1^2=g_2^2=1$ and $g_1g_2=g_2g_1$. For convenience, we denote $g_{12}=g_1g_2$.
Note that the symmetry realization $W^{g_i}$ and $\delta \phi^{g_i}$ should satisfy
\begin{subequations}
\begin{align}
&(W^{g_1})^TKW^{g_1}=K\\
&(W^{g_2})^TKW^{g_2}=K\\
&(W^{g_{12}})^TKW^{g_{12}}=K
\end{align}
\end{subequations}
and 
\begin{subequations}
\begin{align}
&(W^{g_1})^2=(W^{g_2})^2=(W^{g_{12}})^2=1_{2\times 2} \\
&(W^{g_1}+1_{2 \times 2})\delta \phi^{g_1}=0 \label{phig1} \\
&(W^{g_2}+1_{2 \times 2})\delta \phi^{g_2}=0 \label{phig2}\\
&(W^{g_1}+ W^{g_2})(\delta \phi^{g_2} + W^{g_1} \delta \phi^{g_1})=0. \label{phig1g2}
\end{align}
\end{subequations}
From these relation and the fact that $W^{g}\in GL(2, \Z)$, we have the following solutions
\begin{subequations}
\begin{align}
W^{g_1}=\pm 1_{2\times 2}, \pm \sigma_z,\\
W^{g_2}=\pm 1_{2\times 2}, \pm \sigma_z.
\end{align}
\end{subequations}
$W^{g_1}$ and $W^{g_2}$ can independently take the four choices of solutions, hence there are in total 16 choices of solutions.  However, some pair of choices are related by exchanging the two $\Z_2$ subgroups. So we only need to consider  ten choices and we explicitly discuss the possible $\delta\phi^{g_i}$ for two choices while others are left in Appendix.\ref{append:z2z2z2d}.
\begin{enumerate}
\item For $W^{g_1}=1_{2\times 2}, W^{g_2}=\sigma_z$, from (\ref{phig1}), we have
\begin{align}
\delta \phi^{g_1}=\pi \begin{pmatrix} t_1^{g_1} \\ t_2^{g_1}
\end{pmatrix}, \quad  t_{1,2}^{g_1}=0,1.
\end{align}
From (\ref{phig2}) and via gauge transformation, we have
\begin{align}
\delta \phi^{g_2}=\pi \begin{pmatrix} t_1^{g_2} \\ 0
\end{pmatrix}, \quad  t_{1}^{g_2}=0,1.
\end{align}
 We use $[1_{2\times 2},\sigma_z,(t_1^{g_1}, t_2^{g_1}), t_1^{g_2}]$ to denote these phases and consider them case by case as follows.

\item For $W^{g_1}=W^{g_2}=-1_{2\times 2}$, we can perform the gauge transformation (\ref{gauge_transformation}) to fix either $\delta\phi^{g_1}$ or $\delta \phi^{g_2}$ to be zero, but the condition (\ref{phig1g2}) prevent to fixing both $\delta\phi^{g_1}$ and $\delta \phi^{g_2}$ to be zero. In other words, when  we gauge fixing  one of the two phases $\delta\phi^{g_1}$ and $\delta \phi^{g_2}$ to be zero, from (\ref{phig1g2}), the other one must be quantized to be multiple of  $\pi$. Here we choose to gauge fix $\delta \phi^{g_2}=0$, then we have 
\begin{align}
\delta \phi^{g_1}=\pi \begin{pmatrix} t_1^{g_1} \\t_2^{g_1} \end{pmatrix}, \quad t_{1,2}^{g_1}=0,1
\end{align}
%From (\ref{phig2}), we have
%\begin{align}
%\delta \phi^{g_2}= \begin{pmatrix} \theta_ 1\\ \theta_2 \end{pmatrix} , \theta_{1,2} \in [0, 2\pi)
%\end{align}
%Using the gauge transformation, we can shift $\delta \phi^{g_2}=0$. 
%It is easy to see that (\ref{phig1g2}) have no constraint on $\delta \phi^{g_{1,2}}$.  
Further $W^{g_{12}}=1_{2\times 2}$ and $\delta\phi^{g_{12}}=\pi ( t_1^{g_1},t_2^{g_1})^T$.
We also use $[-1_{2\times 2},-1_{2\times 2},t_1^{g_1},t_2^{g_1}]$   to denote the phases corresponding to symmetry realization with $W^{g_1}= W^{g_2}=-1_{2\times 2}$.
\end{enumerate}

%Instead of discussing cases by cases, we focus on three specific solutions which can give three root states while other solutions will be discussed in Appendix.\ref{append:z2z2z2d}.

The classification of $(\Z_2)^2\times \Z_2^f$ fSPT in two dimension is $\Z_8\times \Z_8\times \Z_4$. Among various symmetry realizations, we identify that the three root ones for the classification come from realization of $g_1,g_2$ as (1) $W^{g_1}=1_{2\times 2}, W^{g_2}=\sigma_z$, (2)
$W^{g_1}=\sigma_z, W^{g_2}=1_{2\times 2}$ and (3) $W^{g_1}=W^{g_2}=-1_{2\times 2}$. The former two contribute to two $Z_8$ classification and the last one is for $Z_4$ classification.

Below we focus on the symmetry realizations with the above cases of $W^{g_1}$ and $W^{g_2}$ and we identify the root ones and also relate other solution to the root ones.  For other realization of $W^{g_1}$ and $W^{g_2}$ and how they relate to the root ones are discussed in Appendix.\ref{append:z2z2z2d}.

\subsubsection{Two root states for $(\Z_8)^2$ classification}
\label{sec_z2z2z2f_root_z8}

Here we will show that these two root states for two $Z_8$ classification are  $[1_{2\times 2}, \sigma_z, (0,0), 0]$ and $[ \sigma_z,1_{2\times 2}, (0,0), 0]$ and then also discuss how other realizations relate to the root ones.

%First consider  solutions with $W^{g_1}=1_{2\times 2}$ and $W^{g_2}=\sigma_z$.
%From (\ref{phig1}), we have
%\begin{align}
%\delta \phi^{g_1}=\pi \begin{pmatrix} t_1^{g_1} \\ t_2^{g_1}
%\end{pmatrix}, \quad  t_{1,2}^{g_1}=0,1.
%\end{align}
%From (\ref{phig2}) and via gauge transformation, we have
%\begin{align}
%\delta \phi^{g_2}=\pi \begin{pmatrix} t_1^{g_2} \\ 0
%\end{pmatrix}, \quad  t_{1}^{g_2}=0,1.
%\end{align}
%%where we have set $\delta \phi_2^{g_2}=0$ via gauge transformation.
%%The condition (\ref{phig1g2}), we have
%%%\begin{align}
%%$t_1^{g_1}+t_1^{g_2}=0 \text{ mod } 2$, and $t_2^{g_1}=0 \text{ mod }2$. 
%%%\end{align}
%%Therefore 
%%\begin{align}
%%\delta \phi^{g_1}=\delta \phi^{g_2}=\pi\begin{pmatrix} t \\ 0 \end{pmatrix}, t=0,1
%%\end{align}
% We use $[1_{2\times 2},\sigma_z,(t_1^{g_1}, t_2^{g_1}), t_1^{g_2}]$ to denote these phases and consider them case by case as follows.

First of all, we consider the solution $[1_{2\times 2}, \sigma_z,(0,0), 0]$ which can give rise to the root state for one $\Z_8$ classification. 
Under symmetry, 
\begin{subequations}
\begin{align}
&g_1:  \phi_1  \rightarrow    \phi_1,\, \phi_2  \rightarrow     \phi_2  \\
&g_2:  \phi_1 \rightarrow   \phi_1 ,\, \phi_2  \rightarrow   -\phi_2 .
\end{align}
\end{subequations}
This behave as if the theory have only one $Z_2$ symmetry generated by $g_2$. Parallel to the discussion in Sec.\ref{z2z2f_majorana}, we can define the majorana fermion $\eta_{R,L}^{1,2}$ as (\ref{refermionization}), and add symmetric mass term $im \eta^1_R\eta_L^2$ to gap out $\eta_R^1, \eta_L^2$, leaving two gapless majorana fermions $\eta_R^2, \eta_L^1$,  which transform under symmetry as 
\begin{subequations}
\begin{align}
&g_{1}: \eta_R^2\rightarrow  \eta_R^2,\,  \eta_L^{1}  \rightarrow \eta_L^1 \label{Z_4z2f_z8_root_symmetry1} \\
&g_{2}:  \eta_R^2 \rightarrow   -\eta_R^2 ,\, \eta_L^{1} \rightarrow \eta_L^1.  \label{Z_4z2f_z8_root_symmetry2}
\end{align}
\end{subequations}
Therefore we can ignore the first $\Z_2$ symmetry, and only the second $\Z_2$ is nontrivial. As the case of $\Z_2\times \Z_2^f$ in Sec.\ref{z2z2f_majorana}, we conclude that 
\begin{align}
[1_{2\times 2}, \sigma_z, (0,0), 0]^{\oplus 8}= 1.
\label{eqn_z2z2z2f_root_relation_1}
\end{align}
Therefore,  $[1_{2\times 2}, \sigma_z, (0,0),0]$ is the root state for one $Z_8$ classification.

Similarly, for the case $W^{g_1}=\sigma_z$, $W^{g_2}=1_{2\times 2}$, we can denote  phases related to different solutions by $[\sigma_z,1_{2\times 2},t_1^{g_1}, (t_1^{g_2}, t_2^{g_2})]$.   The root phase for another $Z_8$ can be obtained just by exchanging the two $\Z_2$ symmetry subgroups, so $[\sigma_z, 1_{2\times 2}, (0,0), 0]$ is another root state for $\Z_8$ classification.

\subsubsection{Root state for $\Z_4$ classification}

We will show that the root state for  $\Z_4$ classification protected by the whole symmetry can be realized when $W^{g_1}=-1_{2\times 2}, W^{g_2}=-1_{2\times 2}$.  In fact, we identify this root as $[-1_{2\times 2},-1_{2\times 2},0 ,1]$.

Now we show that the phase $[-1_{2\times 2},-1_{2\times 2},0,1]$ is the root  for the $Z_4$ classification protected by the whole symmetry.
Under symmetry,  the bosonic edge fields transform as
\begin{subequations}
\begin{align}
&g_1:  \phi_1 \rightarrow   -\phi_1,\, \phi_2 \rightarrow   -\phi_2+\pi  \\
&g_2:  \phi_1  \rightarrow  -\phi_1, \, \phi_2  \rightarrow   -\phi_2 
\end{align}
\end{subequations}
Taking the refermionization, under symmetry the majorana fermion transform as  
\begin{subequations}
\begin{align}
&g_{1}: \eta_R^i \rightarrow (-1)^i\eta_R^i,\, \eta_L^{i} \rightarrow  (-1)^{i-1} \eta_L^i  \\
&g_{2}:  \eta_R^i \rightarrow (-1)^{i-1} \eta_R^i,\, \eta_L^i \rightarrow (-1)^{i-1} \eta_L^i
\end{align}
\end{subequations}
For a single copy of $[-1_{2\times 2},-1_{2\times 2},0 ,1]$, we can not gap out the edge by adding some symmetric mass terms. However, four copies of them can be symmetrically gapped out by the following interaction\cite{ryu_zhang_2012}:
\begin{align}
H_{int}=A(\sum_{a=1}^7 \eta_L^{ a} \eta_R^{a})^2+B(\sum_{a=1}^7 \eta_L^{a} \eta_R^{a}) \eta_L^{8}\eta_R^{8}
\label{interaction}
\end{align}
or by the following symmetric Higgs terms in terms of chiral bosonic fields
\begin{subequations}
\begin{align}
&\text{cos}(\phi_R^1+\phi_R^2+\phi_L^3+\phi_L^4)\nonumber  \\
&\text{cos}(\phi_R^3+\phi_R^4+\phi_L^1+\phi_L^2)\nonumber \\
&\text{cos}(\phi_R^1+\phi_R^3+\phi_L^1+\phi_L^4)\nonumber \\
&\text{cos}(\phi_R^1+\phi_R^4+\phi_L^1+\phi_L^3).\nonumber
\end{align}
\end{subequations}
where we denote the two bosonic edge fields as $\phi_R^i$ and $\phi^i_L$($R,L$ correspond to different chirality)  for the $i$-th copy of $[1_{2\times 2},-1_{2\times 2},0 ,1]$ in the stacking system,  and $\eta_{R}^{2i-1}+i\eta_{R}^{2i}=\frac{1}{\sqrt{\pi}}  e^{i\pm\phi^i_{R,L}}$.
  Therefore, we have
\begin{align}
[-1_{2\times 2},-1_{2\times 2},0 ,1]^{\oplus 4}=1.
\label{eqn_z2z2z2f_relation_root3}
\end{align}

To further comfirm that this state indeed realizes the root state protected by the whole symmetry, we can show that the representation matrices of $g_1$ and $g_2$ form the projective representation on the fermion parity flux. Following the strategy in Sec.\ref{projective_rep}, the ``fermion parity flux" can be created by operator $e^{i\frac{1}{2}(\phi_1-\phi_2)}\sim e^{i\frac{1}{2}(\phi_2-\phi_1)} $ acting on the vaccum. Then on the doublet of fermion parity flux by $(e^{i\frac{1}{2}(\phi_1-\phi_2)}, e^{i\frac{1}{2}(\phi_2-\phi_1)} )$, the matrix forms of $g_1$ and $g_2$ takes $U_{g_1}=\sigma_y$ and $U_{g_2}=\sigma_x$, therefore they form the projective representation of $\Z_2\times \Z_2$.  The state indeed needs both the two $\Z_2$ subgroup to protect since once either   $\Z_2=\{1,g_1\}$ or $\Z_2=\{1,g_2\}$  is broken, this state is trivial since its phase shift under symmetry can all be gauged fixing to zero and then  be symmetrically gapped out by Higgs term $\text{cos}(\phi_1+\phi_2)$.   We can also find that the ``topological spin" of this symmetry flux corresponding to $g_{12}$ is $\frac{\pi}{4}$ according to Sec.\ref{proj_rep}  indicating that  the four copies of them would become $\pi$ which is trivial in ferminic system.

\subsubsection{Group structure of phases}
\label{sec_z2z2z2f_groupstru}
Here we show that the relations between other phases  realized by $K=\sigma_z$ and the two root ones.  In this subsection, we specially focus on the phases with $W^{g_1}=W^{g_2}=-1_{2\times 2}$,  $W^{g_1}=1_{2\times 2}, W^{g_2}=\sigma_z$ and also $W^{g_1}=\sigma_z, W^{g_2}=1_{2\times 2}$ and others are left in Appendix.\ref{append:z2z2z2d}.  

We will use a three-component structure factor of a phase here,  that is $r=(r_1,r_2,r_3)$, to view the how other phases relate to the root ones. We note that $r_{1,2}$ can take 0,1,2,...,7 modulo 8 and $r_3$ can take 0,1,2,3 modulo 4, where  $r_{1,2}$ label the number of the $Z_8$ root phases, and $r_3$ labels the number of the $Z_4$ root.   In particular, 
\begin{subequations}
\begin{align}
&r([\sigma_z, 1_{2\times 2}, (0,0),0])=(1,0,0)\\
&r([ 1_{2\times 2}, \sigma_z,(0,0),0])=(0,1,0)\\
&r([-1_{2\times 2},-1_{2\times 2}, 0,1])=(0,0,1)
\end{align}
\end{subequations}
%Using the strucure factors, the stacking operation becomes the (modular) additive of the three component vector.
%$r([\sigma_z, 1_{2\times 2}, (0,0),0])=(1,0,0)$, $r([ 1_{2\times 2}, \sigma_z,(0,0),0])=(0,1,0)$ and $r([-1_{2\times 2},-1_{2\times 2}, 0,1])=(0,0,1)$. 
 Here we  illustrate the following nontrivial relations between some phases and the root ones:
 \begin{subequations}
\begin{align}
&r([-1_{2\times 2},-1_{2\times 2},1 ,0])=(0,0,3)\label{eqn_z2z2z2f_phase_relation0}\\
&r([1_{2\times 2}, \sigma_z, (1,1),0])=(0,5,1) \label{eqn_z2z2z2f_phase_relation1}\\
&r([1_{2\times 2}, \sigma_z, (0,1),0] )=(2,5,1) \label{eqn_z2z2z2f_phase_relation2}\\
&r([1_{2\times 2}, \sigma_z, (1,0),0])=(6,5,0)\label{eqn_z2z2z2f_phase_relation3}\\
&r([1_{2\times 2}, \sigma_z, (0,1),1])=(6,3,1) \label{eqn_z2z2z2f_root_relation_5_1} \\
&r([1_{2\times 2}, \sigma_z, (1,0),1])=(6,7,2) \label{eqn_z2z2z2f_phase_relation6}
\end{align}
\end{subequations}
We will show them as follows.
Frist of all, we show the relation  (\ref{eqn_z2z2z2f_phase_relation0}). For this purpose, we consider the following stacking system
 \begin{align}
 [-1_{2\times 2},-1_{2\times 2},1 ,0]\oplus[-1_{2\times 2},-1_{2\times 2},0,1]
 \label{eqn_z2z2z2f_root_relation_00}
 \end{align}
 is trivial since their bosonic edge fields, denoted as $\phi_{1,2}$ and $\tilde\phi_{1,2}$ respectively, can be symmetry fully gapped out by Higgs terms $\text{cos}(\phi_1-\tilde \phi_4)$ and $\text{cos}(\phi_2+\phi_3)$.
So the structure factor of $ [-1_{2\times 2},-1_{2\times 2},1 ,0]$ just is $(0,0,3)$.
We  note that it is easy to see that the solutions $[-1_{2\times 2},-1_{2\times 2},t ,t]$ with $t=0,1$ are trivial since we can symmetrically gap out the edge fields via Higgs terms $\text{cos}(\phi_1+\phi_2)$. 

Now  we come to consider how $[1_{2\times 2}, \sigma_z, (t_1,t_2),t_3]$ relate to root phases. 
%Since the first $\Z_2$ is trivial for this case, 
Simiar to Sec.\ref{z2z2f_majorana}, we can show that 
\begin{subequations}
\begin{align}
&[1_{2\times 2}, \sigma_z, (0,0),1]=[1_{2\times 2}, \sigma_z, (0,0),0]^{-1} ,\label{eqn_z2z2z2f_root_relation_1_1}\\
&[1_{2\times 2}, \sigma_z, (1,1),1]=[1_{2\times 2}, \sigma_z, (1,1),0]^{-1}\label{eqn_z2z2z2f_root_relation_1_1}.
\end{align} 
\end{subequations}

To show (\ref{eqn_z2z2z2f_phase_relation1}), we  prove the following relation
%Then we show that the phase $[1_{2\times 2}, \sigma_z, (1,1),0]$ can be related to $[1_{2\times 2}, \sigma_z, (0,0),0]$ by 
\begin{align}
&[1_{2\times 2}, \sigma_z, (1,1),0] \nonumber \\
=& [1_{2\times 2}, \sigma_z, (0,0),0]^{-1}\nonumber \\
 \oplus &[-1_{2\times 2}, 1_{2\times 2}, (0,1)]^{-1}.
 \label{eqn_z2z2z2f_root_relation_2_11}
\end{align} 
As in Appendix.\ref{append:z2z2z2d}, we can show that the structure factor of  $[ 1_{2\times 2},-1_{2\times 2},  (1,0)]$ is $(2,0,3)$ (i.e.,\ref{eqn_append_z2z2z2f_relation_6}). Just by exchanging the two subgroups, we can obtain the structure factor of $[-1_{2\times 2}, 1_{2\times 2}, (0,1)]$ immediately, that is $(0,2,3)$, so that the structure factor of its inverse is just $(0,6,1)$.  Therefore, the structure factor of $[1_{2\times 2}, \sigma_z, (1,1),0]$ is $(0,-1,0)+(0,6,1)=(0,5,1)$, which   is indeed given by (\ref{eqn_z2z2z2f_phase_relation1}).
%\begin{align}
%r([1_{2\times 2}, \sigma_z, (1,1),0])=(0,5,1).
%%(0,-1,0)+(0,-2,-2).
%\end{align}

The remaining thing to do is to prove (\ref{eqn_z2z2z2f_root_relation_2_11}), which is equivalent to show the  stacking system 
\begin{align}
&[1_{2\times 2}, \sigma_z, (1,1),0] \nonumber \\
\oplus&  [1_{2\times 2}, \sigma_z, (0,0),0]\nonumber \\
 \oplus& [-1_{2\times 2}, 1_{2\times 2}, (0,1)]
 \label{eqn_z2z2z2f_root_relation_2_1}
\end{align} 
is trivial. Then we denote the bosonic edge fields of these three phases by $\phi_{1,2}$, $\varphi_{1,2}$ and $\tilde \varphi_{1,2}$ respectively, which transform under symmetry as 
\begin{subequations}
\begin{align}
g_1:& \phi_i\rightarrow \phi_i+\pi,\, \varphi_{i}\rightarrow \varphi_i, \tilde\varphi_i\rightarrow -\tilde \varphi_i \\
g_2: & \phi_i\rightarrow (-1)^{i-1}\phi_i,\, \varphi_i\rightarrow (-1)^{i-1}\varphi_i,\,\nonumber \\
&\tilde\varphi_i\rightarrow \tilde \varphi_i+\delta_{1,i}\pi
\end{align}
\end{subequations}

Similar to (\ref{refermionization}), we define the majorana fermions $\eta_{R,L}^{1,2}$, $\xi_{R,L}^{1,2}$, $\chi_{R,L}^{1,2}$ using $\phi_{1,2},\varphi_{1,2},\tilde \varphi_{1,2}$ respectively. The edge fields can be fully gapped out by  the following mass terms
\begin{align}
&im_1\eta_R^1\eta_L^1+im_2 \xi_R^1\xi_L^1+im_3\chi_R^1\xi_L^2\nonumber \\
+&im_4\chi_R^2\eta_L^2+im_5\xi_R^2\chi_L^1+im_6 \eta_R^2\chi_L^2.
\end{align}
The symmetry properties of these majorana fermions can be inherited from those of bosonic edge fields, and it turns out that the above mass terms are symmetric. Therefore, the stacking system (\ref{eqn_z2z2z2f_root_relation_2_11}) is trivial.

In a similar way, we can also show (\ref{eqn_z2z2z2f_phase_relation2})-(\ref{eqn_z2z2z2f_phase_relation6}).  More explicitly, the relations (\ref{eqn_z2z2z2f_phase_relation2})-(\ref{eqn_z2z2z2f_phase_relation6}) can be obtained through showing the following stacking systems 
 \begin{align}
 S_1=&[1_{2\times 2}, \sigma_z, (0,1),0] 
\oplus  [1_{2\times 2}, \sigma_z, (1,1),0] \nonumber \\
 \oplus& [1_{2\times 2}, 1_{2\times 2},(1,0), (1,0)]\nonumber
% \label{eqn_z2z2z2f_root_relation_3_1}
\\
  S_2 =&[1_{2\times 2}, \sigma_z, (1,0),0] 
\oplus  [1_{2\times 2}, \sigma_z, (0,0),0]\nonumber \\
 \oplus& [1_{2\times 2}, 1_{2\times 2},(0,1), (1,0)]\nonumber
% \label{eqn_z2z2z2f_root_relation_3_1}
\\
 S_3 =&[1_{2\times 2}, \sigma_z, (0,1),1] 
\oplus  [1_{2\times 2}, \sigma_z, (1,1),0]\nonumber \\
 \oplus& [1_{2\times 2}, 1_{2\times 2},(1,0), (1,1)]\nonumber
% \label{eqn_z2z2z2f_root_relation_5_2}
 \\
S_4=&[1_{2\times 2}, \sigma_z, (1,0),1] 
\oplus  [1_{2\times 2}, \sigma_z, (0,0),0]\nonumber \\
 \oplus& [1_{2\times 2}, 1_{2\times 2},(0,1), (1,1)]\nonumber
% \label{eqn_z2z2z2f_root_relation_9_2}
 \end{align}
are trivial, respectively. Assuming that the stacking phases $S_1$-$S_4$ are trivial,
 we attempt to derive (\ref{eqn_z2z2z2f_phase_relation2})-(\ref{eqn_z2z2z2f_phase_relation6}).  As in Appendix.\ref{append:z2z2z2d}, we show that the structure factor of  $[1_{2\times 2}, 1_{2\times 2},(0,1), (0,1)]$, which is  the inverse $[1_{2\times 2}, 1_{2\times 2},(1,0), (1,0)]$, is $(2,2,2)$ (see \ref{eqn_z2z2z2f_append_relation_root3_1} and \ref{eqn_append_z2z2z2f_relation_100}).   From (\ref{eqn_z2z2z2f_phase_relation1}) that just is proved above, the structure factor of the inverse of $[1_{2\times 2}, \sigma_z, (1,1),0]$ is $(0,3,3)$. Therefore, the structure factor of 
$[1_{2\times 2}, \sigma_z, (0,1),0]$ is $(2,5,1)$, that is indeed the same as (\ref{eqn_z2z2z2f_phase_relation2}).  As for (\ref{eqn_z2z2z2f_phase_relation3}), according to  Appendix.\ref{append:z2z2z2d}(see \ref{eqn_z2z2z2f_append_relation_root3_1} and \ref{eqn_append_z2z2z2f_relation_2_00}), the structure factor of the inverse of $[1_{2\times 2}, 1_{2\times 2},(0,1), (1,1)]$ is $(6,6,0)$, therefore, the structure factor of $[1_{2\times 2}, \sigma_z, (1,0),0]$ is $(0,-1,0)+(6,6,0)=(6,5,0)$. As for (\ref{eqn_z2z2z2f_root_relation_5_1}),  since  the structure factor of the inverse of $[1_{2\times 2}, \sigma_z, (1,1),0]$ is $(0,3,3)$ according to (\ref{eqn_z2z2z2f_phase_relation1}),  and that of the inverse of  $[1_{2\times 2}, 1_{2\times 2},(1,0), (1,1)]$ is $(2,0,2)$ according to (\ref{eqn_z2z2z2f_append_relation_8}) in Appendix.\ref{append:z2z2z2d}, the structure factor of $[1_{2\times 2}, \sigma_z, (0,1),1] $ is $(0,3,3)+(2,0,2)=(2,3,1)$. Finally for (\ref{eqn_z2z2z2f_phase_relation6}), as from (\ref{eqn_z2z2z2f_append_relation_8}), the structure factor of the inverse of $[1_{2\times 2}, 1_{2\times 2},(0,1), (1,1)]$ and $[1_{2\times 2}, \sigma_z, (0,0),0]$ is $(6,0,2)$ and $(0,7,0)$, therefore the structure factor of $[1_{2\times 2}, \sigma_z, (1,0),1]$ is $(6,0,2)+(0,7,0)=(6,7,2)$.

Now we are going to prove that $S_1$-$S_4$ are  trivial. For any $S_i$
we always denote its bosonic edge fields  by $\phi_{1,2}$, $\varphi_{1,2}$ and $\tilde \varphi_{1,2}$ respectively and   we can always define the majorana fermions $\eta_{R,L}^{1,2}$, $\xi_{R,L}^{1,2}$, and $\chi_{R,L}^{1,2}$ in terms of  $\phi_{1,2},\varphi_{1,2}$, and $\tilde \varphi_{1,2}$ respectively, similar to (\ref{refermionization}). We note that the symmetry properties of these bosonic edge fields can be obtained by reviewing the notation of these phase and those of majorana fermions can be inherited from the bosonic ones straightforwardly.

Now we show these four stacking systems can all be fully gapped out without breaking symmetry. In particular, for $S_1$ and $S_2$, all the edge majorana fermions can be fully gapped out by the following symmetric mass terms
\begin{align}
&im_1\eta_R^1\chi_L^2+im_2 \eta_R^2\chi_L^2+im_{3i}\chi_R^i\eta_L^i\nonumber \\
+&im_4\xi_R^1\xi_L^1+im_5\xi_R^2\chi_L^1.
\label{eqn_z2z2z2f_root_relation_3_1_mass}
\end{align}
where the repeated $i$ is summed. On the other hand, for $S_3$ and $S_4$, all the edge majorana fermions can be fully gapped out by the symmetric mass terms
\begin{align}
&im_1\eta_R^1\xi_L^2+im_2 \eta_R^2\chi_L^2+im_{3i}\chi_R^i\eta_L^i\nonumber \\
+&im_4\xi_R^1\xi_L^1+im_5\xi_R^2\chi_L^1.
\label{eqn_z2z2z2f_root_relation_3_1_mass}
\end{align}
where the repeated $i$ is summed.

\section{$G_b\times_{\omega_2} \Z_2^f$ type of symmetry group}

\label{nontrivial_extension}

In this section, we consider the total symmetry  $G_f$ of fermionic system to be nontrivial extension of $G_b$ by $\Z_2^f$. In particular, we consider the examples $\Z_8^f$, $\Z_4^f\times \Z_2$ and $\Z_4^f\times \Z_4$, and fSPT protected by them are classified by $Z_2$, $Z_4$ and $Z_8\times Z_2$, respectively. We also discuss the exmaple $\Z_4^f$ in the Appendix.\ref{z4f}.

%%%
\subsection{$\Z_8^f$ Symmetry}
\label{z8f}
\subsubsection{Symmetry realization}
 In this case, we denote the generator of $\Z_8^f$ is $g$, besides the group relation $g^8=1$, there is one more important group relation  as  $g^4=P_f$, which indicates $(W^g)^4=1$ and 
 \begin{align}
%&(W^g)^4=1\label{z8f_relation_1}\\
%&(W^g)^TK W^g=K \label{z8f_relation_2}\\
&\sum_{i=0}^3 (W^g)^i \delta \phi^g= \pi \begin{pmatrix} 1 \\ 1 \end{pmatrix}  \label{z8f1}
\end{align}
Further, from the constraint (\ref{symmetrytransf2}), we get $W^g= \pm 1_{2\times 2}, \pm \sigma_z$. 
%the conditions on $W^g$ and $\delta \phi^g$:
%From (\ref{z8f_relation_1}) and (\ref{z8f_relation_2}), we get $W^g= \pm 1_{2\times 2}, \pm \sigma_z$.
However, from (\ref{z8f1}), when $W^{g}=-1_{2\times 2}, \pm \sigma_z$, there is no consistent solution for $\delta \phi^g$. Therefore, $W^g$ can only take $1_{2\times 2}$. 
From (\ref{z8f1}), we can get
\begin{align}
4\delta \phi^g=\pi \begin{pmatrix} 1 \\ 1 \end{pmatrix} 
\end{align}
Therefore $\delta\phi^g=\frac{\pi}{4}(2t_1+1,2t_2+1)^T$ mod $2\pi$ where $t_1,t_2=0,1,2,3$.
We denote the phases as $[1_{2\times 2}, (t_1,t_2)]$
where $t_{1,2}=0,1,2,3$.

\subsubsection{Root phase}

 Here
we identify the root  phase to be $[1_{2\times 2}, (0, 1)]$. The physics of the root phase of $\Z_8^f$ fSPT is that  topological  spin of symmetry flux labeled  by $g$ is $\frac{\pi}{8}$ or $-\frac{\pi}{8}$ modulo $\frac{\pi}{4}$\cite{Chenjie2016}. According to Sec.\ref{proj_rep}, we obtain that the fractional vector that represents the $g$ symmetry flux is $l_g=(1/8,-3/8)^T$. Therefore, the topological spin of $g$ symmetry flux can be computed by $\theta_g=\pi l_g^TK^{-1}l_g=-\frac{\pi}{4}$.  So $[1_{2\times 2}, (0, 1)]$ can indeed by identified as  root phase for the $Z_2$ classification.

We can further justify this identification by checking the Higgs terms allowed by symmetry. As for 
$[1_{2\times 2}, (0, 1)]$, under symmetry, the bosonic edge fields $\phi_{1,2}$ transform as
\begin{align}
g:\phi_1\rightarrow \phi_1+\frac{\pi}{4},\,\phi_2\rightarrow \phi_2+\frac{3\pi}{4}.
\end{align}
Therefore, the lowest order Higgs terms $\text{cos}(\phi_1\pm \phi_2)$ explicitly break the symmtry. The lowes order Higgs terms that preserve the symmetry is  $\text{cos}(2\phi_1+ 2\phi_2)$, which, however, leads to the spontaneously symmetry-broken scondensation of $(\phi_1+\phi_2)=0,\pi$.  These simple observation is consistent with the fact the state $[1_{2\times 2},0,1]$ is  the root state for $Z_2$ classification.  

Alternatively, we show that the phase with two root stacking is a trivial phase by showing its bosonic edge fields can be fully gapped out without breaking symmetry.  Equivalently, we   prove that
\begin{align}
[1, (0,1)]^{\oplus 2}\oplus [1, (0,0)] \oplus [1, (1, 1)]=1\label{z8f_phase_relation1}
\end{align}
since the latter two are trivial due to (\ref{eqn_z8f_phase_relation_1}) shown below.
To show this, we denote the edge fields for the above four solutions as
$\phi_{1,2}$, $\tilde{\phi}_{1,2}$, $\phi_{1,2}'$, $\tilde{\phi}'_{1,2}$. We can symmetrically  gap out all the edge fields by the following Higgs terms
\begin{align}
&\text{cos}(\phi_1+\tilde \phi_1+\phi_2'-\tilde \phi_2') \nonumber \\
&\text{cos}(\phi_1+\tilde \phi_2+\phi_2'-\tilde \phi_1') \nonumber \\
&\text{cos}(\phi_2+\tilde \phi_1-\phi_1'-\tilde \phi_2') \nonumber \\
&\text{cos}(\phi_2+\tilde \phi_1+\phi_2'-\tilde \phi_1'). \nonumber 
\end{align}
which satisfy the null vector criterion in Sec.\ref{null_vector}.

\subsubsection{Group structure of phases}
Here we will show how other phases are related to the root phases.
First of all, similar to (\ref{eqn_z4z2f_phase_relation_1}) in  Sec.\ref{sec_z4z2f_groupstru},
it is easy to prove that
\begin{align}
&[1, (t_1,t_2)] \oplus [1, (t_2, t_1)]=1,\\
&[1, (t,t)]=1. \label{eqn_z8f_phase_relation_1}
\end{align}
Therefore, we can only consider $t_1< t_2$, i.e. $(t_1,t_2)=$ (0, 1), (0, 2), (0, 3), (1, 2), (1, 3) and (2, 3). 
%Note that those with $t_1=t_2$ are all trivial. 
The phases $[1,(0,3)]$ and $[1,(1,2)]$ are  also trivial due to the existence of symmetric Higgs terms $\text{cos}(\phi_1+\phi_2)$. 

 Below we show the following phase relations
\begin{align}
%&[1, (0,1)]^{\oplus 2}=1 \label{z8f_phase_relation1}\\
&[1,(0,2)]=[1,(0,1)] \label{z8f_phase_relation2}\\
&[1,(1,3)]=[1,(0,1)]  \label{z8f_phase_relation3}\\
&[1,(2,3)]=[1,(0,1)] \label{z8f_phase_relation4}
\end{align}

%To prove (\ref{z8f_phase_relation1}),
%which is equivalent to  prove that
%\begin{align}
%[1, (0,1)]^{\oplus 2}\oplus [1, (0,0)] \oplus [1, (1, 1)]=1
%\end{align}
%since the latter two are trivial.
%To show this, we denote the edge fields for the above four solutions as
%$\phi_{1,2}$, $\tilde{\phi}_{1,2}$, $\phi_{1,2}'$, $\tilde{\phi}'_{1,2}$. We can symmetrically  gap out all the edge fields by the following Higgs terms
%\begin{align}
%&\text{cos}(\phi_1+\tilde \phi_1+\phi_2'-\tilde \phi_2') \nonumber \\
%&\text{cos}(\phi_1+\tilde \phi_2+\phi_2'-\tilde \phi_1') \nonumber \\
%&\text{cos}(\phi_2+\tilde \phi_1-\phi_1'-\tilde \phi_2') \nonumber \\
%&\text{cos}(\phi_2+\tilde \phi_1+\phi_2'-\tilde \phi_1'). \nonumber 
%\end{align}
%or equivalently, by 
%\begin{align}
%\text{cos}(\phi_1+\tilde \phi_1+\phi_2'-\tilde \phi_2') \nonumber \\
%\text{cos}(\phi_2+\tilde \phi_2-\phi_1'+\tilde \phi_1') \nonumber \\
%\text{cos}(\phi_1+\tilde \phi_2-\phi_1'-\tilde \phi_2') \nonumber \\
%\text{cos}(\phi_1+\tilde \phi_2+\phi_2'+\tilde \phi_1'). 
%\end{align}

To prove (\ref{z8f_phase_relation2}), we can show that the stacking 
\begin{align}
[1,(0,2)]\oplus[1,(0,1)]=1
\end{align}
is trivial where we have used (\ref{z8f_phase_relation1}).
We denote the edge fields of these two solution as $\phi_{1,2}$ and $\tilde \phi_{1,2}$ which under symmetry transform as
 \begin{align}
   g: &   \phi_i  \rightarrow    \phi_i + (5-2i) {\pi }/{4}\\
   &\tilde \phi_i\rightarrow \tilde \phi_i + (5-4i) {\pi }/{4}
\end{align} 
%and
% \begin{align}
%   &g: \begin{pmatrix}   \tilde \phi_1 \\ \tilde \phi_2 \end{pmatrix} \rightarrow   \begin{pmatrix}  \tilde \phi_1 + \frac{\pi}{4}\\ \tilde \phi_2- \frac{3\pi}{4} \end{pmatrix} 
%\end{align} 
Therefore, 
we can symmetrically gap out all the edge fields by adding the Higgs terms
%\begin{align}
$\text{cos}(\phi_1-\tilde \phi_2)$ and 
$\text{cos}(\phi_2+\tilde \phi_1)$. 
%\end{align}
Similarly, we can also show the relations (\ref{z8f_phase_relation3}) and (\ref{z8f_phase_relation4}) is true.

%To summarize, the root state for a $Z_2$ class for $Z_8^f$ is $[1,(0,1)]$.

On the other hand, we can also compute the topological spin of $g$ symmetry flux of various phases, which would also lead to the above relations (\ref{z8f_phase_relation2})-(\ref{z8f_phase_relation2}). For $[1_{2\times 2},(0,2)]$, according to Sec.\ref{proj_rep}, the fractional vector of $g$ symmetry flux is $l_g=(\frac{1}{8},-\frac{5}{8})$, whose  topological  spin can be computed, that is $\pi l_g^TK^{-1}l_g=-\frac{3\pi}{8}$, equivalent to $\frac{\pi}{8}$ mod $\frac{\pi}{4}$. Similarly, the fractional vectors of $g$ symmetry flux corresponding to phases $[1_{2\times 2},(1,3)]$ and $[1_{2\times 2},(2,3)]$ are $(\frac{3}{8},-\frac{7}{8})$ and 
$(\frac{5}{8},-\frac{7}{8})$ respectively. Therefore, the topological spins of $g$ flux are $-\frac{5\pi}{8}$ and $-\frac{3\pi}{8}$ which both are equivalent to $\frac{\pi}{8}$ modulo $\frac{\pi}{4}$.
We remark that, in fact, we can identify any phase in (\ref{z8f_phase_relation1})-(\ref{z8f_phase_relation4}) as the root phase.

\subsection{$\Z_4^f\times \Z_2$ symmetry} 
\label{z4fz2}

\subsubsection{Symmetry realization}
We denote the two generators as $g_1$ and $g_2$, which satisfy
$g_1^2=P_f$ and 
$g_2^2=e$, indicating that
$(W^{g_1})^2=1$ and $
(W^{g_2})^2=1 $. Considering the constraints (\ref{symmetrytransf2}), 
 in general, $W^{g_{1,2}}$ can take $\pm 1$ and $\pm \sigma_z$.
However, $W^{g_1}$ and $\delta \phi^{g_1}$ need to satisfy 
$(1+W^{g_1})\delta \phi^{g_1} =\pi (1, 1)^T$
which constraint that $W^{g_1}$ can only take 1.
Similarly, consider 
$(g_1g_2)^2=P_f$. 
When acting on $\phi$, we have
\begin{align}
(1+W^{g_1g_2}) \delta \phi^{g_1g_2}=\pi(1, 1)^T \label{z4fz21}
\end{align}
Note that $W^{g_1g_2}=W^{g_1}W^{g_2}$, If $W^{g_2}=-1, \pm \sigma_z$, then  $W^{g_1g_2}=-1, \pm \sigma_z$, the equility (\ref{z4fz21}) can not be satisfied.
Therefore, $W^{g_{1,2}}$ can only take 1, and then we have the solutions of $\delta \phi^{g_1,g_2}$ as $\phi^{g_1}=\pi (t_{11}+ \frac{1}{2}, t_{12}+\frac{1}{2})^T$ mod  $2\pi$ and $\phi^{g_2}=\pi (t_{21}, t_{22})^T$  mod $2\pi$,
%\begin{align}
%&\phi^{g_1}=\pi \begin{pmatrix} t_{11}+ \frac{1}{2}\\ t_{12}+\frac{1}{2} \end{pmatrix} \text{ mod } 2\pi \\
%&\phi^{g_2}=\pi \begin{pmatrix} t_{21}\\ t_{22} \end{pmatrix} \text{ mod } 2\pi \end{align}
where $t_{ij}=0,1$. Hence,  we denote different phases by 
% \begin{align}
$ [1,1,(t_{11}, t_{12}), (t_{21}, t_{22})]$.
% \end{align}

\subsubsection{Root phase}
Here we identify the root phase as $[1,1,(0,0),(0,1)]$.
The physics of the root phases of $\Z_4^f\times \Z_2$ fSPT is that the topological spin of symmetry flux labeled by $g_2$
is $\frac{\pi}{4}$ or $-\frac{\pi}{4}$ mod $\pi$ and also $g_2$ symmetry flux  carries half unit  of fermion charge\cite{Chenjie2016}.  In stacking two root phases, the $g_2$ symmetry flux carries integer fermion charge which is trivial, therefore,  from the view of fermion charge, it cannot tell the $\Z_4$ classification.  Therefore, the essential feature of the root phase is the $\pm\frac{\pi}4$ topological spin of $g_2$ symmetry flux.

According to Sec.\ref{proj_rep}, the fractional vector that represents the  $g_2$ symmetry flux is $l_{g_2}=(0,-1/2)^T$ and therefore, its topological spin is $\pi l_{g_2}^TK^{-1}l_{g_2}=-\pi/4$.  Therefore, the phase $[1,1,(0,0),(0,1)]$ is indeed the root one. We can also justify it by checking the symmetric Higgs terms. Under symmetry, the bosonic edge fields $\phi_{1,2}$ of $[1,1,(0,0),(0,1)]$ transform as
 \begin{align}
   &g_1:  \phi_i  \rightarrow \phi_i + \frac{\pi}{2} \nonumber  \\
   &g_2:   \phi_i \rightarrow   \phi_i  +\delta_{2,i}\pi  \label{eqn_z4fz2_symmetry_root}
\end{align} 
Therefore, the lowes order Higgs terms that preserve the symmetry is $\text{cos}(2\phi_1+ 2\phi_2)$, which, however, would lead to spontanesly symmetry breaking.  These simple observation is consistent with the fact the state $[1_{2\times 2},0,1]$ is indeed the root state for $Z_2$ classification.

More rigorously, we  show that the phase  stacking  four copies of root phase can admit symmetric gapped edge. Namely, we prove 
\begin{align}
[1,1,(0,0),(1,0)]^{\oplus 4}=1 \label{z4fz2_phase_relation0}
\end{align}
  Recall that the bosonic edge fields of  the root phase  transform according to (\ref{eqn_z4fz2_symmetry_root}).
 Now we denote the  edge fields for these four root phases by $\phi_{1,2}$,$\tilde \phi_{1,2}$,$\phi_{1,2}'$ and $\tilde \phi_{1,2}'$.
Thereby, we can fully gap out all the edge fields without breaking symmetry by the following symmetric Higgs terms
 \begin{align}
&\text{cos}(\phi_1+\tilde \phi_2+\phi_1'+\tilde \phi_2') \nonumber \\
&\text{cos}(\phi_2+\tilde \phi_1+\phi_1'+\tilde \phi_2') \nonumber \\
&\text{cos}(\phi_1+\tilde \phi_2+\phi_2'+\tilde \phi_1') \nonumber \\
&\text{cos}(\phi_2+\tilde \phi_2+\phi_1'+\tilde \phi_1'),
\end{align}
according to the null vector criterion in Sec.\ref{null_vector}.

\subsubsection{Group structure of phases}

Here we will show that how other phases are related to the root one.
First of all, similar to (\ref{eqn_z4z2f_phase_relation_1}) in  Sec.\ref{sec_z4z2f_groupstru}, 
 it is easy to show that 
$ [1,1,(t_{11}, t_{12}), (t_{21}, t_{22})]\oplus [1,1,(t_{12}, t_{11}), (t_{22}, t_{21})]=1$
 and all solutions with  $t_{21}=t_{22}$ are trivial.
%  Furthermore, solutions $[1,1,(0,1),(0,0)]$ and $[1,1,(0,1),(1,1)]$ are easy to see that they are trivial.
 Therefore, we only need to consider 
 the following solutions: $[1,1,(0,0),(0,1)]$,  $[1,1,(0,1),(0,1)]$, $[1,1,(0,1),(1, 0)]$, $[1,1,(1,1),(0,1)]$.
 
 We will show  the following phase relations 
 \begin{subequations}
 \begin{align}
% &[1,1,(0,0),(1,0)]^{\oplus 4}=1 \label{z4fz2_phase_relation0}\\
&[1,1,(0,1),(0,1)]\oplus [1,1,(0,0),(1,0)]=1 \label{z4fz2_phase_relation1}\\
&[1,1,(0,1),(1,0)]\oplus [1,1,(0,0),(0,1)]=1\label{z4fz2_phase_relation2}\\
&[1,1,(1,1),(0,1)]\oplus [1,1,(0,0),(1,0)]=1\label{z4fz2_phase_relation3}
\end{align}
\end{subequations}
 which tell only $[1,1(0,0),(0,1)]$ is the fundamental root state.
 
%  First, we show  (\ref{z4fz2_phase_relation0}). In view point of topological spin and fermion charge of $g_2$ symmetry flux as discussed above, stacking four root phases will indeed lead to a trivial phase (i.e., phase with both trivial topological spin and fermion charge). Alternatively, we show that we can symmetrically gap out all the edge fields of the stacking phases. Recall that the bosonic edge fields of  the root phase  transform according to (\ref{eqn_z4fz2_symmetry_root}).
% Now we denote the  edge fields for these four root phases by $\phi_{1,2}$,$\tilde \phi_{1,2}$,$\phi_{1,2}'$ and $\tilde \phi_{1,2}'$.
%Thereby, we can fully gap out all the edge fields without breaking symmetry by the following symmetric Higgs terms
% \begin{align}
%&\text{cos}(\phi_1+\tilde \phi_2+\phi_1'+\tilde \phi_2') \nonumber \\
%&\text{cos}(\phi_2+\tilde \phi_1+\phi_1'+\tilde \phi_2') \nonumber \\
%&\text{cos}(\phi_1+\tilde \phi_2+\phi_2'+\tilde \phi_1') \nonumber \\
%&\text{cos}(\phi_2+\tilde \phi_2+\phi_1'+\tilde \phi_1'),
%\end{align}
%according to the null vector criterion in Sec.\ref{null_vector}.

To show (\ref{z4fz2_phase_relation1}), we denote the edge fields as $\phi_{1,2}$ and $\tilde\phi_{1,2}$ of 
$[1,1,(0,1),(0,1)]$ and $ [1,1,(0,0),(1,0)]$, respectively. Then we can fully gap out the edge field without breaking symmetry by the Higgs terms
$\text{cos}(\phi_1-\tilde \phi_2)$ and
$\text{cos}(\phi_2+\tilde \phi_1)$.
Similarly, we can also show (\ref{z4fz2_phase_relation2}) and (\ref{z4fz2_phase_relation3}).

%%%%%%%%%%%%%%%%
\subsection{$\Z_4^f\times \Z_4$ symmetry}
\subsubsection{Symmetry realization}
We denote the generators as $g_1$ and $g_2$ which satisfy
$g_1^2=P_f$ and
$g_2^4=e $,  so that
$
(W^{g_1})^2=1$ and $
(W^{g_2})^4=1$.
Considering the constraint (\ref{symmetrytransf2}), $W^{g_{1,2}}$ may take $\pm 1$ and $\pm \sigma_z$. 
Besides, we also need to consider $\delta \phi^{g_{1,2}}$, which, via the above group relations, have to satisfy
\begin{align}
(1+W^{g_1})\delta \phi^{g_1}=\pi (1,1)^T
\label{eqn_z4fz4_symmetry_constraint_relation_1}
\end{align} and 
\begin{align}
\sum_{i=0}^3( W^{g_2})^i\delta \phi^{g_2}=0.
\label{eqn_z4fz4_symmetry_constraint_relation_2}
\end{align}
Then we can see that $W^{g_1}$ can only take 1. Below, we can also show that $W^{g_2}$ can only take 1.
For this purpose, we consider another group relation $
(g_1g_2)^2=g_2^2 P_f$.
Acting on $\phi^{g_i}$, we get
$(1+W^{g_1g_2}) \delta \phi^{g_1g_2}=\pi (1, 1)^T  +\phi^{g_2g_2}$ and it can be simplified to be 
\begin{align}
(1+W^{g_2})\delta \phi^{g_1}=\pi (1, 1)^T
\end{align}
where we have used 
 the relations $\delta \phi^{g_1g_2}=\delta \phi^{g_1} + W^{g_1}\delta \phi^{g_2}$ and 
$\delta \phi^{g_2g_2}=\delta \phi^{g_2} +  W^{g_2}\delta \phi^{g_2}$. Therefore, Similar to Sec.\ref{z8f} and Sec.\ref{z4fz2}, $W^{g_2}$ can also only take $1$.

Then from (\ref{eqn_z4fz4_symmetry_constraint_relation_1}) and (\ref{eqn_z4fz4_symmetry_constraint_relation_2}), the allowed values of $\delta \phi^{g_{1,2}}$ take
%\begin{align}
$\delta \phi^{g_1}=\pi (t_{11}+\frac{1}{2}, t_{12}+\frac{1}{2})^T$  mod  $2\pi$ and
$\delta \phi^{g_2}=\frac{\pi}{2} ( t_{21}, t_{22})^T$ mod  $2\pi$
%\end{align}
where $t_{11}$ and $t_{12}$ can take 0, 1 while $t_{21}$ and $t_{22}$ can take 0, 1, 2 and  3.
Therefore, there are in total $2^2 \times 4^2=64$ different solutons, which we denote as $[1,1,(t_{11}, t_{12}), (t_{21}, t_{22})]$.

\subsubsection{ Root phase for  $\Z_2$ classification}

The first  root phase which generates a $\Z_2$ classification is  identified as $[1,1,(0,1), (1, 1)]$.
The  physics of the $\Z_2$ root phase is that the symmetry flux $g_2$ carries one fourth unit of fermion charge, which means that the mutual statistics between the symmetry flux $g_1$ and symmetry flux $g_2$ is $\pm \frac{\pi}{4}$ modulo $\frac{\pi}{2}$\cite{Chenjie2016}.  Meanwhile, the topological spins of both symmetry flux $g_1$ and  $g_2$ are zero modulo $\pi$ and $\frac{\pi}{2}$, respetively\cite{Chenjie2016}. The modular phase factors can be understood as the charge attachment to the symmetry fluxes. According to Sec.\ref{proj_rep}, the fractional vectors corresponding to symmetry flux $g_1$ and $g_2$ are $l_{g_1}=(\frac{1}{4},\frac{1}{4})$ and $l_{g_2}=(\frac{1}{4},-\frac{1}{4})$. Therefore we can compute that the topological spins of the two symmetry flux $g_1, g_2$ are $\pi l_{g_1}^TK^{-1}l_{g_1}=\pi l_{g_2}^TK^{-1}l_{g_2}=0$ and the mutual statistics between these fluxes is $2\pi l_{g_1}^TK^{-1}l_{g_2}=\frac{\pi}{4}$.

We now check the symmetric Higgs terms. Under symmetry, the bosonic edge fields $\phi_{1,2}$ of $[1,1,(0,1),(1,1)]$ transform as
\begin{subequations}
 \label{eqn_z4fz2_symmetry_root}
 \begin{align}
   &g_1:  \phi_i  \rightarrow \phi_i -(-1)^i{\pi}/{2}   \\
   &g_2:   \phi_i \rightarrow   \phi_i  +\delta_{2,i}\pi 
\end{align} 
\end{subequations}
Therefore, the lowes order Higgs terms that preserve the symmetry is $\text{cos}(2\phi_1+ 2\phi_2)$, which, however, would lead to spontanesly symmetry breaking.  These simple observation is consistent with the fact the state $[1,1,(0,1),(1,1)]$ is indeed the root state for $Z_2$ classification.

Furthermore, we show that the edge fields of the phase that is a stacking of two root phases can be fully gapped out without breaking symmetry. For convenience, we stack two trivial phases to it, namely, we will show
\begin{align}
[1,1,(0,1),(1,1)]^{\oplus 2} \oplus [1,1,(0,1),(0,0)]^{\oplus 2}=1\label{z4fz4_phase_relation1}
\end{align}
We denote the edge fields for these for solutions as $\phi_{1,2}$, $\tilde \phi_{1,2}$,  $\phi_{1,2}'$ and $\tilde \phi_{1,2}'$. 
We can symmetrically gap out all the edge fields by 
\begin{align}
\text{cos}(\phi_1-\tilde \phi_1+\phi_2'-\tilde \phi_2') \nonumber \\
\text{cos}(\phi_2-\tilde \phi_2+\phi_1'-\tilde \phi_1') \nonumber \\
\text{cos}(\phi_1-\tilde \phi_2+\phi_1'-\tilde \phi_2') \nonumber \\
\text{cos}(\phi_1-\tilde \phi_2+\phi_2'-\tilde \phi_1') \nonumber
\end{align}
according to the null vector criterion in Sec.\ref{null_vector}.
Therefore, we prove (\ref{z4fz4_phase_relation1}), which $[1,1,(0,1),(1,1)]$  generates a $\Z_2$ classification.

\subsubsection{Root phase for $\Z_8$ classification}

Next we consider another root state for $\Z_8$ classification, that is identified as $[1,1,(0,0), (0,1)]$. 
The physics of this root state is that the topological spin of the symmetry flux $g_2$ is $\pm \frac{\pi}{16}$ modulo $\frac{\pi}{2}$, and also the mutual statistics between the symmetry fluxes $g_1$ and $g_2$ is $\pm \frac{\pi}{8}$ modulo $\frac{\pi}{2}$\cite{Chenjie2016}. According to Sec.\ref{proj_rep}, the fractional vectors corresponding to symmetry flux $g_1$ and $g_2$ are $l_{g_1}=(\frac{1}{4},-\frac{1}{4})$ and $l_{g_2}=(0,-\frac{1}{4})$, thereby the topological  spin of  symmetry flux $g_2$ is $\pi l_{g_2}^TK^{-1}l_{g_2}=-\frac{\pi}{16}$ and the mutual statistics between $g_1$ and $g_2$ fluxes is $2\pi l_{g_1}^TK^{-1}l_{g_2}=-\frac{\pi}{8}$. Therefore, this phase is indeed the root phase for $\Z_8$ classification.

More  straightforwadly, we will show  that all the edge fields of the phase stacking  the  eight copies of the root phases can all be gapped out without symmetry breaking.
 To show this, we note the following phase relations
 \begin{subequations}
\begin{align}
&[1,1,(0,0), (0,2)]^{\oplus 2} = 1\label{z4z2f_phase_relation4}\\
&[1, 1, (0,0), (3,2)]= [1,1,(0,0), (0,1)]^{\oplus 3} \label{z4z2f_phase_relation3}\\
&[1,1,(0,0),(0,2)]=[1,1,(0,0), (0,1)]\oplus [1, 1, (0,0), (1,2)] \label{z4z2f_phase_relation2}\\
&[1,1,(0,0), (1,2)]= [1,1,(0,0), (3, 2)] \oplus [1,1, (0,1), (1,1)] \label{z4z2f_phase_relation5} 
%&[1,1,(0,0), (0,1)]\oplus [1,1, (0,0), (1,2)] = [1,1, (0,0), (0,2)] \label{z4z2f_phase_relation6}
\end{align}
 \end{subequations}
 via studying their edge theories, which are shown in Sec.\ref{sec_z4fz4_stru}.
Therefore, we can achieve the conclusion that 
\begin{align}
[1,1,(0,0),(0,1)]^{\oplus 8}=1
\end{align}
where we have used the fact that $[1,1, (0,1), (1,1)]$ is a $\Z_2$ root phase.

%To justify that the phase $[1,1,(0,1), (1, 1)]$ and $[1,1,(0,0), (0, 1)]$ are indeed two different root states, it is sufficient to show that one topological invariant in this two theory is indeed different. The topological invariant defines as $\Theta_{g_2}=\frac{\pi}{4} q^T_{g_2}\cdot K^{-1}\cdot q_{g_2}$ mod $2\pi$ where $q_{g_2}=\frac{4}{2\pi}\delta \phi^{g_2}$.  This topological invariant is related to the topological spin of the $g_2$ flux upon gauging.  For $[1,1,(0,1), (1, 1)]$, $\Theta_{g_2}=0$ while for $[1,1,(0,0), (0, 1)]$, $\Theta_{g_2}=-\frac{\pi}{4}$. According to Ref.\cite{Chenjie2016}, they are indeed different root state. One byproduct of this invariant is that the only stacking eight copies of $[1,1,(0,0), (0, 1)]$ will give a trivial value of this invariant, further comfirming that $[1,1,(0,0),(0,1)]$ generates a $Z_8$ classification.

\subsubsection{Group structure of phases}
\label{sec_z4fz4_stru}

Here we will show how other phases  are related to the root ones.  Similar to Sec.\ref{sec_z4z2f_groupstru} and Sec.\ref{sec_z2z2z2f_groupstru}, we use a two-component structure factor $r=(r_1,r_2)$ to manifest the structure of a phase, 
which means the phase contain $r_1$-copy of $[1,1,(0,1),(1,1)]$ and $r_1$-copy of $[1,1,(0,0),(0,1)]$. 
 In particular, the  fundamental phases correspond to the basic structure factors
 \begin{subequations}
\begin{align}
&r([1,1,(0,1),(1,1)])=(1,0)\\
&r([1,1,(0,0),(0,1)])=(0,1)
%&r([\sigma_z, 0])=(6,1).
\end{align}
 \end{subequations}

First of all, similar to (\ref{eqn_z4z2f_phase_relation_1}) in  Sec.\ref{sec_z4z2f_groupstru},
it is easy to prove that  
%\begin{align}
 \begin{subequations}
\begin{align}
   &  [1,1, (t_{1}, t_{1}), (t_{2}, t_{2})] =1,\\
      & [1,1, (t_{11}, t_{12}), (t_{21}, t_{22})]  \oplus  [1,1, (t_{12}, t_{11}), (t_{22}, t_{21})] = 1.
    \end{align}
     \end{subequations}
%\end{align}
Therefore, we only need to consider the cases with (1) $t_{11}< t_{12}$, and (2) $t_{11}=t_{12}$ and $t_{21}<t_{22}$, which in total contain 28 different cases, including 16 cases with $(t_{11},t_{12})=(0,1)$,  6 cases with $(t_{11},t_{12})=(0,0)$ and 6 cases with $(t_{11},t_{12})=(1,1)$.

Now we claim the following relations
 \begin{subequations}
\begin{align}
&r([1,1,(0,0),(t_1,t_2)])=(\frac{(A-1)B}{2},-AB)
\label{eqn_z4fz4_phase_structure_1}\\
&r([1,1,(1,1),(t_1,t_2)])=(\frac{(A+1)B}{2},-AB)\label{eqn_z4fz4_phase_structure_2}\\
&r([1,1,(0,1),(t_1,t_2)])=(\frac{A(B-1)}{2},-AB)\label{eqn_z4fz4_phase_structure_3}
\end{align}
 \end{subequations}
where $A=t_1+t_2, B=t_1-t_2$. Instead of enumerate all the 28 cases, here we utilize the physics of topological spin and mutual statistics between two different symmetry fluxes to prove these three relations for general cases. In particular, we show these relation for some special cases by studying their edge theories. 

Now we are going to prove (\ref{eqn_z4fz4_phase_structure_1})-(\ref{eqn_z4fz4_phase_structure_3}). To unify to proof for  (\ref{eqn_z4fz4_phase_structure_1})-(\ref{eqn_z4fz4_phase_structure_3}), we first denote the phase generally by $[1,1,(s_1,s_2),(t_1,t_2)]$.  According to Sec.\ref{proj_rep}, the fractional vectors of $g_1$ and $g_2$ fluxes are $l_{g_1}=\frac{1}{4}(1+2s_1,-1-2s_2)$ and $l_{g_1}=\frac{1}{4}(t_1,-t_2)$.  Thereby, the topological spin of $g_2$ flux is $\pi l_{g_2}^TK^{-1}l_{g_2}=\frac{\pi}{16}(t_1^2-t_2^2)$ and the mutual statistics between the two fluxes $g_1,g_2$ is $2\pi l_{g_1}^TK^{-1}l_{g_2}=\frac{\pi}{8}(t_1-t_2+2t_1s_1-2t_2s_2)$. Now we assume that the structure factor of the general phase by $r=(r_1,r_2)$. We recall that for the first root state $r=(1,0)$, the topological spin  of $g_2$ flux is zero modulo $\frac{\pi}{2}$ and the mutual statistics between  the two fluxes $g_1$ and $g_2$ is -$\frac{\pi}{4}$ modulo $\frac{\pi}{2}$ while for the second root state $r=(0,1)$, the topological spin  of $g_2$ flux is $-\frac{\pi}{16}$ modulo $\frac{\pi}{2}$ and the mutual statistics between  the two fluxes $g_1$ and $g_2$ is -$\frac{\pi}{8}$ modulo $\frac{\pi}{2}$. Therefore, for phase $[1,1,(s_1,s_2),(t_1,t_2)]$ whose structure factor is assumed to be $r=(r_1,r_2)$, we should have the following  equations
\begin{align}
-r_1&=t_1^2-t_2^2 \text{ mod } 8\\
-r_1-2r_2&=(t_1-t_2)+2(t_1s_1-t_2s_2) \text{ mod } 4.
\end{align}
By solving these equations, we have
\begin{align}
r_1&=t_2^2-t_1^2\text{ mod }8\\
r_2&=\frac{(t_1+t_2-1)(t_1-t_2)}{2}+(t_1s_1-t_2s_2)
\end{align}
More explicitly, when $(s_1,s_2)=(0,0)$, we obtain (\ref{eqn_z4fz4_phase_structure_1}); when  $(s_1,s_2)=(1,1)$, we obtain (\ref{eqn_z4fz4_phase_structure_2}); when $(s_1,s_2)=(0,1)$, we obtain (\ref{eqn_z4fz4_phase_structure_3}).

Alternatively, now we study the structure of phases  for several examples by studying their edge fields.
First, we study the relations (\ref{z4z2f_phase_relation4})-(\ref{z4z2f_phase_relation5}).

%Before proceding, we note that the solution $[1,1,(0,1),(0,0)]$ is trivial since we can symmetrically gap out the edge fields by $\text{cos}(\phi_1+\phi_2)$. 
%
%Note that it is easy to show that
%%\begin{align}
%$[1,1,(0,0), (3,1)]=[1,1,(0,1), (1,1)]$,
%%\end{align}
%therefore we can also take $[1,1,(0,0), (3,1)]$ as the $Z_2$ root state.
%
%
%It is very easy to show that
%%\begin{align}
%$[1,1, (t_{11}, t_{12}), (t_{21}, t_{22})] $$ \oplus  [1,1, (t_{12}, t_{11}), (t_{22}, t_{21})] = 1$.
%%\end{align}
%Therefore, we only need to consider the cases where $t_{11}< t_{12}$ and the cases  where $t_{11}=t_{12}$ and $t_{21}<t_{22}$(note that the cases where  $t_{11}=t_{12}$ and $t_{21}=t_{22}$ are trivial.) Even taking into these consideration, there are still  28 choices of solutions left. Therefore, instead of discussing them case by case, we here only show the two root states while others are left in the Appendix.\ref{appendix_z4fz4}
% For the last two relations, their bosonic edge fields can be symmetrically gapped out similarly to (\ref{z4101b}) and (\ref{z4101a}). 
For proof of (\ref{z4z2f_phase_relation4}), we first note that the phase $[1,1,(1,0), (1,3)]$ is trivial since its edge fields can be fully gapped out without breaking symmetry by Higgs term $\text{cos}(\phi_1+\phi_2)$. Thereby to prove (\ref{z4z2f_phase_relation4}) can be equivelant to prove 
\begin{align}
[1,1,(0,0), (0,2)]^{\oplus 2}\oplus [1,1,(1,0), (1,3)]^{\oplus 2} = 1.\label{z4z2f_phase_relation41} 
\end{align}
 We use $\phi_{1,2}$ and $\tilde \phi_{1,2}$ to denote the edge fields of two $[1,1,(0,0),(0,2)]$ respectively and similarly   $\phi'_{1,2}$ and $\tilde \phi'_{1,2}$ for two $[1,1,(1,0),(1,3)]$ respcetively.
All the edge fields can be symmetrically gapped out by the following Higgs terms
\begin{align}
\text{cos}(\phi_1-\tilde \phi_1+\phi_2'-\tilde \phi_2') \nonumber \\
\text{cos}(\phi_2-\tilde \phi_2+\phi_1'-\tilde \phi_1') \nonumber \\
\text{cos}(\phi_1+ \phi_2+\phi_1'- \tilde \phi_2') \nonumber \\
\text{cos}(\phi_1+ \phi_2+\phi_2'- \phi_1') \nonumber
\end{align}
according to the null vector criterion in Sec.\ref{null_vector}.

 To prove (\ref{z4z2f_phase_relation3})  is equivalent to prove 
\begin{align}
 [1,1,(0,0),( 3, 2)]\oplus[1,1,(0,0), (1, 0)]^{\oplus 3} =1\label{z4101c}
\end{align}
We denote the edge fields for $[1_{2\times 2}, (0,0), (3,2)]$ as $\phi_R^1, \phi_L^1$ and those for the three $[1_{2\times 2}, (0,0), (1,1)]$ as $\phi_R^a, \phi_L^a$ $(a=2,3,4)$. Therefore, the following Higgs terms will symmetrically gap out the edge fields wiouth breaking symmetry:
\begin{align}
&\text{cos}(\phi_L^1+\phi_L^2+ \phi_R^3+ \phi_R^4) \nonumber\\
&\text{cos}(\phi_R^1+\phi_R^2+ \phi_L^3+ \phi_L^4) \nonumber\\
&\text{cos}(\phi_L^1+\phi_R^2+ \phi_R^3+ \phi_L^4) \nonumber\\
&\text{cos}(\phi_L^1+\phi_R^2+ \phi_L^3+\phi_R^4).\nonumber
\end{align}
Therefore, we prove (\ref{z4z2f_phase_relation3}).

To show (\ref{z4z2f_phase_relation2}), we can equivalently to show that the stacking system
\begin{align}
[1,1, (0,0), (0,1)]\oplus [1,1,(0,0),(1, 2)] \oplus [1,1,(0,0), (2, 0)]\nonumber 
%\label{eqn_z2z2f_stacking_1}
\end{align}
is trivial. This  can be shown  by adding the following symmetric Higgs terms $
\text{cos}(\phi_R^1-\phi_L^3),
\text{cos}(\phi_R^2-\phi_L^1)$ and 
$\text{cos}(\phi_R^3-\phi_L^2) $
which fully gap out the edge fields withtout breaking symmetry and where $\phi^{1,2,3}_{\alpha}$ $(\alpha=R,L)$ denote the right and left moving fields of $[1_{2\times 2}, 0, 1], [1_{2\times 2}, 1, 2]$ and $[1_{2\times 2}, 2,0]$ respectively.

Finally, to show (\ref{z4z2f_phase_relation5}), is equivalent to show the following stacking system 
\begin{align}
[1,1,(0,1),(1,1)]\oplus [1,1,(0,0),(3,2)]\oplus [1,1,(0,0),(2,1)]\nonumber 
\end{align}
is trivial. We can see that all the edge fields of the stacking system can be gapped out without breaking symmetry by considering the following  Higgs terms $
\text{cos}(\phi_L^1-\phi_R^3),
\text{cos}(\phi_R^1+\phi_L^2)$ and 
$\text{cos}(\phi_R^2-\phi_L^3) $
 where $\phi^{1,2,3}_{\alpha}$ $(\alpha=R,L)$ denote the right and left moving fields of $[1,1,(0,1),(1,1)],  [1,1,(0,0),(3,2)]$ and $[1,1,(0,0),(2,1)]$ respectively.

\begin{figure}[t]
   \centering
   \includegraphics[width=8cm]{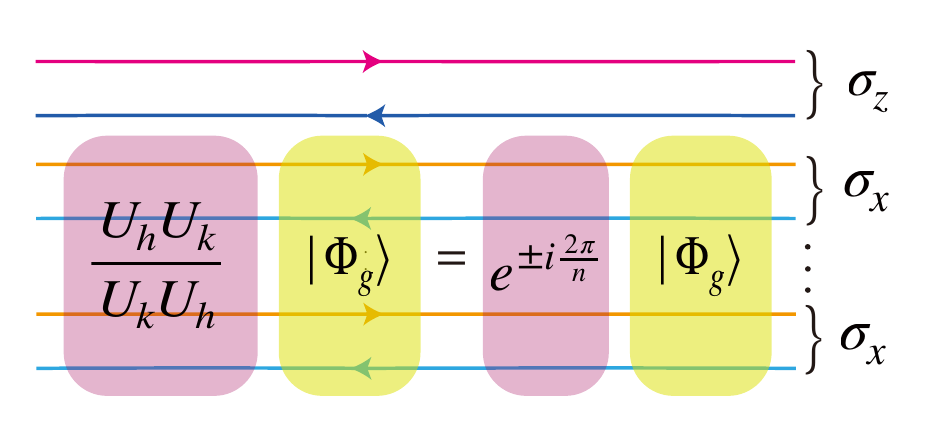} % requires the graphicx package
   \caption{ The boundary of the bulk with K matrix $K_f=\sigma_z\oplus K_b$ where $K_b$ is the K matrix for Type-III bosonic SPT phases in bosonic systems. The  bosonic symmetry $G_b=(\Z_n)^3$ with $n\ge 2$ only acts on the $K_b$ nontrivially. $|\Phi_g\rangle$, necessary being a multiplet, represent flux of one $\Z_n$ symmetry generated by $g$. $U_h,U_k$ are the matrix representations of the elements $h,k$, the generators of the remaining $(\Z_n)^2$ symmetry and they form the fundamental projective representation of $(\Z_n)^2$.  In our consruction, we conjecture $K_b=(\sigma_x){\oplus n-1}$ for $n
   \ge 2$. The shaded regions indicate that the symmetry realization of $h,k$ may permute different edge fields, as see (\ref{eqn_z3cubic_matrix}), (\ref{eqn_z5cubic_matrix}) and (\ref{eqn_z4cubic_matrix})  and the creation operator for $g$ symmetry flux involves multiple edge fields. }
   \label{fig_type_III}
\end{figure}

\section{Type-III bosonic SPT embedded phases}
\label{type-III}
Type-III bosonic SPT phases protected by $G_b$ would not be trivialized when embedding in fermionic system no matter how the $G_b$ extension by $\Z_2^f$. Using this fact, we can realize this kind of fSPT phase by considering the K matrix to be in the form of $K=\sigma_z\oplus \sigma_x\oplus...\oplus \sigma_x$. We also require that the realization of $G_f$ symmetry on  the first fermionic block is trivial, namely we can symmetrically gap out the edge fields corresponding to the first block. Therefore, the symmetry anomaly of the edge fields would come from the bosonic block, namely the $K_b=\sigma_x\oplus...\oplus \sigma_x$ (see Fig.\ref{fig_type_III}). For this consideration, in the following, we will only consider the realization of symmetry on $K_b$ which may give rise to the type-III anomaly.

We here come up with a construction which can realize the Type-III bosonic SPT  phases, instead of exausting all possible solutions. For $G_b=\Z_n\times \Z_n\times \Z_n$, we denote the generators of this group as $g_1,g_2,g_3$. We take  $K_b$ to be the one consisting of $n-1$ block of $\sigma_x$, namely $K_b=\sigma_x\oplus ...\oplus \sigma_x$ where there are $(n-1)$ $\sigma_x$ in the direct sum.  The construction takes
$$W^{g_1}=1_{2(n-1)\times 2(n-1)}, W^{g_2}=A, W^{g_3}=A^{n-1}.$$  The definition of $A$ is 
%$A:=(U\tilde{U})^{-1}MU\tilde{U}$ 
$$A=\tilde U U M (\tilde U  U)^{-1}$$
where 
$M=1_{2\times2} \otimes \tilde M$ with 
\begin{align}
\tilde M=\begin{pmatrix} 0_{\tilde n\times 1}&1_{\tilde n\times \tilde n} \\-1 & -1_{1\times \tilde n} \end{pmatrix}
\end{align}
for $\tilde n \equiv n-2\ge1$ and $\tilde M=-1$ for $\tilde n=0$,
and
\begin{align}
U=\begin{pmatrix} 1_{\bar  n\times \bar n}& 1_{\bar n\times \bar n} \\1_{\bar n\times \bar n}-O_{\bar n\times \bar n} & O_{\bar n\times \bar n}-1_{\bar n\times \bar n} \end{pmatrix}
\end{align}
with $\bar n\equiv n-1$ and $\tilde{U}$ is the integer unimodular matrix  that transforms $K_b'=\sigma_x\otimes 1_{(n-1)\times (n-1)}$ in $K_b=(\tilde U^{-1})^T K_b' \tilde U^{-1}$. It can be checked that $M$ is an order-$n$ matrix, therefore $A^n=1_{2(n-1)\times 2(n-1)}$. $O_{\bar n\times \bar n}$ is an off-diagonal matrix whose elements $O_{i,j}=\delta_{i+1,j}$.
The second part of the construction is the values of $\delta \phi^{g_{1,2,3}}$, which depend on  $n$. To obtain the proper values of  $\delta \phi^{g_i}=\frac{2\pi}{n} t_i$, we need to take into account the following sufficient conditions

(1)  $t_i+W^{g_i}\cdot t_j=t_j+W^{g_j}\cdot t_i$  mod $n$, for all $i>j$ 

(2)$|(K_b^{-1}\cdot t_1)^T\cdot (A^k-1_{2\bar n\times 2\bar n})\cdot t_3 |= k $ mod $n^2$ for all $k=0,1,...,n-1$.

The first condition ensures that the generator $g_i$ and $g_j$ commute when acting on the basis of local Hilbert space, which is generated by $e^{i\phi_i},i=1,...,2n-2$. The second condition ensures that when acting on the symmetry flux corresponding to $g_1$, the $\Z_n\times \Z_n$ subgroup generated by $g_{2,3}$ form the fundamental projective representation. More precisely, 
\begin{align}
U_{g_2}=\begin{pmatrix}0 & 1_{n-1\times n-1} \\ 1 &0 \end{pmatrix},  U_{g_3}=\alpha \begin{pmatrix} 0 &0&...&0&1 \\ w &0&...&0&0 \\ 0&w^2 &0&\cdots &0 \\...&...&...&...&... \\ 0&0&...&w^{n-1} & 0 \end{pmatrix}
\end{align}
where $\alpha$ is a constant phase factor and $w=e^{\pm i\frac{2\pi}{n}}$ (see Fig.\ref{fig_type_III}).
 We believe that the  solution of $t_i$ always exists for general $n$, even though we can not prove here. 

\subsection{$G_b=\Z_n\times \Z_n\times \Z_n$: $n$ is odd}
For $n=3$, $K_b=\sigma_x\oplus \sigma_x$, $W^{g_1}=1_{4\times 4}$, $W^{g_2}=A_3$, $W^{g_3}=(A_3)^2$ with 
\begin{align}
A_3=\begin{pmatrix} 
0&0&-1&0 \\
0&-1&0&-1\\
1&0&-1&0\\
0&1&0&0
\end{pmatrix},
\label{eqn_z3cubic_matrix}
\end{align}
and $\phi^{g_1}=\frac{2\pi}{3}(1,0,-1,0)^T$ mod $2\pi$, $\delta \phi^{g_2}=0$, $\delta \phi^{g_3}=\frac{2\pi}{3}(0,1,0,1)^T$ mod $2\pi$. Thereby $t_1=(1,0,-1,0)^T$, $t_2=0$, and $t_3=(0,1,0,1)^T$. As discused above, to check that the symmetry $g_i,g_j$ acting on the local excitation $e^{i\phi_i}$ with $i=1,2,...,2n-2$ commute, we  calculate
\begin{align}
&(1_{4\times 4}-W^{g_2})t_1-(1_{4\times 4}-W^{g_1})t_2=(0,0,-3,0)^T\nonumber \\
&(1_{4\times 4}-W^{g_3})t_1-(1_{4\times 4}-W^{g_1})t_3=(3,0,0,0)^T\nonumber\\
&(1_{4\times 4}-W^{g_2})t_3-(1_{4\times 4}-W^{g_3})t_2=(0,3,0,0)^T\nonumber
\end{align}
which indicates that $g_i,g_j$ indeed commute.

We now calculate  the matrix representation of $\Z_3\times \Z_3$ generated by $g_{2,3}$   on the symmetry flux of the subgroup $\Z_3$ generated $g_1$, which is represented by the fractional vector $l_g=\frac{1}{3}(0,1,0,-1)^T$.  Under symmetry $g_2$ or $g_3$, the $g_1$ flux represented by $l_{g_1}$ transform to $A_3^Tl_{g_1}$ or $(A_3^T)^2l_{g_1}$ which are equivalent to $l_{g_1}$; under $(g_2)^2$ or $(g_3)^2$, $l_{g_1}$ transform to $(A_3^T)^2l_{g_1}$ or $A_3^Tl_{g_1}$. Therefore, the $g_1$ symmetry flux forms a triplet $\{ e^{il_{g_1}^T\cdot  \phi },  e^{il_{g_1}^T\cdot A_3\cdot \phi },  e^{il_{g_1}^T\cdot (A_3)^2\cdot  \phi }\}$ that $g_2,g_3$ act on.  
We  compute the phase factors of the triplet under symmetry by
\begin{align}
&l_{g_1}^T(A_3-1)t_3=1\nonumber\\
&l_{g_1}^T[(A_3)^2-1]t_3=-1\nonumber.
\end{align}
Correspondingly, the matrix representation of $g_{2,3}$ on the triplet take
\begin{align}
U_{g_2}=\begin{pmatrix} 0&1&0\\ 
0&0&1 \\
1&0&0
\end{pmatrix}, U_{g_3}=\begin{pmatrix} 0&0&1\\ 
w&0&0 \\
0&w^2&0
\end{pmatrix}
\end{align} where $w=e^{-i\frac{2\pi}{3}}$. It is easy to check that $U_{g_2}U_{g_3}=w U_{g_3}U_{g_2}$ which suffices to show that they indeed form the fundamental projective representation of $\Z_3\times \Z_3$. It can also be checked that breaking this whole symmetry to any of it subgroup, the corresponding symmetry realization inherited can not protected any nontrivial phase. Therefore, we can conclude that this solution realize the pure Type-III root state.

For $n=5$, $K_b=(\sigma_x)^{\oplus4} $, $W^{g_1}=1_{8\times 8}$, $W^{g_2}=A_5$ and $W^{g_3}=(A_5)^4$, 
with
\begin{align}
A_5=\begin{pmatrix}
 0 & 0 & 0 & 1 & -1 & 0 & 0 & 0 \\
 0 & 0 & 1 & 0 & 0 & 0 & 0 & 0 \\
 0 & 0 & 0 & 0 & 0 & 0 & 1 & 0 \\
 0 & 0 & 0 & 0 & -1 & 0 & 0 & 1 \\
 1 & 0 & 0 & 0 & -1 & 0 & 0 & 0 \\
 0 & 1 & 0 & 0 & 0 & 0 & 0 & 0 \\
 0 & -1 & -1 & 0 & 0 & -1 & -1 & 0 \\
 0 & 0 & 0 & 0 & -1 & 0 & 0 & 0 \\
\end{pmatrix}
\label{eqn_z5cubic_matrix}
\end{align}
 and $\delta\phi^{g_2}=\frac{2\pi}{5} (3,0,0,2,-1,0,0,1)^T$, $\delta \phi^{g_2}=0$, and $\delta \phi^{g_3}=\frac{2\pi}{5}(0,1,1,0,0,1,1,0)$.
 Accordingly,  we have $t_1=(3,0,0,2,-1,0,0,1)^T$,  $t_3=(0,1,1,0,0,1,1,0)^T$ and $t_2=0$. As discused above, to check that the symmetry $g_i,g_j$ acting on the local excitation $e^{i\phi_i}$ with $i=1,2,...,8$ commute, we  calculate
\begin{align}
&(1_{4\times 4}-W^{g_2})t_1-(1_{4\times 4}-W^{g_1})t_2=(0,0,0,0,-5,0,0,0)^T\nonumber\\
&(1_{4\times 4}-W^{g_3})t_1-(1_{4\times 4}-W^{g_1})t_3=(5,0,0,0,0,0,0,0)^T\nonumber\\
&(1_{4\times 4}-W^{g_2})t_3-(1_{4\times 4}-W^{g_3})t_2=(0,0,0,0,0,5,0,0)^T\nonumber
\end{align}
which indicates that $g_i,g_j$ indeed commute.
 We  compute the phase factors of the quintet under symmetry by
\begin{align}
&l_{g_1}^T(A_5-1)t_3=-1\nonumber\\
&l_{g_1}^T[(A_5)^2-1]t_3=-2\nonumber\\
&l_{g_1}^T[(A_5)^3-1]t_3=-3\nonumber\\
&l_{g_1}^T[(A_5)^4-1]t_3=1.\nonumber
\end{align}
Therefore,  on the basis $\{e^{i l_{g_1}^T\cdot (A_5)^i \cdot \phi },i=0,1,...,4 \}$ generated by the $g_1$ symmetry fluxes, $g_{2,3}$ act as
 \begin{align}
 U_{g_2}=\begin{pmatrix}
 0 & 1& 0 & 0 &0 \\
 0 & 0 & 1 & 0 & 0 \\
 0 & 0 & 0 & 1 & 0  \\
 0 & 0 & 0 & 0 & 1  \\
 1 & 0 & 0 & 0 &  0 
 \end{pmatrix},  
 U_{g_3}=e^{i\frac{2\pi}{5}}\begin{pmatrix}
 0 & 0& 0 & 0 &1 \\
 w & 0 & 0 & 0 & 0 \\
 0 & w^2 & 0 & 0 & 0  \\
 0 & 0 & w^3 & 0 & 0  \\
 0 & 0 & 0 & w^4 &  0 
 \end{pmatrix}
 \end{align}
 where $w=e^{-i\frac{2\pi}{5}}$. It is easy to check that $U_{g_2}U_{g_3}=wU_{g_3}U_{g_2}$ which suffices to show that $U_{g_{2,3}}$ form the fundamental projective representation of $\Z_5\times \Z_5$. 
 We do not show that the constrution realizes the pure Type-III bosonic SPT phase, and it may contain the other component(s) of bosonic SPT phase(s) protected by subgroups of $\Z_5\times \Z_5\times \Z_5$. However, it does not matter since we can stacking the inverse of extra phase to cancel it, leaving the pure Type-III root phase.

\subsection{$G_b=\Z_n\times \Z_n\times \Z_n$: $n$ is even}
For $n=2$, $ K_b=\sigma_x$, $W^{g_1}=1_{2\times 2}$, $W^{g_{2,3}}=-1_{2\times 2}$, and $\delta \phi^{g_1}=\pi (1,0)^T$, $\delta \phi^{g_2}=0$, $\delta \phi^{g_3}=\pi(0,1)^T$. In fact, for this realization of symmetry any subgroup of $\Z_2\times \Z_2\times \Z_2$ can not protect a nontrivial phase, but the whole the symmetry can.  This fact already indicates that this solution realizes the Type-III bosonic SPT phase protected by $\Z_2\times \Z_2\times \Z_2$. To further comfirm this fact, we can alse check that subgroup $\Z_2\times \Z_2$ generated by $g_{2,3}$ forms the projective representation when acting on the symmetry flux of $g_1$. The symmetry flux of $g_1$ is related to the "fractionalied" vetex operator $e^{i\frac{1}{2}\phi_1}$. Then on the basis of $(e^{i\frac{1}{2}\phi_2}, e^{-i\frac{1}{2}\phi_2})$, $g_2$ acts as $\sigma_x$ while $g_2$ acts as $-\sigma_y$,  therefore, they form the projective representation of $\Z_2\times \Z_2$ group.  We make a remark about the construction of this case, that is the same construction can also be obtained from a free fermion construction.

For $n=4$, $K_b=\sigma_x\oplus \sigma_x \oplus \sigma_x$, and $W^{g_1}=1$, $W^{g_2}=A_4, W^{g_3}=(A_4)^3$ with 
\begin{align}
A_4= 
\left(
\begin{array}{cccccc}
 0 & 0 & -1 & 0 & 0 & 0 \\
 0 & -1 & 0 & -1 & -1 & 0 \\
 0 & 0 & -1 & 0 & 0 & 1 \\
 0 & 0 & 0 & 0 & 1 & 0 \\
 0 & 1 & 0 & 0 & 0 & 0 \\
 1 & 0 & -1 & 0 & 0 & 0 \\
\end{array}
\right),
\label{eqn_z4cubic_matrix}
\end{align}
and $\delta \phi^{g_1}=\pi/2(1,0,-1,0,0,2)^T$ mod $2\pi$, $\delta \phi^{g_2}=0$ and $\delta \phi^{g_3}=\pi/2(0,1,0,1,1,0)^T$ mod $2\pi$. 
 Accordingly,  we have $t_1=(1,0,-1,0,0,2)^T$,  $t_3=(0,1,0,1,1,0)^T$ and $t_2=0$. As discused above, to check that the symmetry $g_i,g_j$ acting on the local excitation $e^{i\phi_i}$ with $i=1,2,...,6$ commute, we  calculate
\begin{align}
&(1_{4\times 4}-W^{g_2})t_1-(1_{4\times 4}-W^{g_1})t_2=(0,0,-4,0,0,0)^T\nonumber\\
&(1_{4\times 4}-W^{g_3})t_1-(1_{4\times 4}-W^{g_1})t_3=(0,0,0,0,0,4)^T\nonumber\\
&(1_{4\times 4}-W^{g_2})t_3-(1_{4\times 4}-W^{g_3})t_2=(0,0,0,4,0,0)^T\nonumber
\end{align}
which indicates that $g_i,g_j$ indeed commute. The phase factors of the quartet under symmetry is given by
\begin{align}
&l_{g_1}^T(A_4-1)t_3=-1\nonumber\\
&l_{g_1}^T[(A_4)^2-1]t_3=-2\nonumber\\
&l_{g_1}^T[(A_4)^3-1]t_3=1\nonumber.
\end{align}

To see that the solution realizes the phase that contain the type-III root state, we check that the matrix representation of $\Z_4\times \Z_4$ generated by $g_{2,3}$ forms the fundamental projective representation when acting on the symmetry flux of the subgroup $\Z_4$ generated $g_1$, which is represented by the fractional vector $l_g=\frac{1}{4}(0,1,0,-1,2,0)^T$.   More explicitly, $g_{2,3}$ take
\begin{align}
U_{g_2}=\begin{pmatrix} 0&1&0&0\\ 
0&0&1&0 \\
0&0&0&1\\
1&0&0&0
\end{pmatrix}, U_{g_3}=e^{i\frac{\pi}{4}}\begin{pmatrix} 0&0&0&1\\ 
w&0&0&0 \\
0&w^2&0&0\\
0&0&w^3&0
\end{pmatrix}
\end{align} where $w=e^{-i\frac{\pi}{2}}$. It is easy to check that $U_{g_2}U_{g_3}=w U_{g_3}U_{g_2}$ which suffice to show that they indeed form the fundamental projective representation of $\Z_4\times \Z_4$.

Though we do not show that the constrution realizes the pure Type-III bosonic SPT phase,  it does not matter since we can stacking the inverse of extra phase to cancel other component which is more easy to obtain, leaving the pure Type-III root phase.\\

For more other $n$,  using the above construction, the solution of Type-III bosonic SPT phase can be constructed. Though lacking rigorous proof, we believe that the above construction can be used to find any value of $n$. 

\section{Intrinsically interacting fSPT}
\label{sec_intrinsic}
\begin{figure}[t]
   \centering
   \includegraphics[width=8cm]{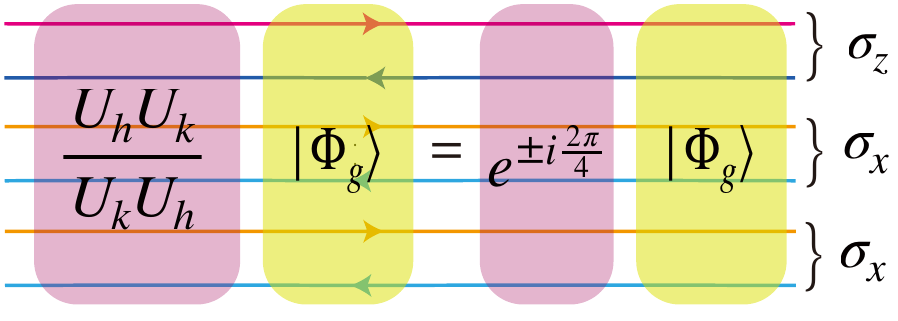} % requires the graphicx package
   \caption{The K matrix for intrinsically interacting fSPT protected by $\Z_4^f\times \Z_4\times \Z_4$ symmetry is $K=\sigma_z\oplus \sigma_x\oplus \sigma_x$. $|\Phi_g\rangle$ represents $\frac{\pi}{2}$ flux of $\Z_4^f$ whose generator is denoted as $g$ and $U_h,U_k$ are the matrix representation of the generators $h,k$ of the remaining  $\Z_4\times \Z_4$. The shaded regions indicate that the symmetry realization of $h,k$ may permute different edge fields, including the fermionic ones, as see (\ref{eqn_z4fz4z4_intrinsic_matrix}) and the $\frac{\pi}{2}$ symmetry flux is a multiplet that invovles different edge fields.}
   \label{fig_intrinsic}
\end{figure}

In this section, we first construct a solution that can realize   root state of intrinsically interacting phases of $\Z_4^f\times \Z_4\times \Z_4$ fSPT.  The fingerprint of the root state is that  $\frac{\pi}{2}$ flux of $\Z_4^f$ carries  fundamental projective representation of the remaining $\Z_4\times \Z_4$ (see Fig.\ref{fig_intrinsic}). We also show that it is the square root of the fundamental phase of Type-III bosonic SPT protected by symmetry $\Z_2\times \Z_4\times \Z_4$.

The K matrix takes
\begin{align}
K_f=\sigma_z\oplus \sigma_x \oplus \sigma_x 
\label{eqn_intrinsic_K_1}
\end{align}

The symmetry generators for  $\Z_4^f\times \Z_4\times \Z_4$ is $g, h,k$ with $g^2=P_f$ and $h^4=k^4=1$. The symmetry realization on the edge fields for the intrinsic interacting fSPT phase takes $W^{g}=1_{6\times 6}$, $W^h=\tilde A_4, W^k=\tilde A_4^{-1}=(\tilde A_4)^3$ 
where 
\begin{align}
\tilde A_{4}=
%\left(
%\begin{array}{cccccc}
% 1 & 0 & 0 & -1 & 0 & -2 \\
% 0 & 1 & 0 & -1 & 0 & 2 \\
% 0 & 0 & 0 & 0 & 0 & 1 \\
% 2 & 2 & 0 & -2 & 1 & 0 \\
% 0 & 0 & 0 & 1 & 0 & 0 \\
% 1 & -1 & 1 & 0 & 0 & -2 \\
%\end{array}
%\right)
\left(
\begin{array}{cccccc}
 1 & 0 & 0 & -2 & 0 & -1 \\
 0 & 1 & 0 & 2 & 0 & -1 \\
 0 & 0 & 0 & 0 & 0 & 1 \\
 1 & -1 & 0 & -2 & 1 & 0 \\
 0 & 0 & 0 & 1 & 0 & 0 \\
 2 & 2 & 1 & 0 & 0 & -2 \\
\end{array}
\right),
\label{eqn_z4fz4z4_intrinsic_matrix}
\end{align}
and $\delta \phi^g=\frac{\pi}{2}(1,1,0,0,0,0)^T$, $\delta \phi^h=0$ and $\delta \phi^k=\frac{\pi}{2}(2,0,2,1,1,2)^T$.
To see it indeed corresponds to the edge theory of intrinsically interacting fSPT phase,  now we check the following points.
First of all, the symmetry realization of $g$ is consistent with the fact that $g^2=P_f$ where $P_f$ is conventionally realized as $W^{P_f}=1$ and  $\delta \phi^{P_f}=\pi (1,1,0,0,0,0)^T$. Namely the fermion get $\pi$ phase under $P_f$ while boson is invariant up to $2\pi$ phase shift. Secondly, this symmetry realization of $g_1,g_2,g_3$ on the local excitaitons $e^{i\phi_i}$, with $i=1,2,...,6$ commute with each other, which can be easily confirmed by calculating 
\begin{align}
&(1_{4\times 4}-W^{h})\delta \phi^g-(1_{4\times 4}-W^{g})\delta \phi^h=(0,0,0,0,0,-2\pi)^T\nonumber\\
&(1_{4\times 4}-W^{k})\delta \phi^g-(1_{4\times 4}-W^{g})\delta \phi^k=(0,0,2\pi,0,0,0)^T\nonumber\\
&(1_{4\times 4}-W^{h})\delta \phi^k-(1_{4\times 4}-W^{k})\delta \phi^h=(2\pi,0,0,0,0,0)^T.\nonumber
\end{align}
Finally, we check that the $g$-flux (i.e., $\frac{\pi}{2}$ flux defects of $\Z_4^f$) carries the projective representation of $\Z_4\times \Z_4$ generated by $h,k$.  The $g$-flux excitation corresponds to the fractional vector $l_g=\frac{K_f}{2\pi} \delta \phi^g=\frac{1}{4}(1,-1,0,0,0,0)^T$. The $g$-flux excitations form a quartet $\{e^{il_g^T\cdot \phi},e^{il_g^T\cdot \tilde A_4 \cdot \phi}, e^{il_g^T\cdot (\tilde A_4)^2\cdot \phi}, e^{il_g^T\cdot (\tilde A_4)^3 \cdot \phi}\}$ under the symmetry of $h,k$. 
We compute the phase factors of the quartet under symmetry action $k$ as
\begin{align}
&l_{g}^T(\tilde A_4-1)\delta \phi^k=-\pi/2\nonumber\\
&l_{g}^T[(\tilde A_4)^2-1]\delta \phi^k=-\pi\nonumber\\
&l_{g}^T[(\tilde A_4)^3-1]\delta \phi^k=\pi/2\nonumber.
\end{align}
From the realization and relation above, we have the represenation on the quartet of $\Z_4\times \Z_4$ generated by $h,k$ as
\begin{align}
U^h=\begin{pmatrix}0 & 1 &0 & 0 \\ 0&0&1&0\\ 0&0&0&1\\ 1&0&0&0 \end{pmatrix},\,
U^k=e^{i\frac{\pi}{4}}\begin{pmatrix}0 & 0 &0 & 1 \\ w&0&0&0\\ 0&w^2&0&0\\ 0&0&w^3&0 \end{pmatrix}
\label{eqn_intrinsic_rep_1}
\end{align}
with $w=e^{-i\frac{\pi}{2}}$. Therefore, $U^{h,k}$ form the fundemental projective rerpresentation of $\Z_4\times \Z_4$.

Now we argue that  coupled system stacking two these root phases is equivalent to the bosonic Type-III SPT embedded phase protected by $\Z_2\times \Z_4\times \Z_4$. To show the argument, we check check the projective representation of $\Z_4\times \Z_4$ carried by $g$-defect. K matrix of the stacking system  now is 
\begin{align}
K_{s}=K_f\oplus K_f
\end{align}
where $K_f$ is the one in (\ref{eqn_intrinsic_K_1}). The symmetry realization on the edge theory of this system takes
$W_s^g=W^g\oplus W^g$, $W_s^h=W^h\oplus W^h$ and $W_s^k=W^k\oplus W^k$ and $\delta \phi_s^g=\delta \phi^g\oplus \delta \phi^g$, $\delta \phi_s^h=\delta \phi^h\oplus \delta \phi^h$ and $\delta \phi_s^k=\delta \phi^k\oplus \delta \phi^k$. Similrly, the $g$-defect corresponds to the fractional vector $\tilde l_g=\frac{K_s}{2\pi} \delta \phi_s^g=\frac{1}{4}(1,-1,0,0,0,0,1,-1,0,0,0,0)^T$ and also form a quartet $\{e^{i\tilde l_g^T\cdot \phi},e^{i \tilde l_g^T\cdot A_s \cdot \phi}, e^{i \tilde l_g^T\cdot A_s^2\cdot \phi}, e^{i\tilde l_g^T\cdot A_s^3 \cdot \phi}\}$
where $A_s=A\oplus A$. Therefore,
 the projective representation of $\Z_4\times \Z_4$ carried by $g$-defect takes
\begin{align}
U_s^h=\begin{pmatrix}0 & 1 &0 & 0 \\ 0&0&1&0\\ 0&0&0&1\\ 1&0&0&0 \end{pmatrix},\,
U_s^k=e^{\tilde il_g^T\cdot \delta \phi_s^k}\begin{pmatrix}0 & 0 &0 & 1 \\ -1&0&0&0\\ 0&1&0&0\\ 0&0&-1&0 \end{pmatrix}
\end{align}
It is easy to check that 
\begin{align}
\tilde U_h\tilde U_k=-\tilde U_k\tilde U_h
\end{align}
which indicates that $\tilde U_h$ and $\tilde U_k$ form the projective representation of $\Z_4\times \Z_4$ and this projective representation is the square of the root one, as in (\ref{eqn_intrinsic_rep_1}). 
 It can also be calculated similarly that the representation of $\Z_4\times \Z_4$ carried by $\pi$-flux(i.e. $P_f$ flux) is linear.  Therefore, the representation carried by $g$-defect is the same as that in the bosonic Type-III SPT embedded phases, indicating the phase realized in this stacking system is consistent with  the bosonic Type-III SPT embedded phases. 
 
 \section{Conclusion and Discussion}
\label{discussion}
In this paper, we obtain the edge theories of 2D fSPT protected by unitary Abelian total symmetry group $G_f$ by utilizing the K matrix formulation of Abelian Chern-Simions theory. These edge theories admit the anomalous symmetry actions that prevent the edge being a symmetric gapped state. 
%which reflects the nontrivial topology in the bulk of these fermionic SPT. 
In fact, for a specific $K$ matrix and total symmetry $G_f$, we can obtain many different realizations of symmetry action for the edge fields. Some of them are anomaly-free, while some are anomalous.  Among various anomalous edge theories, we identify the root state(s) and also show how others are related to the root one(s). Although without having a general formulation, we consider some representative unitary Abelian total symmetries $G_f$ including both trivial or nontrivial central extension over $\Z_2^f$. These discussion can be generalized into arbitrary unitary Abelian total symmetry $G_f$. 

Moreover, we also construct the Luttinger liquid edge theories of Type-III bosonic SPT protected by $(\Z_n)^3$ for the first time. 
%For this case, we come up with construction for both even and odd $n$. 
We note that an edge theory can be described by three data $[K, \{W^{g_i}, \delta \phi^{g^i}\}]$. In our construction, the $K$ matrix is chosen as $K=(\sigma_x)^{\oplus (n-1)}$, which indicates the central charge of edge theory is $n-1$. The key construction is the general expression of $W^{g_i}$ for general $n$, which can be used to obtain the proper $\{\delta \phi^{g_i}\}$ that give rise to the correct anomalous symmetry action for any Type-III bosonic SPT root states. We explicitly show that the solution for $\delta \phi^{g_i}$ with $n=2,3,4,5$ and we believe that our construction indeed can be applied to arbitrary $n$. Technically, one shall pay attention to the constraints from the linear representation of group on the basis of local excitations and derive the projective representation of two $\Z_n$s  on symmetry flux of the third subgroup. We stress that even though 1+1D $SU(n)_1$ WZW model can be constructed from the $n-1$ components Luttinger liquid with a fine-tuning radius, our construction does not have constraint on the radius,  which implies that the edge theory can flow away from the previously conjectured $SU(n)_1$ point by some relevant symmetric interaction.   It is interesting to study the stability of the $SU(n)_1$ WZW model with such symmetry realization on the lattice model. Related to this question, Ref.\cite{Massaim_2019} study the similar question in a 1+1D spin chain with a similar anomaly, the so-called LSM anomaly.

More interestingly, we also discuss the edge theory of the so-called intrinsically interacting fSPT.
The minimal symmetry protecting this kind of phases is $\Z_4^f \times \Z_4\times \Z_2^T$, whose edge theories are studied in Ref.\cite{Meng2019_04}. For the unitary Abelian $G_b$, the minimal one is $\Z_4^f\times \Z_4 \times Z_4$. We construct the corresponding edge theory for the root state of $\Z_4^f\times \Z_4 \times Z_4$  intrinsically interacting fSPT. It is very interesting to compare this case with the Type-III bosonic SPT state. 
%and $\Z_4^f\times \Z_4 \times Z_4$ intrinsically interacting fSPT. 
Upon gauging one  $\Z_4$  in the former or $\Z_4^f$ in the latter, we both obtain  $\Z_4$ toric code enriched by $\Z_4\times \Z_4$ symmetry, but with different symmetry fractionalization on anyons. For the $\Z_4$ gauged Type-III bosonic SPT state, the $\Z_4$ gauge charge is bosonic, and $\Z_4$ flux carries the fundamental projective representation of remaining $\Z_4\times \Z_4$ symmetry. In contrast, for the $\Z_4^f$ gauged fSPT, the $\Z_4^f$ gauge charge is fermionic and $\Z_4^f$ flux carries the fundamental projective representation of remaining $\Z_4\times \Z_4$ symmetry. Based these observations,  starting from the $\Z_4$ toric code 
enriched by $\Z_4\times \Z_4$ symmetry, to obtain the   Type-III bosonic SPT, one may condense some bosons(such as the nuetral bosonic gauge charge), while for intrinsically interacting fSPT, one should condense the fermion that does not carries projective representation of $\Z_4\times \Z_4$.  Such understanding can be generalized to other intrinsically interacting fSPT. In fact, the basic concept of boson/fermion condensation is also very useful to understand the gapless edge theory of symmetry enriched topological(SET) phases, and we will study more examples in our future works.

SQN especially acknowledges Meng Cheng for sharing his theory on anomaly of 1+1D CFT and insightful discussion on the construction on edge theories of Type-III bosonic SPT. SQN and CW are supported by the Research Grant Council of Hong Kong (ECS 21301018, GRF 11300819) and URC, HKU (Grant No. 201906159002). QRW and ZCG are supported by  the Research Grant Council of Hong Kong (GRF No.14306918, ANR/RGC Joint Research Scheme No. A-CUHK402/18) , and Direct Grant No. 4053346 from The Chinese University of Hong Kong.
% and funding from Hong Kong's Research Grants Council (GRF No.14306918, ANR/RGC Joint Research Scheme No. A-CUHK402/18). 

\bibliography{edge2DfSPT}

\begin{thebibliography}{66}%
\makeatletter
\providecommand \@ifxundefined [1]{%
 \@ifx{#1\undefined}
}%
\providecommand \@ifnum [1]{%
 \ifnum #1\expandafter \@firstoftwo
 \else \expandafter \@secondoftwo
 \fi
}%
\providecommand \@ifx [1]{%
 \ifx #1\expandafter \@firstoftwo
 \else \expandafter \@secondoftwo
 \fi
}%
\providecommand \natexlab [1]{#1}%
\providecommand \enquote  [1]{``#1''}%
\providecommand \bibnamefont  [1]{#1}%
\providecommand \bibfnamefont [1]{#1}%
\providecommand \citenamefont [1]{#1}%
\providecommand \href@noop [0]{\@secondoftwo}%
\providecommand \href [0]{\begingroup \@sanitize@url \@href}%
\providecommand \@href[1]{\@@startlink{#1}\@@href}%
\providecommand \@@href[1]{\endgroup#1\@@endlink}%
\providecommand \@sanitize@url [0]{\catcode `\\12\catcode `\$12\catcode
  `\&12\catcode `\#12\catcode `\^12\catcode `\_12\catcode `\%12\relax}%
\providecommand \@@startlink[1]{}%
\providecommand \@@endlink[0]{}%
\providecommand \url  [0]{\begingroup\@sanitize@url \@url }%
\providecommand \@url [1]{\endgroup\@href {#1}{\urlprefix }}%
\providecommand \urlprefix  [0]{URL }%
\providecommand \Eprint [0]{\href }%
\providecommand \doibase [0]{https://doi.org/}%
\providecommand \selectlanguage [0]{\@gobble}%
\providecommand \bibinfo  [0]{\@secondoftwo}%
\providecommand \bibfield  [0]{\@secondoftwo}%
\providecommand \translation [1]{[#1]}%
\providecommand \BibitemOpen [0]{}%
\providecommand \bibitemStop [0]{}%
\providecommand \bibitemNoStop [0]{.\EOS\space}%
\providecommand \EOS [0]{\spacefactor3000\relax}%
\providecommand \BibitemShut  [1]{\csname bibitem#1\endcsname}%
\let\auto@bib@innerbib\@empty
%</preamble>
\bibitem [{\citenamefont {Gu}\ and\ \citenamefont {Wen}(2009)}]{gu09}%
  \BibitemOpen
  \bibfield  {author} {\bibinfo {author} {\bibfnamefont {Z.-C.}\ \bibnamefont
  {Gu}}\ and\ \bibinfo {author} {\bibfnamefont {X.-G.}\ \bibnamefont {Wen}},\
  }\bibfield  {title} {\bibinfo {title} {Tensor-entanglement-filtering
  renormalization approach and symmetry-protected topological order},\ }\href
  {https://doi.org/10.1103/PhysRevB.80.155131} {\bibfield  {journal} {\bibinfo
  {journal} {Phys. Rev. B}\ }\textbf {\bibinfo {volume} {80}},\ \bibinfo
  {pages} {155131} (\bibinfo {year} {2009})}\BibitemShut {NoStop}%
\bibitem [{\citenamefont {Chen}\ \emph {et~al.}(2012)\citenamefont {Chen},
  \citenamefont {Gu}, \citenamefont {Liu},\ and\ \citenamefont
  {Wen}}]{chenScience2012}%
  \BibitemOpen
  \bibfield  {author} {\bibinfo {author} {\bibfnamefont {X.}~\bibnamefont
  {Chen}}, \bibinfo {author} {\bibfnamefont {Z.-C.}\ \bibnamefont {Gu}},
  \bibinfo {author} {\bibfnamefont {Z.-X.}\ \bibnamefont {Liu}},\ and\ \bibinfo
  {author} {\bibfnamefont {X.-G.}\ \bibnamefont {Wen}},\ }\bibfield  {title}
  {\bibinfo {title} {Symmetry-protected topological orders in interacting
  bosonic systems},\ }\href {https://doi.org/10.1126/science.1227224}
  {\bibfield  {journal} {\bibinfo  {journal} {Science}\ }\textbf {\bibinfo
  {volume} {338}},\ \bibinfo {pages} {1604} (\bibinfo {year}
  {2012})}\BibitemShut {NoStop}%
\bibitem [{\citenamefont {Chen}\ \emph {et~al.}(2013)\citenamefont {Chen},
  \citenamefont {Gu}, \citenamefont {Liu},\ and\ \citenamefont {Wen}}]{chen13}%
  \BibitemOpen
  \bibfield  {author} {\bibinfo {author} {\bibfnamefont {X.}~\bibnamefont
  {Chen}}, \bibinfo {author} {\bibfnamefont {Z.-C.}\ \bibnamefont {Gu}},
  \bibinfo {author} {\bibfnamefont {Z.-X.}\ \bibnamefont {Liu}},\ and\ \bibinfo
  {author} {\bibfnamefont {X.-G.}\ \bibnamefont {Wen}},\ }\bibfield  {title}
  {\bibinfo {title} {Symmetry protected topological orders and the group
  cohomology of their symmetry group},\ }\href
  {https://doi.org/10.1103/PhysRevB.87.155114} {\bibfield  {journal} {\bibinfo
  {journal} {Phys. Rev. B}\ }\textbf {\bibinfo {volume} {87}},\ \bibinfo
  {pages} {155114} (\bibinfo {year} {2013})}\BibitemShut {NoStop}%
\bibitem [{\citenamefont {Gu}\ and\ \citenamefont {Wen}(2014)}]{GuWen2014}%
  \BibitemOpen
  \bibfield  {author} {\bibinfo {author} {\bibfnamefont {Z.-C.}\ \bibnamefont
  {Gu}}\ and\ \bibinfo {author} {\bibfnamefont {X.-G.}\ \bibnamefont {Wen}},\
  }\bibfield  {title} {\bibinfo {title} {Symmetry-protected topological orders
  for interacting fermions: Fermionic topological nonlinear
  $\ensuremath{\sigma}$ models and a special group supercohomology theory},\
  }\href {https://doi.org/10.1103/PhysRevB.90.115141} {\bibfield  {journal}
  {\bibinfo  {journal} {Phys. Rev. B}\ }\textbf {\bibinfo {volume} {90}},\
  \bibinfo {pages} {115141} (\bibinfo {year} {2014})}\BibitemShut {NoStop}%
\bibitem [{\citenamefont {{Kapustin}}(2014)}]{kapustin14a}%
  \BibitemOpen
  \bibfield  {author} {\bibinfo {author} {\bibfnamefont {A.}~\bibnamefont
  {{Kapustin}}},\ }\bibfield  {title} {\bibinfo {title} {{Symmetry Protected
  Topological Phases, Anomalies, and Cobordisms: Beyond Group Cohomology}},\
  }\href@noop {} {\bibfield  {journal} {\bibinfo  {journal} {ArXiv e-prints}\ }
  (\bibinfo {year} {2014})},\ \Eprint {https://arxiv.org/abs/1403.1467}
  {arXiv:1403.1467} \BibitemShut {NoStop}%
\bibitem [{\citenamefont {{Kapustin}}\ \emph {et~al.}(2014)\citenamefont
  {{Kapustin}}, \citenamefont {{Thorngren}}, \citenamefont {{Turzillo}},\ and\
  \citenamefont {{Wang}}}]{kapustin14}%
  \BibitemOpen
  \bibfield  {author} {\bibinfo {author} {\bibfnamefont {A.}~\bibnamefont
  {{Kapustin}}}, \bibinfo {author} {\bibfnamefont {R.}~\bibnamefont
  {{Thorngren}}}, \bibinfo {author} {\bibfnamefont {A.}~\bibnamefont
  {{Turzillo}}},\ and\ \bibinfo {author} {\bibfnamefont {Z.}~\bibnamefont
  {{Wang}}},\ }\bibfield  {title} {\bibinfo {title} {{Fermionic Symmetry
  Protected Topological Phases and Cobordisms}},\ }\href@noop {} {\bibfield
  {journal} {\bibinfo  {journal} {arXiv e-prints}\ } (\bibinfo {year}
  {2014})},\ \Eprint {https://arxiv.org/abs/1406.7329} {arXiv:1406.7329}
  \BibitemShut {NoStop}%
\bibitem [{\citenamefont {{Freed}}(2014)}]{freed14}%
  \BibitemOpen
  \bibfield  {author} {\bibinfo {author} {\bibfnamefont {D.~S.}\ \bibnamefont
  {{Freed}}},\ }\bibfield  {title} {\bibinfo {title} {{Short-range entanglement
  and invertible field theories}},\ }\href@noop {} {\bibfield  {journal}
  {\bibinfo  {journal} {arXiv e-prints}\ } (\bibinfo {year} {2014})},\ \Eprint
  {https://arxiv.org/abs/1406.7278} {arXiv:1406.7278} \BibitemShut {NoStop}%
\bibitem [{\citenamefont {Wen}(2015)}]{wen15}%
  \BibitemOpen
  \bibfield  {author} {\bibinfo {author} {\bibfnamefont {X.-G.}\ \bibnamefont
  {Wen}},\ }\bibfield  {title} {\bibinfo {title} {Construction of bosonic
  symmetry-protected-trivial states and their topological invariants via
  $g\ifmmode\times\else\texttimes\fi{}so(\ensuremath{\infty})$ nonlinear
  $\ensuremath{\sigma}$ models},\ }\href
  {https://doi.org/10.1103/PhysRevB.91.205101} {\bibfield  {journal} {\bibinfo
  {journal} {Phys. Rev. B}\ }\textbf {\bibinfo {volume} {91}},\ \bibinfo
  {pages} {205101} (\bibinfo {year} {2015})}\BibitemShut {NoStop}%
\bibitem [{\citenamefont {{Wang}}\ \emph {et~al.}(2014)\citenamefont {{Wang}},
  \citenamefont {{Potter}},\ and\ \citenamefont {{Senthil}}}]{wangc-science}%
  \BibitemOpen
  \bibfield  {author} {\bibinfo {author} {\bibfnamefont {C.}~\bibnamefont
  {{Wang}}}, \bibinfo {author} {\bibfnamefont {A.~C.}\ \bibnamefont
  {{Potter}}},\ and\ \bibinfo {author} {\bibfnamefont {T.}~\bibnamefont
  {{Senthil}}},\ }\bibfield  {title} {\bibinfo {title} {{Classification of
  Interacting Electronic Topological Insulators in Three Dimensions}},\ }\href
  {https://doi.org/10.1126/science.1243326} {\bibfield  {journal} {\bibinfo
  {journal} {Science}\ }\textbf {\bibinfo {volume} {343}},\ \bibinfo {pages}
  {629} (\bibinfo {year} {2014})},\ \Eprint {https://arxiv.org/abs/1306.3238}
  {arXiv:1306.3238} \BibitemShut {NoStop}%
\bibitem [{\citenamefont {Wang}\ and\ \citenamefont {Senthil}(2014)}]{chong14}%
  \BibitemOpen
  \bibfield  {author} {\bibinfo {author} {\bibfnamefont {C.}~\bibnamefont
  {Wang}}\ and\ \bibinfo {author} {\bibfnamefont {T.}~\bibnamefont {Senthil}},\
  }\bibfield  {title} {\bibinfo {title} {Interacting fermionic topological
  insulators/superconductors in three dimensions},\ }\href
  {https://doi.org/10.1103/PhysRevB.89.195124} {\bibfield  {journal} {\bibinfo
  {journal} {Phys. Rev. B}\ }\textbf {\bibinfo {volume} {89}},\ \bibinfo
  {pages} {195124} (\bibinfo {year} {2014})}\BibitemShut {NoStop}%
\bibitem [{\citenamefont {{Cheng}}\ \emph {et~al.}(2015)\citenamefont
  {{Cheng}}, \citenamefont {{Bi}}, \citenamefont {{You}},\ and\ \citenamefont
  {{Gu}}}]{cheng15}%
  \BibitemOpen
  \bibfield  {author} {\bibinfo {author} {\bibfnamefont {M.}~\bibnamefont
  {{Cheng}}}, \bibinfo {author} {\bibfnamefont {Z.}~\bibnamefont {{Bi}}},
  \bibinfo {author} {\bibfnamefont {Y.-Z.}\ \bibnamefont {{You}}},\ and\
  \bibinfo {author} {\bibfnamefont {Z.-C.}\ \bibnamefont {{Gu}}},\ }\bibfield
  {title} {\bibinfo {title} {{Towards a Complete Classification of
  Symmetry-Protected Phases for Interacting Fermions in Two Dimensions}},\
  }\href@noop {} {\bibfield  {journal} {\bibinfo  {journal} {arXiv e-prints}\ }
  (\bibinfo {year} {2015})},\ \Eprint {https://arxiv.org/abs/1501.01313}
  {arXiv:1501.01313} \BibitemShut {NoStop}%
\bibitem [{\citenamefont {Freed}\ and\ \citenamefont
  {Hopkins}(2016)}]{freed16}%
  \BibitemOpen
  \bibfield  {author} {\bibinfo {author} {\bibfnamefont {D.~S.}\ \bibnamefont
  {Freed}}\ and\ \bibinfo {author} {\bibfnamefont {M.~J.}\ \bibnamefont
  {Hopkins}},\ }\bibfield  {title} {\bibinfo {title} {{Reflection positivity
  and invertible topological phases}},\ }\href@noop {} {\bibfield  {journal}
  {\bibinfo  {journal} {arXiv e-prints}\ } (\bibinfo {year} {2016})},\ \Eprint
  {https://arxiv.org/abs/1604.06527} {arXiv:1604.06527} \BibitemShut {NoStop}%
\bibitem [{\citenamefont {Wang}\ and\ \citenamefont {Gu}(2018)}]{WangGu2017}%
  \BibitemOpen
  \bibfield  {author} {\bibinfo {author} {\bibfnamefont {Q.-R.}\ \bibnamefont
  {Wang}}\ and\ \bibinfo {author} {\bibfnamefont {Z.-C.}\ \bibnamefont {Gu}},\
  }\bibfield  {title} {\bibinfo {title} {Towards a complete classification of
  symmetry-protected topological phases for interacting fermions in three
  dimensions and a general group supercohomology theory},\ }\href
  {https://doi.org/10.1103/PhysRevX.8.011055} {\bibfield  {journal} {\bibinfo
  {journal} {Phys. Rev. X}\ }\textbf {\bibinfo {volume} {8}},\ \bibinfo {pages}
  {011055} (\bibinfo {year} {2018})}\BibitemShut {NoStop}%
\bibitem [{\citenamefont {Wang}\ and\ \citenamefont {Gu}(2020)}]{WangGu2020}%
  \BibitemOpen
  \bibfield  {author} {\bibinfo {author} {\bibfnamefont {Q.-R.}\ \bibnamefont
  {Wang}}\ and\ \bibinfo {author} {\bibfnamefont {Z.-C.}\ \bibnamefont {Gu}},\
  }\href {https://doi.org/10.1103/PhysRevX.10.031055} {\bibfield  {journal}
  {\bibinfo  {journal} {Phys. Rev. X}\ }\textbf {\bibinfo {volume} {10}},\
  \bibinfo {pages} {031055} (\bibinfo {year} {2020})},\ \Eprint
  {https://arxiv.org/abs/1811.00536} {arXiv:1811.00536 [cond-mat.str-el]}
  \BibitemShut {NoStop}%
\bibitem [{\citenamefont {Kapustin}\ and\ \citenamefont
  {Thorngren}(2017)}]{Kapustin2017}%
  \BibitemOpen
  \bibfield  {author} {\bibinfo {author} {\bibfnamefont {A.}~\bibnamefont
  {Kapustin}}\ and\ \bibinfo {author} {\bibfnamefont {R.}~\bibnamefont
  {Thorngren}},\ }\bibfield  {title} {\bibinfo {title} {Fermionic spt phases in
  higher dimensions and bosonization},\ }\href
  {https://doi.org/10.1007/JHEP10(2017)080} {\bibfield  {journal} {\bibinfo
  {journal} {Journal of High Energy Physics}\ }\textbf {\bibinfo {volume}
  {2017}},\ \bibinfo {pages} {80} (\bibinfo {year} {2017})}\BibitemShut
  {NoStop}%
\bibitem [{\citenamefont {{Wang}}\ \emph
  {et~al.}(2018{\natexlab{a}})\citenamefont {{Wang}}, \citenamefont {{Ohmori}},
  \citenamefont {{Putrov}}, \citenamefont {{Zheng}}, \citenamefont {{Wan}},
  \citenamefont {{Guo}}, \citenamefont {{Lin}}, \citenamefont {{Gao}},\ and\
  \citenamefont {{Yau}}}]{Juven2018}%
  \BibitemOpen
  \bibfield  {author} {\bibinfo {author} {\bibfnamefont {J.}~\bibnamefont
  {{Wang}}}, \bibinfo {author} {\bibfnamefont {K.}~\bibnamefont {{Ohmori}}},
  \bibinfo {author} {\bibfnamefont {P.}~\bibnamefont {{Putrov}}}, \bibinfo
  {author} {\bibfnamefont {Y.}~\bibnamefont {{Zheng}}}, \bibinfo {author}
  {\bibfnamefont {Z.}~\bibnamefont {{Wan}}}, \bibinfo {author} {\bibfnamefont
  {M.}~\bibnamefont {{Guo}}}, \bibinfo {author} {\bibfnamefont
  {H.}~\bibnamefont {{Lin}}}, \bibinfo {author} {\bibfnamefont
  {P.}~\bibnamefont {{Gao}}},\ and\ \bibinfo {author} {\bibfnamefont {S.-T.}\
  \bibnamefont {{Yau}}},\ }\bibfield  {title} {\bibinfo {title} {{Tunneling
  topological vacua via extended operators: (Spin-)TQFT spectra and boundary
  deconfinement in various dimensions}},\ }\href
  {https://doi.org/10.1093/ptep/pty051} {\bibfield  {journal} {\bibinfo
  {journal} {Progress of Theoretical and Experimental Physics}\ }\textbf
  {\bibinfo {volume} {2018}},\ \bibinfo {eid} {053A01} (\bibinfo {year}
  {2018}{\natexlab{a}})},\ \Eprint {https://arxiv.org/abs/1801.05416}
  {arXiv:1801.05416 [cond-mat.str-el]} \BibitemShut {NoStop}%
\bibitem [{\citenamefont {Levin}\ and\ \citenamefont
  {Gu}(2012)}]{levin_gu_model}%
  \BibitemOpen
  \bibfield  {author} {\bibinfo {author} {\bibfnamefont {M.}~\bibnamefont
  {Levin}}\ and\ \bibinfo {author} {\bibfnamefont {Z.-C.}\ \bibnamefont {Gu}},\
  }\bibfield  {title} {\bibinfo {title} {Braiding statistics approach to
  symmetry-protected topological phases},\ }\href
  {https://doi.org/10.1103/PhysRevB.86.115109} {\bibfield  {journal} {\bibinfo
  {journal} {Phys. Rev. B}\ }\textbf {\bibinfo {volume} {86}},\ \bibinfo
  {pages} {115109} (\bibinfo {year} {2012})}\BibitemShut {NoStop}%
\bibitem [{\citenamefont {Chen}\ and\ \citenamefont
  {Wen}(2012)}]{Xie12chiralsymm}%
  \BibitemOpen
  \bibfield  {author} {\bibinfo {author} {\bibfnamefont {X.}~\bibnamefont
  {Chen}}\ and\ \bibinfo {author} {\bibfnamefont {X.-G.}\ \bibnamefont {Wen}},\
  }\bibfield  {title} {\bibinfo {title} {Chiral symmetry on the edge of
  two-dimensional symmetry protected topological phases},\ }\href
  {https://doi.org/10.1103/PhysRevB.86.235135} {\bibfield  {journal} {\bibinfo
  {journal} {Phys. Rev. B}\ }\textbf {\bibinfo {volume} {86}},\ \bibinfo
  {pages} {235135} (\bibinfo {year} {2012})}\BibitemShut {NoStop}%
\bibitem [{\citenamefont {Cheng}\ and\ \citenamefont {Gu}(2014)}]{cheng2014}%
  \BibitemOpen
  \bibfield  {author} {\bibinfo {author} {\bibfnamefont {M.}~\bibnamefont
  {Cheng}}\ and\ \bibinfo {author} {\bibfnamefont {Z.-C.}\ \bibnamefont {Gu}},\
  }\bibfield  {title} {\bibinfo {title} {Topological response theory of abelian
  symmetry-protected topological phases in two dimensions},\ }\href
  {https://doi.org/10.1103/PhysRevLett.112.141602} {\bibfield  {journal}
  {\bibinfo  {journal} {Phys. Rev. Lett.}\ }\textbf {\bibinfo {volume} {112}},\
  \bibinfo {pages} {141602} (\bibinfo {year} {2014})},\ \Eprint
  {https://arxiv.org/abs/arXiv:1302.4803} {arXiv:1302.4803} \BibitemShut
  {NoStop}%
\bibitem [{\citenamefont {Wang}\ and\ \citenamefont {Levin}(2014)}]{threeloop}%
  \BibitemOpen
  \bibfield  {author} {\bibinfo {author} {\bibfnamefont {C.}~\bibnamefont
  {Wang}}\ and\ \bibinfo {author} {\bibfnamefont {M.}~\bibnamefont {Levin}},\
  }\bibfield  {title} {\bibinfo {title} {Braiding statistics of loop
  excitations in three dimensions},\ }\href
  {https://doi.org/10.1103/PhysRevLett.113.080403} {\bibfield  {journal}
  {\bibinfo  {journal} {Phys. Rev. Lett.}\ }\textbf {\bibinfo {volume} {113}},\
  \bibinfo {pages} {080403} (\bibinfo {year} {2014})}\BibitemShut {NoStop}%
\bibitem [{\citenamefont {Jiang}\ \emph {et~al.}(2014)\citenamefont {Jiang},
  \citenamefont {Mesaros},\ and\ \citenamefont {Ran}}]{ran14}%
  \BibitemOpen
  \bibfield  {author} {\bibinfo {author} {\bibfnamefont {S.}~\bibnamefont
  {Jiang}}, \bibinfo {author} {\bibfnamefont {A.}~\bibnamefont {Mesaros}},\
  and\ \bibinfo {author} {\bibfnamefont {Y.}~\bibnamefont {Ran}},\ }\bibfield
  {title} {\bibinfo {title} {Generalized modular transformations in
  $(3+1)\mathrm{D}$ topologically ordered phases and triple linking invariant
  of loop braiding},\ }\href {https://doi.org/10.1103/PhysRevX.4.031048}
  {\bibfield  {journal} {\bibinfo  {journal} {Phys. Rev. X}\ }\textbf {\bibinfo
  {volume} {4}},\ \bibinfo {pages} {031048} (\bibinfo {year}
  {2014})}\BibitemShut {NoStop}%
\bibitem [{\citenamefont {Wang}\ and\ \citenamefont {Levin}(2015)}]{wangcj15}%
  \BibitemOpen
  \bibfield  {author} {\bibinfo {author} {\bibfnamefont {C.}~\bibnamefont
  {Wang}}\ and\ \bibinfo {author} {\bibfnamefont {M.}~\bibnamefont {Levin}},\
  }\bibfield  {title} {\bibinfo {title} {Topological invariants for gauge
  theories and symmetry-protected topological phases},\ }\href
  {https://doi.org/10.1103/PhysRevB.91.165119} {\bibfield  {journal} {\bibinfo
  {journal} {Phys. Rev. B}\ }\textbf {\bibinfo {volume} {91}},\ \bibinfo
  {pages} {165119} (\bibinfo {year} {2015})}\BibitemShut {NoStop}%
\bibitem [{\citenamefont {Wang}\ and\ \citenamefont {Wen}(2015)}]{wangj15}%
  \BibitemOpen
  \bibfield  {author} {\bibinfo {author} {\bibfnamefont {J.~C.}\ \bibnamefont
  {Wang}}\ and\ \bibinfo {author} {\bibfnamefont {X.-G.}\ \bibnamefont {Wen}},\
  }\bibfield  {title} {\bibinfo {title} {Non-abelian string and particle
  braiding in topological order: Modular $\mathrm{SL}(3,\mathbb{Z})$
  representation and $(3+1)$-dimensional twisted gauge theory},\ }\href
  {https://doi.org/10.1103/PhysRevB.91.035134} {\bibfield  {journal} {\bibinfo
  {journal} {Phys. Rev. B}\ }\textbf {\bibinfo {volume} {91}},\ \bibinfo
  {pages} {035134} (\bibinfo {year} {2015})}\BibitemShut {NoStop}%
\bibitem [{\citenamefont {Wang}\ \emph
  {et~al.}(2015{\natexlab{a}})\citenamefont {Wang}, \citenamefont {Gu},\ and\
  \citenamefont {Wen}}]{Juven2015}%
  \BibitemOpen
  \bibfield  {author} {\bibinfo {author} {\bibfnamefont {J.~C.}\ \bibnamefont
  {Wang}}, \bibinfo {author} {\bibfnamefont {Z.-C.}\ \bibnamefont {Gu}},\ and\
  \bibinfo {author} {\bibfnamefont {X.-G.}\ \bibnamefont {Wen}},\ }\bibfield
  {title} {\bibinfo {title} {Field-theory representation of gauge-gravity
  symmetry-protected topological invariants, group cohomology, and beyond},\
  }\href {https://doi.org/10.1103/PhysRevLett.114.031601} {\bibfield  {journal}
  {\bibinfo  {journal} {Phys. Rev. Lett.}\ }\textbf {\bibinfo {volume} {114}},\
  \bibinfo {pages} {031601} (\bibinfo {year} {2015}{\natexlab{a}})}\BibitemShut
  {NoStop}%
\bibitem [{\citenamefont {Lin}\ and\ \citenamefont {Levin}(2015)}]{lin15}%
  \BibitemOpen
  \bibfield  {author} {\bibinfo {author} {\bibfnamefont {C.-H.}\ \bibnamefont
  {Lin}}\ and\ \bibinfo {author} {\bibfnamefont {M.}~\bibnamefont {Levin}},\
  }\bibfield  {title} {\bibinfo {title} {Loop braiding statistics in exactly
  soluble three-dimensional lattice models},\ }\href
  {https://doi.org/10.1103/PhysRevB.92.035115} {\bibfield  {journal} {\bibinfo
  {journal} {Phys. Rev. B}\ }\textbf {\bibinfo {volume} {92}},\ \bibinfo
  {pages} {035115} (\bibinfo {year} {2015})}\BibitemShut {NoStop}%
\bibitem [{\citenamefont {Wang}\ \emph
  {et~al.}(2017{\natexlab{a}})\citenamefont {Wang}, \citenamefont {Lin},\ and\
  \citenamefont {Gu}}]{wanggu16}%
  \BibitemOpen
  \bibfield  {author} {\bibinfo {author} {\bibfnamefont {C.}~\bibnamefont
  {Wang}}, \bibinfo {author} {\bibfnamefont {C.-H.}\ \bibnamefont {Lin}},\ and\
  \bibinfo {author} {\bibfnamefont {Z.-C.}\ \bibnamefont {Gu}},\ }\bibfield
  {title} {\bibinfo {title} {Interacting fermionic symmetry-protected
  topological phases in two dimensions},\ }\href
  {https://doi.org/10.1103/PhysRevB.95.195147} {\bibfield  {journal} {\bibinfo
  {journal} {Phys. Rev. B}\ }\textbf {\bibinfo {volume} {95}},\ \bibinfo
  {pages} {195147} (\bibinfo {year} {2017}{\natexlab{a}})}\BibitemShut
  {NoStop}%
\bibitem [{\citenamefont {Putrov}\ \emph {et~al.}(2016)\citenamefont {Putrov},
  \citenamefont {Wang},\ and\ \citenamefont {Yau}}]{juven16}%
  \BibitemOpen
  \bibfield  {author} {\bibinfo {author} {\bibfnamefont {P.}~\bibnamefont
  {Putrov}}, \bibinfo {author} {\bibfnamefont {J.}~\bibnamefont {Wang}},\ and\
  \bibinfo {author} {\bibfnamefont {S.-T.}\ \bibnamefont {Yau}},\ }\bibfield
  {title} {\bibinfo {title} {{Braiding Statistics and Link Invariants of
  Bosonic/Fermionic Topological Quantum Matter in 2+1 and 3+1 dimensions}},\
  }\href@noop {} {\bibfield  {journal} {\bibinfo  {journal} {arXiv e-prints}\ }
  (\bibinfo {year} {2016})},\ \Eprint {https://arxiv.org/abs/1612.09298}
  {arXiv:1612.09298} \BibitemShut {NoStop}%
\bibitem [{\citenamefont {{Wang}}\ \emph
  {et~al.}(2018{\natexlab{b}})\citenamefont {{Wang}}, \citenamefont {{Cheng}},
  \citenamefont {{Wang}},\ and\ \citenamefont {{Gu}}}]{WCWG2018}%
  \BibitemOpen
  \bibfield  {author} {\bibinfo {author} {\bibfnamefont {Q.-R.}\ \bibnamefont
  {{Wang}}}, \bibinfo {author} {\bibfnamefont {M.}~\bibnamefont {{Cheng}}},
  \bibinfo {author} {\bibfnamefont {C.}~\bibnamefont {{Wang}}},\ and\ \bibinfo
  {author} {\bibfnamefont {Z.-C.}\ \bibnamefont {{Gu}}},\ }\bibfield  {title}
  {\bibinfo {title} {{Topological Quantum Field Theory for Abelian Topological
  Phases and Loop Braiding Statistics in $(3+1)$-Dimensions}},\ }\href@noop {}
  {\bibfield  {journal} {\bibinfo  {journal} {ArXiv e-prints}\ } (\bibinfo
  {year} {2018}{\natexlab{b}})},\ \Eprint {https://arxiv.org/abs/1810.13428}
  {arXiv:1810.13428 [cond-mat.str-el]} \BibitemShut {NoStop}%
\bibitem [{\citenamefont {Vishwanath}\ and\ \citenamefont
  {Senthil}(2013)}]{vishwanath13}%
  \BibitemOpen
  \bibfield  {author} {\bibinfo {author} {\bibfnamefont {A.}~\bibnamefont
  {Vishwanath}}\ and\ \bibinfo {author} {\bibfnamefont {T.}~\bibnamefont
  {Senthil}},\ }\bibfield  {title} {\bibinfo {title} {Physics of
  three-dimensional bosonic topological insulators: Surface-deconfined
  criticality and quantized magnetoelectric effect},\ }\href
  {https://doi.org/10.1103/PhysRevX.3.011016} {\bibfield  {journal} {\bibinfo
  {journal} {Phys. Rev. X}\ }\textbf {\bibinfo {volume} {3}},\ \bibinfo {pages}
  {011016} (\bibinfo {year} {2013})}\BibitemShut {NoStop}%
\bibitem [{\citenamefont {Wang}\ and\ \citenamefont {Senthil}(2013)}]{wangc13}%
  \BibitemOpen
  \bibfield  {author} {\bibinfo {author} {\bibfnamefont {C.}~\bibnamefont
  {Wang}}\ and\ \bibinfo {author} {\bibfnamefont {T.}~\bibnamefont {Senthil}},\
  }\bibfield  {title} {\bibinfo {title} {Boson topological insulators: A window
  into highly entangled quantum phases},\ }\href
  {https://doi.org/10.1103/PhysRevB.87.235122} {\bibfield  {journal} {\bibinfo
  {journal} {Phys. Rev. B}\ }\textbf {\bibinfo {volume} {87}},\ \bibinfo
  {pages} {235122} (\bibinfo {year} {2013})}\BibitemShut {NoStop}%
\bibitem [{\citenamefont {Chen}\ \emph {et~al.}(2015)\citenamefont {Chen},
  \citenamefont {Burnell}, \citenamefont {Vishwanath},\ and\ \citenamefont
  {Fidkowski}}]{chen14}%
  \BibitemOpen
  \bibfield  {author} {\bibinfo {author} {\bibfnamefont {X.}~\bibnamefont
  {Chen}}, \bibinfo {author} {\bibfnamefont {F.~J.}\ \bibnamefont {Burnell}},
  \bibinfo {author} {\bibfnamefont {A.}~\bibnamefont {Vishwanath}},\ and\
  \bibinfo {author} {\bibfnamefont {L.}~\bibnamefont {Fidkowski}},\ }\bibfield
  {title} {\bibinfo {title} {Anomalous symmetry fractionalization and surface
  topological order},\ }\href {https://doi.org/10.1103/PhysRevX.5.041013}
  {\bibfield  {journal} {\bibinfo  {journal} {Phys. Rev. X}\ }\textbf {\bibinfo
  {volume} {5}},\ \bibinfo {pages} {041013} (\bibinfo {year}
  {2015})}\BibitemShut {NoStop}%
\bibitem [{\citenamefont {Bonderson}\ \emph {et~al.}(2013)\citenamefont
  {Bonderson}, \citenamefont {Nayak},\ and\ \citenamefont {Qi}}]{bonderson13}%
  \BibitemOpen
  \bibfield  {author} {\bibinfo {author} {\bibfnamefont {P.}~\bibnamefont
  {Bonderson}}, \bibinfo {author} {\bibfnamefont {C.}~\bibnamefont {Nayak}},\
  and\ \bibinfo {author} {\bibfnamefont {X.-L.}\ \bibnamefont {Qi}},\
  }\bibfield  {title} {\bibinfo {title} {A time-reversal invariant topological
  phase at the surface of a 3d topological insulator},\ }\href
  {http://stacks.iop.org/1742-5468/2013/i=09/a=P09016} {\bibfield  {journal}
  {\bibinfo  {journal} {Journal of Statistical Mechanics: Theory and
  Experiment}\ }\textbf {\bibinfo {volume} {2013}},\ \bibinfo {pages} {P09016}
  (\bibinfo {year} {2013})}\BibitemShut {NoStop}%
\bibitem [{\citenamefont {Wang}\ \emph {et~al.}(2013)\citenamefont {Wang},
  \citenamefont {Potter},\ and\ \citenamefont {Senthil}}]{wangc13b}%
  \BibitemOpen
  \bibfield  {author} {\bibinfo {author} {\bibfnamefont {C.}~\bibnamefont
  {Wang}}, \bibinfo {author} {\bibfnamefont {A.~C.}\ \bibnamefont {Potter}},\
  and\ \bibinfo {author} {\bibfnamefont {T.}~\bibnamefont {Senthil}},\
  }\bibfield  {title} {\bibinfo {title} {Gapped symmetry preserving surface
  state for the electron topological insulator},\ }\href
  {https://doi.org/10.1103/PhysRevB.88.115137} {\bibfield  {journal} {\bibinfo
  {journal} {Phys. Rev. B}\ }\textbf {\bibinfo {volume} {88}},\ \bibinfo
  {pages} {115137} (\bibinfo {year} {2013})}\BibitemShut {NoStop}%
\bibitem [{\citenamefont {Fidkowski}\ \emph {et~al.}(2013)\citenamefont
  {Fidkowski}, \citenamefont {Chen},\ and\ \citenamefont
  {Vishwanath}}]{fidkowski13}%
  \BibitemOpen
  \bibfield  {author} {\bibinfo {author} {\bibfnamefont {L.}~\bibnamefont
  {Fidkowski}}, \bibinfo {author} {\bibfnamefont {X.}~\bibnamefont {Chen}},\
  and\ \bibinfo {author} {\bibfnamefont {A.}~\bibnamefont {Vishwanath}},\
  }\bibfield  {title} {\bibinfo {title} {Non-abelian topological order on the
  surface of a 3d topological superconductor from an exactly solved model},\
  }\href {https://doi.org/10.1103/PhysRevX.3.041016} {\bibfield  {journal}
  {\bibinfo  {journal} {Phys. Rev. X}\ }\textbf {\bibinfo {volume} {3}},\
  \bibinfo {pages} {041016} (\bibinfo {year} {2013})}\BibitemShut {NoStop}%
\bibitem [{\citenamefont {Chen}\ \emph {et~al.}(2014)\citenamefont {Chen},
  \citenamefont {Fidkowski},\ and\ \citenamefont {Vishwanath}}]{chen14a}%
  \BibitemOpen
  \bibfield  {author} {\bibinfo {author} {\bibfnamefont {X.}~\bibnamefont
  {Chen}}, \bibinfo {author} {\bibfnamefont {L.}~\bibnamefont {Fidkowski}},\
  and\ \bibinfo {author} {\bibfnamefont {A.}~\bibnamefont {Vishwanath}},\
  }\bibfield  {title} {\bibinfo {title} {Symmetry enforced non-abelian
  topological order at the surface of a topological insulator},\ }\href
  {https://doi.org/10.1103/PhysRevB.89.165132} {\bibfield  {journal} {\bibinfo
  {journal} {Phys. Rev. B}\ }\textbf {\bibinfo {volume} {89}},\ \bibinfo
  {pages} {165132} (\bibinfo {year} {2014})}\BibitemShut {NoStop}%
\bibitem [{\citenamefont {{Metlitski}}\ \emph {et~al.}(2014)\citenamefont
  {{Metlitski}}, \citenamefont {{Fidkowski}}, \citenamefont {{Chen}},\ and\
  \citenamefont {{Vishwanath}}}]{metlitski14}%
  \BibitemOpen
  \bibfield  {author} {\bibinfo {author} {\bibfnamefont {M.~A.}\ \bibnamefont
  {{Metlitski}}}, \bibinfo {author} {\bibfnamefont {L.}~\bibnamefont
  {{Fidkowski}}}, \bibinfo {author} {\bibfnamefont {X.}~\bibnamefont
  {{Chen}}},\ and\ \bibinfo {author} {\bibfnamefont {A.}~\bibnamefont
  {{Vishwanath}}},\ }\bibfield  {title} {\bibinfo {title} {{Interaction effects
  on 3D topological superconductors: surface topological order from vortex
  condensation, the 16 fold way and fermionic Kramers doublets}},\ }\href@noop
  {} {\bibfield  {journal} {\bibinfo  {journal} {ArXiv e-prints}\ } (\bibinfo
  {year} {2014})},\ \Eprint {https://arxiv.org/abs/1406.3032} {arXiv:1406.3032}
  \BibitemShut {NoStop}%
\bibitem [{\citenamefont {Metlitski}\ \emph {et~al.}(2015)\citenamefont
  {Metlitski}, \citenamefont {Kane},\ and\ \citenamefont
  {Fisher}}]{metlitski15}%
  \BibitemOpen
  \bibfield  {author} {\bibinfo {author} {\bibfnamefont {M.~A.}\ \bibnamefont
  {Metlitski}}, \bibinfo {author} {\bibfnamefont {C.~L.}\ \bibnamefont
  {Kane}},\ and\ \bibinfo {author} {\bibfnamefont {M.~P.~A.}\ \bibnamefont
  {Fisher}},\ }\bibfield  {title} {\bibinfo {title} {Symmetry-respecting
  topologically ordered surface phase of three-dimensional electron topological
  insulators},\ }\href {https://doi.org/10.1103/PhysRevB.92.125111} {\bibfield
  {journal} {\bibinfo  {journal} {Phys. Rev. B}\ }\textbf {\bibinfo {volume}
  {92}},\ \bibinfo {pages} {125111} (\bibinfo {year} {2015})}\BibitemShut
  {NoStop}%
\bibitem [{\citenamefont {Witten}(2016)}]{witten15}%
  \BibitemOpen
  \bibfield  {author} {\bibinfo {author} {\bibfnamefont {E.}~\bibnamefont
  {Witten}},\ }\bibfield  {title} {\bibinfo {title} {Fermion path integrals and
  topological phases},\ }\href {https://doi.org/10.1103/RevModPhys.88.035001}
  {\bibfield  {journal} {\bibinfo  {journal} {Rev. Mod. Phys.}\ }\textbf
  {\bibinfo {volume} {88}},\ \bibinfo {pages} {035001} (\bibinfo {year}
  {2016})}\BibitemShut {NoStop}%
\bibitem [{\citenamefont {Wang}\ \emph {et~al.}(2016)\citenamefont {Wang},
  \citenamefont {Lin},\ and\ \citenamefont {Levin}}]{wangPRX2016}%
  \BibitemOpen
  \bibfield  {author} {\bibinfo {author} {\bibfnamefont {C.}~\bibnamefont
  {Wang}}, \bibinfo {author} {\bibfnamefont {C.-H.}\ \bibnamefont {Lin}},\ and\
  \bibinfo {author} {\bibfnamefont {M.}~\bibnamefont {Levin}},\ }\bibfield
  {title} {\bibinfo {title} {Bulk-boundary correspondence for three-dimensional
  symmetry-protected topological phases},\ }\href
  {https://doi.org/10.1103/PhysRevX.6.021015} {\bibfield  {journal} {\bibinfo
  {journal} {Phys. Rev. X}\ }\textbf {\bibinfo {volume} {6}},\ \bibinfo {pages}
  {021015} (\bibinfo {year} {2016})}\BibitemShut {NoStop}%
\bibitem [{\citenamefont {{Fidkowski}}\ \emph {et~al.}(2018)\citenamefont
  {{Fidkowski}}, \citenamefont {{Vishwanath}},\ and\ \citenamefont
  {{Metlitski}}}]{Fidkowski2018}%
  \BibitemOpen
  \bibfield  {author} {\bibinfo {author} {\bibfnamefont {L.}~\bibnamefont
  {{Fidkowski}}}, \bibinfo {author} {\bibfnamefont {A.}~\bibnamefont
  {{Vishwanath}}},\ and\ \bibinfo {author} {\bibfnamefont {M.~A.}\ \bibnamefont
  {{Metlitski}}},\ }\bibfield  {title} {\bibinfo {title} {{Surface Topological
  Order and a new 't Hooft Anomaly of Interaction Enabled 3+1D Fermion SPTs}},\
  }\href@noop {} {\bibfield  {journal} {\bibinfo  {journal} {ArXiv e-prints}\ }
  (\bibinfo {year} {2018})},\ \Eprint {https://arxiv.org/abs/1804.08628}
  {arXiv:1804.08628 [cond-mat.str-el]} \BibitemShut {NoStop}%
\bibitem [{\citenamefont {Chen}\ \emph {et~al.}(2011)\citenamefont {Chen},
  \citenamefont {Gu},\ and\ \citenamefont {Wen}}]{chen11b}%
  \BibitemOpen
  \bibfield  {author} {\bibinfo {author} {\bibfnamefont {X.}~\bibnamefont
  {Chen}}, \bibinfo {author} {\bibfnamefont {Z.-C.}\ \bibnamefont {Gu}},\ and\
  \bibinfo {author} {\bibfnamefont {X.-G.}\ \bibnamefont {Wen}},\ }\bibfield
  {title} {\bibinfo {title} {Classification of gapped symmetric phases in
  one-dimensional spin systems},\ }\href
  {https://doi.org/10.1103/PhysRevB.83.035107} {\bibfield  {journal} {\bibinfo
  {journal} {Phys. Rev. B}\ }\textbf {\bibinfo {volume} {83}},\ \bibinfo
  {pages} {035107} (\bibinfo {year} {2011})}\BibitemShut {NoStop}%
\bibitem [{\citenamefont {Fidkowski}\ and\ \citenamefont
  {Kitaev}(2011)}]{fidkowski11}%
  \BibitemOpen
  \bibfield  {author} {\bibinfo {author} {\bibfnamefont {L.}~\bibnamefont
  {Fidkowski}}\ and\ \bibinfo {author} {\bibfnamefont {A.}~\bibnamefont
  {Kitaev}},\ }\bibfield  {title} {\bibinfo {title} {Topological phases of
  fermions in one dimension},\ }\href
  {https://doi.org/10.1103/PhysRevB.83.075103} {\bibfield  {journal} {\bibinfo
  {journal} {Phys. Rev. B}\ }\textbf {\bibinfo {volume} {83}},\ \bibinfo
  {pages} {075103} (\bibinfo {year} {2011})}\BibitemShut {NoStop}%
\bibitem [{\citenamefont {Gaiotto}\ and\ \citenamefont
  {Kapustin}(2016)}]{Gaiotto2016}%
  \BibitemOpen
  \bibfield  {author} {\bibinfo {author} {\bibfnamefont {D.}~\bibnamefont
  {Gaiotto}}\ and\ \bibinfo {author} {\bibfnamefont {A.}~\bibnamefont
  {Kapustin}},\ }\bibfield  {title} {\bibinfo {title} {Spin tqfts and fermionic
  phases of matter},\ }\href {https://doi.org/10.1142/S0217751X16450445}
  {\bibfield  {journal} {\bibinfo  {journal} {International Journal of Modern
  Physics A}\ }\textbf {\bibinfo {volume} {31}},\ \bibinfo {pages} {1645044}
  (\bibinfo {year} {2016})}\BibitemShut {NoStop}%
\bibitem [{\citenamefont {Tarantino}\ and\ \citenamefont
  {Fidkowski}(2016)}]{Fidkowski1604}%
  \BibitemOpen
  \bibfield  {author} {\bibinfo {author} {\bibfnamefont {N.}~\bibnamefont
  {Tarantino}}\ and\ \bibinfo {author} {\bibfnamefont {L.}~\bibnamefont
  {Fidkowski}},\ }\bibfield  {title} {\bibinfo {title} {Discrete spin
  structures and commuting projector models for two-dimensional fermionic
  symmetry-protected topological phases},\ }\href
  {https://doi.org/10.1103/PhysRevB.94.115115} {\bibfield  {journal} {\bibinfo
  {journal} {Phys. Rev. B}\ }\textbf {\bibinfo {volume} {94}},\ \bibinfo
  {pages} {115115} (\bibinfo {year} {2016})}\BibitemShut {NoStop}%
\bibitem [{\citenamefont {Wang}\ \emph {et~al.}(2018)\citenamefont {Wang},
  \citenamefont {Ning},\ and\ \citenamefont {Chen}}]{Chen17}%
  \BibitemOpen
  \bibfield  {author} {\bibinfo {author} {\bibfnamefont {Z.}~\bibnamefont
  {Wang}}, \bibinfo {author} {\bibfnamefont {S.-Q.}\ \bibnamefont {Ning}},\
  and\ \bibinfo {author} {\bibfnamefont {X.}~\bibnamefont {Chen}},\ }\bibfield
  {title} {\bibinfo {title} {Exactly solvable model for two-dimensional
  topological superconductors},\ }\href
  {https://doi.org/10.1103/PhysRevB.98.094502} {\bibfield  {journal} {\bibinfo
  {journal} {Phys. Rev. B}\ }\textbf {\bibinfo {volume} {98}},\ \bibinfo
  {pages} {094502} (\bibinfo {year} {2018})}\BibitemShut {NoStop}%
\bibitem [{\citenamefont {{Bhardwaj}}\ \emph {et~al.}(2016)\citenamefont
  {{Bhardwaj}}, \citenamefont {{Gaiotto}},\ and\ \citenamefont
  {{Kapustin}}}]{gaiotto16}%
  \BibitemOpen
  \bibfield  {author} {\bibinfo {author} {\bibfnamefont {L.}~\bibnamefont
  {{Bhardwaj}}}, \bibinfo {author} {\bibfnamefont {D.}~\bibnamefont
  {{Gaiotto}}},\ and\ \bibinfo {author} {\bibfnamefont {A.}~\bibnamefont
  {{Kapustin}}},\ }\bibfield  {title} {\bibinfo {title} {{State sum
  constructions of spin-TFTs and string net constructions of fermionic phases
  of matter}},\ }\href@noop {} {\bibfield  {journal} {\bibinfo  {journal}
  {ArXiv e-prints}\ } (\bibinfo {year} {2016})},\ \Eprint
  {https://arxiv.org/abs/1605.01640} {arXiv:1605.01640} \BibitemShut {NoStop}%
\bibitem [{\citenamefont {{Brumfiel}}\ and\ \citenamefont
  {{Morgan}}(2016)}]{morgan16}%
  \BibitemOpen
  \bibfield  {author} {\bibinfo {author} {\bibfnamefont {G.}~\bibnamefont
  {{Brumfiel}}}\ and\ \bibinfo {author} {\bibfnamefont {J.}~\bibnamefont
  {{Morgan}}},\ }\bibfield  {title} {\bibinfo {title} {{The Pontrjagin Dual of
  3-Dimensional Spin Bordism}},\ }\href@noop {} {\bibfield  {journal} {\bibinfo
   {journal} {ArXiv e-prints}\ } (\bibinfo {year} {2016})},\ \Eprint
  {https://arxiv.org/abs/1612.02860} {arXiv:1612.02860 [math.AT]} \BibitemShut
  {NoStop}%
\bibitem [{\citenamefont {{Brumfiel}}\ and\ \citenamefont
  {{Morgan}}(2018)}]{Morgan2018}%
  \BibitemOpen
  \bibfield  {author} {\bibinfo {author} {\bibfnamefont {G.}~\bibnamefont
  {{Brumfiel}}}\ and\ \bibinfo {author} {\bibfnamefont {J.}~\bibnamefont
  {{Morgan}}},\ }\bibfield  {title} {\bibinfo {title} {{The Pontrjagin Dual of
  4-Dimensional Spin Bordism}},\ }\href@noop {} {\bibfield  {journal} {\bibinfo
   {journal} {ArXiv e-prints}\ } (\bibinfo {year} {2018})},\ \Eprint
  {https://arxiv.org/abs/1803.08147} {arXiv:1803.08147 [math.GT]} \BibitemShut
  {NoStop}%
\bibitem [{\citenamefont {{Chen}}\ \emph {et~al.}(2018)\citenamefont {{Chen}},
  \citenamefont {{Kapustin}}, \citenamefont {{Turzillo}},\ and\ \citenamefont
  {{You}}}]{Kapustin2018}%
  \BibitemOpen
  \bibfield  {author} {\bibinfo {author} {\bibfnamefont {Y.-A.}\ \bibnamefont
  {{Chen}}}, \bibinfo {author} {\bibfnamefont {A.}~\bibnamefont {{Kapustin}}},
  \bibinfo {author} {\bibfnamefont {A.}~\bibnamefont {{Turzillo}}},\ and\
  \bibinfo {author} {\bibfnamefont {M.}~\bibnamefont {{You}}},\ }\bibfield
  {title} {\bibinfo {title} {{Free and Interacting Short-Range Entangled Phases
  of Fermions: Beyond the Ten-Fold Way}},\ }\href@noop {} {\bibfield  {journal}
  {\bibinfo  {journal} {ArXiv e-prints}\ } (\bibinfo {year} {2018})},\ \Eprint
  {https://arxiv.org/abs/1809.04958} {arXiv:1809.04958 [cond-mat.str-el]}
  \BibitemShut {NoStop}%
\bibitem [{\citenamefont {Kane}\ and\ \citenamefont
  {Mele}(2005)}]{kanemele2005}%
  \BibitemOpen
  \bibfield  {author} {\bibinfo {author} {\bibfnamefont {C.~L.}\ \bibnamefont
  {Kane}}\ and\ \bibinfo {author} {\bibfnamefont {E.~J.}\ \bibnamefont
  {Mele}},\ }\bibfield  {title} {\bibinfo {title} {${Z}_{2}$ topological order
  and the quantum spin hall effect},\ }\href
  {https://doi.org/10.1103/PhysRevLett.95.146802} {\bibfield  {journal}
  {\bibinfo  {journal} {Phys. Rev. Lett.}\ }\textbf {\bibinfo {volume} {95}},\
  \bibinfo {pages} {146802} (\bibinfo {year} {2005})}\BibitemShut {NoStop}%
\bibitem [{\citenamefont {Ning}\ \emph {et~al.}(2015)\citenamefont {Ning},
  \citenamefont {Jiang},\ and\ \citenamefont {Liu}}]{Ning14}%
  \BibitemOpen
  \bibfield  {author} {\bibinfo {author} {\bibfnamefont {S.-Q.}\ \bibnamefont
  {Ning}}, \bibinfo {author} {\bibfnamefont {H.-C.}\ \bibnamefont {Jiang}},\
  and\ \bibinfo {author} {\bibfnamefont {Z.-X.}\ \bibnamefont {Liu}},\
  }\bibfield  {title} {\bibinfo {title} {Fermionic symmetry-protected
  topological phase induced by interactions},\ }\href
  {https://doi.org/10.1103/PhysRevB.91.241105} {\bibfield  {journal} {\bibinfo
  {journal} {Phys. Rev. B}\ }\textbf {\bibinfo {volume} {91}},\ \bibinfo
  {pages} {241105} (\bibinfo {year} {2015})}\BibitemShut {NoStop}%
\bibitem [{\citenamefont {Liu}\ \emph {et~al.}(2014)\citenamefont {Liu},
  \citenamefont {Gu},\ and\ \citenamefont
  {Wen}}]{bosonic_TI_zheng_xin_liu_2014_prl}%
  \BibitemOpen
  \bibfield  {author} {\bibinfo {author} {\bibfnamefont {Z.-X.}\ \bibnamefont
  {Liu}}, \bibinfo {author} {\bibfnamefont {Z.-C.}\ \bibnamefont {Gu}},\ and\
  \bibinfo {author} {\bibfnamefont {X.-G.}\ \bibnamefont {Wen}},\ }\bibfield
  {title} {\bibinfo {title} {Microscopic realization of two-dimensional bosonic
  topological insulators},\ }\href
  {https://doi.org/10.1103/PhysRevLett.113.267206} {\bibfield  {journal}
  {\bibinfo  {journal} {Phys. Rev. Lett.}\ }\textbf {\bibinfo {volume} {113}},\
  \bibinfo {pages} {267206} (\bibinfo {year} {2014})}\BibitemShut {NoStop}%
\bibitem [{\citenamefont {He}\ \emph {et~al.}(2015)\citenamefont {He},
  \citenamefont {Bhattacharjee}, \citenamefont {Moessner},\ and\ \citenamefont
  {Pollmann}}]{yin_chen_he_bqhe_2015_prl}%
  \BibitemOpen
  \bibfield  {author} {\bibinfo {author} {\bibfnamefont {Y.-C.}\ \bibnamefont
  {He}}, \bibinfo {author} {\bibfnamefont {S.}~\bibnamefont {Bhattacharjee}},
  \bibinfo {author} {\bibfnamefont {R.}~\bibnamefont {Moessner}},\ and\
  \bibinfo {author} {\bibfnamefont {F.}~\bibnamefont {Pollmann}},\ }\bibfield
  {title} {\bibinfo {title} {Bosonic integer quantum hall effect in an
  interacting lattice model},\ }\href
  {https://doi.org/10.1103/PhysRevLett.115.116803} {\bibfield  {journal}
  {\bibinfo  {journal} {Phys. Rev. Lett.}\ }\textbf {\bibinfo {volume} {115}},\
  \bibinfo {pages} {116803} (\bibinfo {year} {2015})}\BibitemShut {NoStop}%
\bibitem [{\citenamefont {Slagle}\ \emph {et~al.}(2015)\citenamefont {Slagle},
  \citenamefont {You},\ and\ \citenamefont
  {Xu}}]{bilayer_graphen_bosonic_SPT_model_cenke_2014_prb}%
  \BibitemOpen
  \bibfield  {author} {\bibinfo {author} {\bibfnamefont {K.}~\bibnamefont
  {Slagle}}, \bibinfo {author} {\bibfnamefont {Y.-Z.}\ \bibnamefont {You}},\
  and\ \bibinfo {author} {\bibfnamefont {C.}~\bibnamefont {Xu}},\ }\bibfield
  {title} {\bibinfo {title} {Exotic quantum phase transitions of strongly
  interacting topological insulators},\ }\href
  {https://doi.org/10.1103/PhysRevB.91.115121} {\bibfield  {journal} {\bibinfo
  {journal} {Phys. Rev. B}\ }\textbf {\bibinfo {volume} {91}},\ \bibinfo
  {pages} {115121} (\bibinfo {year} {2015})}\BibitemShut {NoStop}%
\bibitem [{\citenamefont {Wang}\ \emph
  {et~al.}(2015{\natexlab{b}})\citenamefont {Wang}, \citenamefont {Santos},\
  and\ \citenamefont {Wen}}]{Juven14edge}%
  \BibitemOpen
  \bibfield  {author} {\bibinfo {author} {\bibfnamefont {J.~C.}\ \bibnamefont
  {Wang}}, \bibinfo {author} {\bibfnamefont {L.~H.}\ \bibnamefont {Santos}},\
  and\ \bibinfo {author} {\bibfnamefont {X.-G.}\ \bibnamefont {Wen}},\
  }\bibfield  {title} {\bibinfo {title} {Bosonic anomalies, induced fractional
  quantum numbers, and degenerate zero modes: The anomalous edge physics of
  symmetry-protected topological states},\ }\href
  {https://doi.org/10.1103/PhysRevB.91.195134} {\bibfield  {journal} {\bibinfo
  {journal} {Phys. Rev. B}\ }\textbf {\bibinfo {volume} {91}},\ \bibinfo
  {pages} {195134} (\bibinfo {year} {2015}{\natexlab{b}})}\BibitemShut
  {NoStop}%
\bibitem [{\citenamefont {Hasan}\ and\ \citenamefont {Kane}(2010)}]{KaneRMP}%
  \BibitemOpen
  \bibfield  {author} {\bibinfo {author} {\bibfnamefont {M.~Z.}\ \bibnamefont
  {Hasan}}\ and\ \bibinfo {author} {\bibfnamefont {C.~L.}\ \bibnamefont
  {Kane}},\ }\href
  {https://journals.aps.org/rmp/abstract/10.1103/RevModPhys.82.3045} {\bibfield
   {journal} {\bibinfo  {journal} {Rev. Mod. Phys.}\ }\textbf {\bibinfo
  {volume} {82}},\ \bibinfo {pages} {3045} (\bibinfo {year}
  {2010})}\BibitemShut {NoStop}%
\bibitem [{\citenamefont {Lu}\ and\ \citenamefont {Vishwanath}(2012)}]{Lu2012}%
  \BibitemOpen
  \bibfield  {author} {\bibinfo {author} {\bibfnamefont {Y.-M.}\ \bibnamefont
  {Lu}}\ and\ \bibinfo {author} {\bibfnamefont {A.}~\bibnamefont
  {Vishwanath}},\ }\bibfield  {title} {\bibinfo {title} {Theory and
  classification of interacting integer topological phases in two dimensions: A
  chern-simons approach},\ }\href {https://doi.org/10.1103/PhysRevB.86.125119}
  {\bibfield  {journal} {\bibinfo  {journal} {Phys. Rev. B}\ }\textbf {\bibinfo
  {volume} {86}},\ \bibinfo {pages} {125119} (\bibinfo {year}
  {2012})}\BibitemShut {NoStop}%
\bibitem [{\citenamefont {Heinrich}\ and\ \citenamefont
  {Levin}(2018)}]{Levin2018}%
  \BibitemOpen
  \bibfield  {author} {\bibinfo {author} {\bibfnamefont {C.}~\bibnamefont
  {Heinrich}}\ and\ \bibinfo {author} {\bibfnamefont {M.}~\bibnamefont
  {Levin}},\ }\bibfield  {title} {\bibinfo {title} {Criteria for protected edge
  modes with ${\mathbb{z}}_{2}$ symmetry},\ }\href
  {https://doi.org/10.1103/PhysRevB.98.035101} {\bibfield  {journal} {\bibinfo
  {journal} {Phys. Rev. B}\ }\textbf {\bibinfo {volume} {98}},\ \bibinfo
  {pages} {035101} (\bibinfo {year} {2018})}\BibitemShut {NoStop}%
\bibitem [{\citenamefont {Lu}\ and\ \citenamefont
  {Vishwanath}(2016)}]{Lu13SET}%
  \BibitemOpen
  \bibfield  {author} {\bibinfo {author} {\bibfnamefont {Y.-M.}\ \bibnamefont
  {Lu}}\ and\ \bibinfo {author} {\bibfnamefont {A.}~\bibnamefont
  {Vishwanath}},\ }\bibfield  {title} {\bibinfo {title} {Classification and
  properties of symmetry-enriched topological phases: Chern-simons approach
  with applications to ${Z}_{2}$ spin liquids},\ }\href
  {https://doi.org/10.1103/PhysRevB.93.155121} {\bibfield  {journal} {\bibinfo
  {journal} {Phys. Rev. B}\ }\textbf {\bibinfo {volume} {93}},\ \bibinfo
  {pages} {155121} (\bibinfo {year} {2016})}\BibitemShut {NoStop}%
\bibitem [{\citenamefont {Fidkowski}\ and\ \citenamefont
  {Kitaev}(2010)}]{Fidkowski10}%
  \BibitemOpen
  \bibfield  {author} {\bibinfo {author} {\bibfnamefont {L.}~\bibnamefont
  {Fidkowski}}\ and\ \bibinfo {author} {\bibfnamefont {A.}~\bibnamefont
  {Kitaev}},\ }\bibfield  {title} {\bibinfo {title} {Effects of interactions on
  the topological classification of free fermion systems},\ }\href
  {https://doi.org/10.1103/PhysRevB.81.134509} {\bibfield  {journal} {\bibinfo
  {journal} {Phys. Rev. B}\ }\textbf {\bibinfo {volume} {81}},\ \bibinfo
  {pages} {134509} (\bibinfo {year} {2010})}\BibitemShut {NoStop}%
\bibitem [{\citenamefont {Ryu}\ and\ \citenamefont
  {Zhang}(2012)}]{ryu_zhang_2012}%
  \BibitemOpen
  \bibfield  {author} {\bibinfo {author} {\bibfnamefont {S.}~\bibnamefont
  {Ryu}}\ and\ \bibinfo {author} {\bibfnamefont {S.-C.}\ \bibnamefont
  {Zhang}},\ }\bibfield  {title} {\bibinfo {title} {Interacting topological
  phases and modular invariance},\ }\href
  {https://doi.org/10.1103/PhysRevB.85.245132} {\bibfield  {journal} {\bibinfo
  {journal} {Phys. Rev. B}\ }\textbf {\bibinfo {volume} {85}},\ \bibinfo
  {pages} {245132} (\bibinfo {year} {2012})}\BibitemShut {NoStop}%
\bibitem [{\citenamefont {Gu}\ and\ \citenamefont
  {Levin}(2014)}]{Gu_levin_2014}%
  \BibitemOpen
  \bibfield  {author} {\bibinfo {author} {\bibfnamefont {Z.-C.}\ \bibnamefont
  {Gu}}\ and\ \bibinfo {author} {\bibfnamefont {M.}~\bibnamefont {Levin}},\
  }\bibfield  {title} {\bibinfo {title} {Effect of interactions on
  two-dimensional fermionic symmetry-protected topological phases with
  ${Z}_{2}$ symmetry},\ }\href {https://doi.org/10.1103/PhysRevB.89.201113}
  {\bibfield  {journal} {\bibinfo  {journal} {Phys. Rev. B}\ }\textbf {\bibinfo
  {volume} {89}},\ \bibinfo {pages} {201113} (\bibinfo {year}
  {2014})}\BibitemShut {NoStop}%
\bibitem [{\citenamefont {Wang}\ \emph
  {et~al.}(2017{\natexlab{b}})\citenamefont {Wang}, \citenamefont {Lin},\ and\
  \citenamefont {Gu}}]{Chenjie2016}%
  \BibitemOpen
  \bibfield  {author} {\bibinfo {author} {\bibfnamefont {C.}~\bibnamefont
  {Wang}}, \bibinfo {author} {\bibfnamefont {C.-H.}\ \bibnamefont {Lin}},\ and\
  \bibinfo {author} {\bibfnamefont {Z.-C.}\ \bibnamefont {Gu}},\ }\bibfield
  {title} {\bibinfo {title} {Interacting fermionic symmetry-protected
  topological phases in two dimensions},\ }\href
  {https://doi.org/10.1103/PhysRevB.95.195147} {\bibfield  {journal} {\bibinfo
  {journal} {Phys. Rev. B}\ }\textbf {\bibinfo {volume} {95}},\ \bibinfo
  {pages} {195147} (\bibinfo {year} {2017}{\natexlab{b}})}\BibitemShut
  {NoStop}%
\bibitem [{\citenamefont {{Alavirad}}\ and\ \citenamefont
  {{Barkeshli}}(2019)}]{Massaim_2019}%
  \BibitemOpen
  \bibfield  {author} {\bibinfo {author} {\bibfnamefont {Y.}~\bibnamefont
  {{Alavirad}}}\ and\ \bibinfo {author} {\bibfnamefont {M.}~\bibnamefont
  {{Barkeshli}}},\ }\bibfield  {title} {\bibinfo {title} {{Anomalies and
  unnatural stability of multi-component Luttinger liquids in
  $\mathbb{Z}_n\times\mathbb{Z}_n$ spin chains}},\ }\href@noop {} {\bibfield
  {journal} {\bibinfo  {journal} {arXiv e-prints}\ ,\ \bibinfo {eid}
  {arXiv:1910.00589}} (\bibinfo {year} {2019})},\ \Eprint
  {https://arxiv.org/abs/1910.00589} {arXiv:1910.00589 [cond-mat.str-el]}
  \BibitemShut {NoStop}%
\bibitem [{\citenamefont {{Sullivan}}\ and\ \citenamefont
  {{Cheng}}(2019)}]{Meng2019_04}%
  \BibitemOpen
  \bibfield  {author} {\bibinfo {author} {\bibfnamefont {J.}~\bibnamefont
  {{Sullivan}}}\ and\ \bibinfo {author} {\bibfnamefont {M.}~\bibnamefont
  {{Cheng}}},\ }\bibfield  {title} {\bibinfo {title} {{Interacting Edge States
  of Fermionic Symmetry-Protected Topological Phases in Two Dimensions}},\
  }\href@noop {} {\bibfield  {journal} {\bibinfo  {journal} {arXiv e-prints}\
  ,\ \bibinfo {eid} {arXiv:1904.08953}} (\bibinfo {year} {2019})},\ \Eprint
  {https://arxiv.org/abs/1904.08953} {arXiv:1904.08953 [cond-mat.str-el]}
  \BibitemShut {NoStop}%
\bibitem [{\citenamefont {Wang}(2016)}]{Chenjie1610}%
  \BibitemOpen
  \bibfield  {author} {\bibinfo {author} {\bibfnamefont {C.}~\bibnamefont
  {Wang}},\ }\bibfield  {title} {\bibinfo {title} {Braiding statistics and
  classification of two-dimensional charge-$2m$ superconductors},\ }\href
  {https://doi.org/10.1103/PhysRevB.94.085130} {\bibfield  {journal} {\bibinfo
  {journal} {Phys. Rev. B}\ }\textbf {\bibinfo {volume} {94}},\ \bibinfo
  {pages} {085130} (\bibinfo {year} {2016})}\BibitemShut {NoStop}%
\end{thebibliography}%

%\section{ Intrinsic interacting fermionic SPT }
%
%\subsection{$\Z_4^f\times \Z_4 \times \Z_4$ symmetry}
%

\appendix

\section{Other solutions of $\Z_2\times \Z_2 \times \Z_2^f$ symmetry}
\label{append:z2z2z2d}
%We denote $g_1$ and $g_2$ as the two generators of the two symmetry subgroup which satisfy group relations $g_1^2=g_2^2=1$ and $g_1g_2=g_2g_1$ and $g_{12}=g_1g_2$.
%A set of relation for this symmetry:
%\begin{align}
%(W^{g_1})^TKW^{g_1}=K\\
%(W^{g_2})^TKW^{g_2}=K\\
%(W^{g_{12}})^TKW^{g_{12}}=K
%\end{align}
%and 
%\begin{align}
%&(W^{g_1})^2=(W^{g_2})^2=(W^{g_{12}})^2=1_{2\times 2} \\
%&(W^{g_1}+1_{2 \times 2})\delta \phi^{g_1}=0 \label{phig1} \\
%&(W^{g_2}+1_{2 \times 2})\delta \phi^{g_2}=0 \label{phig2}\\
%&(W^{g_1}+ W^{g_2})(\delta \phi^{g_2} + W^{g_1} \delta \phi^{g_1})=0 \label{phig1g2}
%\end{align}
%
%From these relation and the fact that $W^{g}\in GL(2, Z)$, we have the following solutions
%
%\begin{align}
%W^{g_1}=\pm 1_{2\times 2}, \pm \sigma_z,\\
%W^{g_2}=\pm 1_{2\times 2}, \pm \sigma_z.
%\end{align}

\textbf{1. Solutions with $W^{g_1}= W^{g_2}=1_{2\times 2}$}\\

From (\ref{phig1}) and (\ref{phig2}), we have 
\begin{align}
\delta \phi^{g_1}=\pi \begin{pmatrix} t_1^{g_1} \\ t_2^{g_1} \end{pmatrix}, \quad \delta \phi^{g_2}=\pi \begin{pmatrix} t_1^{g_2} \\ t_2^{g_2} \end{pmatrix}, \quad t_{1,2}^{g_{1,2}}=0,1
\end{align}
The (\ref{phig1g2}) does not give any new constraint.  We then denote these solutions as $[1_{2\times 2}, 1_{2\times 2}, (t_1^{g_1},t_2^{g_1}), (t_1^{g_2},t_2^{g_2})]$.
%\begin{align}
%t_1^{g_1}=t_1^{g_2}, \quad t_2^{g_1}=t_2^{g_2}
%\end{align}
%Therefore,
%\begin{align}
%\delta \phi^{g_1}=\delta \phi^{g_2}=\delta \phi=\pi \begin{pmatrix} t_1 \\ t_2 \end{pmatrix}, \quad t_1, t_2=0,1
%\end{align}
%From the symmetry realization, i.e. $W^{g_1}=W^{g_2}=1_{2\times 2}$, $\delta\phi^{g_1}=\delta \phi^{g_2}=\delta \phi= \pi (t_1, t_2)^T$, it seems that the $Z_2 \times Z_2$ symmetry behave as $Z_2$ symmetry, therefore, the edge can not be gapped out for $[W^{g_1}, W^{g_2}, t_1, t_2]$=$[1_{2\times 2}, 1_{2\times 2}, t_1, t_2]$=$[1_{2\times 2},1_{2\times 2},0,1]$ or $[1_{2\times 2},1_{2\times 2},1,0]$.
 Similar to the case discussed in Sec.\ref{sec_z4z2f_groupstru}, we have 
\begin{align}
&[1_{2\times 2}, 1_{2\times 2}, (t_1^{g_1},t_2^{g_1}), (t_1^{g_2},t_2^{g_2})]\nonumber \\
\oplus& [1_{2\times 2}, 1_{2\times 2}, (t_2^{g_1},t_1^{g_1}), (t_2^{g_2},t_1^{g_2})]=1
\label{eqn_z2z2z2f_append_relation_root3_1}
\end{align}
and 
\begin{align}
[1_{2\times 2}, 1_{2\times 2}, (t_1^{g_1},t_1^{g_1}), (t_1^{g_2},t_1^{g_2})]=1
\end{align}
Therefore, we only need to consider the following six cases: $[1_{2\times 2},1_{2\times 2},(0,0),(0,1)]$, $[1_{2\times 2},1_{2\times 2},(0,1),(0,0)]$, $[1_{2\times 2},1_{2\times 2},(0,1),(0,1)]$, $[1_{2\times 2},1_{2\times 2},(0,1),(1,0)]$, $[1_{2\times 2},1_{2\times 2},(0,1),(1,1)]$ and $[1_{2\times 2},1_{2\times 2},(1,1),(0,1)]$.\\

\textit{1.1 Solution: $[1_{2\times 2},1_{2\times 2}, (0,0), (0,1)]$}\\

We can show that 
\begin{align}
[1_{2\times 2},1_{2\times 2},(0,0),(0,1)]=[1_{2\times 2}, \sigma_z, (0,0), 0]^{\oplus 2}
\label{append_z2z2z2f_1_0}
\end{align}
whose structure factor is  
\begin{align}
r([1_{2\times 2},1_{2\times 2},(0,0),(0,1)])=(0,2,0).
\end{align}
%\textit{1.1 Solution: $[1_{2\times 2},1_{2\times 2},0,1]$}\\
To show (\ref{append_z2z2z2f_1_0}), we consider the stacking system
\begin{align}
[1_{2\times 2}, \sigma_z, (0,0), 0]^{\oplus 2}\oplus [1_{2\times 2},1_{2\times 2},(0,0), (1,0)].
\label{append_z2z2z2f_1}
\end{align}
We denote the edge bosonic fields of these three states by $\phi_1^{(1)},\phi_2^{(1)}, \phi_1^{(2)},  \phi_2^{(2)}, \phi_1,\phi_2$.
Under symmetry, 
\begin{align}
&g_{1}:\begin{pmatrix} 
\phi_1^{(\alpha)} \\ 
\phi_2^{(\alpha)}
\end{pmatrix} \rightarrow  \begin{pmatrix} 
\phi_1^{(\alpha)}  \\ 
\phi_2^{(\alpha)} 
\end{pmatrix}\\
&g_{2}:\begin{pmatrix} 
\phi_1^{(\alpha)} \\ 
\phi_2^{(\alpha)}
\end{pmatrix} \rightarrow  \begin{pmatrix} 
\phi_1^{(\alpha)}  \\ 
-\phi_2^{(\alpha)} 
\end{pmatrix}\\
&g_{1}:\begin{pmatrix} 
\phi_1 \\ 
\phi_2
\end{pmatrix} \rightarrow  \begin{pmatrix} 
\phi_1 \\ 
\phi_2
\end{pmatrix}\\
&g_{2}:\begin{pmatrix} 
\phi_1 \\ 
\phi_2
\end{pmatrix} \rightarrow  \begin{pmatrix} 
\phi_1+\pi \\ 
\phi_2
\end{pmatrix}
\end{align}
where $i,\alpha=1,2$.
We define the following majorana fermions  through the bosonic edge fields by
\begin{align}
&\eta_R^1+i \eta_R^2= \frac{1}{\sqrt{\pi}} e^{i\phi_2},\,
\eta_L^1+i \eta_L^2= \frac{1}{\sqrt{\pi}} e^{-i\phi_1},\label{append_z2z2z2f_majorana}\\
&\xi_R^{\alpha 1}+i \xi_R^{\alpha 2}= \frac{1}{\sqrt{\pi}} e^{i\phi_2^{(\alpha)}},\,
\xi_L^{\alpha 1}+i \xi_L^{\alpha2}= \frac{1}{\sqrt{\pi}} e^{-i\phi_1^{(\alpha)}},
\end{align}
which  under symmetry transform as
\begin{align}
% g_1:  \, & \eta_R^i  \rightarrow   \eta_R^i ,\,  \eta_L^{i} \rightarrow \eta_L^i,\\ 
g_{1}:\, & \eta_R^i  \rightarrow   \eta_R^i ,\,  \eta_L^{i} \rightarrow\eta_L^i,\\ 
g_{2}:\, & \eta_R^i  \rightarrow   \eta_R^i ,\,  \eta_L^{i} \rightarrow -\eta_L^i,\\ 
g_{1}:\, 
  &  \xi_R^{\alpha i}  \rightarrow \xi_R^{\alpha j} ,\, \xi_L^{\alpha i } \rightarrow \xi_L^{\alpha i} \\
  g_{2}:\, 
  &  \xi_R^{\alpha i}  \rightarrow(-1)^{i-1} \xi_R^{\alpha i} ,\, \xi_L^{\alpha i } \rightarrow \xi_L^{\alpha i} 
  \end{align}
  where $i,j,\alpha=1,2$ and the repeated $j$ is summed. These twelve majorana fermions can be gapped out by the following symmetric mass terms
  \begin{align}
  im_{i}  \eta_R^i \xi_L^{1i}+ i \tilde {m}_{i}  \xi_R^{i 2} \eta_L^{i }+ i \hat{m}_{ i}  \xi_R^{i 1} \xi_L^{2i} 
  \end{align}
where repreated index $i $ is summed. Therefore, the stacking system (\ref{append_z2z2z2f_1}) is trivial  and then the relation  (\ref{append_z2z2z2f_1_0}) is proved.\\

\textit{1.2 Solution: $[1_{2\times 2},1_{2\times 2}, (0,1), (0,0)]$}\\

Similar to the  $Solution\, 1.1$, we have 
\begin{align}
[\sigma, 1_{2\times 2}, 0]\oplus [\sigma,1_{2\times 2}, 0]=[1_{2\times 2},1_{2\times 2},(0,1),(0,0)]
\label{append_z2z2z2f_1_2}
\end{align}
which means by using the three-component vector, 
\begin{align}
r([1_{2\times 2},1_{2\times 2},(0,1),(0,0)])=(2,0,0).
\end{align}

\textit{1.3 Solution: $[1_{2\times 2},1_{2\times 2}, (0,1), (0,1)]$}\\

For this case, we will show that 
\begin{align}
&[1_{2\times 2},1_{2\times 2}, (0,1), (0,1)]\oplus[1_{2\times 2},1_{2\times 2}, (1,0), (0,0)]\nonumber \\
&[1_{2\times 2},1_{2\times 2}, (0,0), (1,0)]\oplus [-1_{2\times 2},-1_{2\times 2}, 0,1]^{\oplus 2}=1.
\label{eqn_append_z2z2z2f_relation_0}
\end{align}
We denote the two bosonic edge fields of the five phases in the above stacking system as  $\phi_{1,2}$, $\varphi_{1,2}$, $ \tilde{\varphi}_{1,2}$, $ \rho^1_{1,2}$ and $ \rho^2_{1,2}$ respectively, which transform under symmetry as 
\begin{align}
%g_{1,2}\,:& \begin{pmatrix} \phi_1\\ \phi_2 \end{pmatrix}\rightarrow \begin{pmatrix} \phi_1\\ \phi_2+\pi \end{pmatrix}\\
g_1 : \begin{pmatrix} \phi_1\\ \phi_2\\ \varphi_1\\ \varphi_2\\\tilde\varphi_1\\ \tilde\varphi_2 \\ \rho^a_{1}\\\rho^a_{2}\end{pmatrix}\rightarrow \begin{pmatrix}\phi_1\\ \phi_2+\pi \\ \varphi_1+\pi\\ \varphi_2\\\tilde\varphi_1\\ \tilde\varphi_2 \\-\rho^a_{1}\\-\rho^a_{2}+\pi\end{pmatrix},\,
g_2 : \begin{pmatrix} \phi_1\\ \phi_2\\ \varphi_1\\ \varphi_2\\\tilde\varphi_1\\ \tilde\varphi_2\\\rho^a_{1}\\\rho^a_{2} \end{pmatrix}\rightarrow \begin{pmatrix} \phi_1\\ \phi_2+\pi \\\varphi_1\\ \varphi_2\\\tilde\varphi_1+\pi\\ \tilde\varphi_2\\-\rho^a_{1}\\-\rho^a_{2}\end{pmatrix}\label{eqn_append_z2z2z2f_relation_1_1}
\end{align}
Similar to (\ref{append_z2z2z2f_majorana}), we define the following majorana fermions
\begin{align}
&\eta_{R,L}^1+i \eta_{R,L}^2= \frac{1}{\sqrt{\pi}} e^{\pm i\phi_{2,1}},\label{append_z2z2z2f_majorana_3_1}\\
&\xi_{R,L}^1+i \xi_{R,L}^2= \frac{1}{\sqrt{\pi}} e^{\pm i\varphi_{2,1}},\label{append_z2z2z2f_majorana_3_2}\\
&\chi_{R,L}^1+i \chi_{R,L}^2= \frac{1}{\sqrt{\pi}} e^{\pm i\tilde\varphi_{2,1}},\label{append_z2z2z2f_majorana_3_3}\\
&\gamma_{1R,1L}^1+i \gamma_{1R,1L}^2= \frac{1}{\sqrt{\pi}} e^{\pm i\rho^1_{2,1}},\label{append_z2z2z2f_majorana_3_4}\\
&\gamma_{2R,2L}^1+i \gamma_{2R,2L}^2= \frac{1}{\sqrt{\pi}} e^{\pm i\tilde \rho^2_{2,1}}.\label{append_z2z2z2f_majorana_3_5}
\end{align}

We can fully gap out the edge fields of the stacking system by adding the following mass terms
\begin{align}
&im_{1i} \xi_R^i\eta_L^i+im_{2i}\eta_R^i \gamma_{iL}^2+im_{2i}\gamma_{iR}^1\xi_L^i\nonumber \\
&  +im_{4i}\chi_R^i\gamma_{iL}^1+im_{5i}\gamma_{iR}^2\chi_L^i
\end{align}
where repeated $i$ is summed. The symmetry properties of these majorana fermions can be inherited from (\ref{eqn_append_z2z2z2f_relation_1}), and it turns out that all these mass terms are all symmetric. Therefore, the stacking system (\ref{eqn_append_z2z2z2f_relation_0}) is trivial. Further, from (\ref{eqn_z2z2z2f_append_relation_root3_1}),  (\ref{append_z2z2z2f_1_0}) and  (\ref{eqn_z2z2z2f_relation_root3}), we can see the phase $[1_{2\times 2},1_{2\times 2}, (0,1), (0,1)]$ is related to the three root states by
\begin{align}
&[1_{2\times 2},1_{2\times 2}, (0,1), (0,1)]=[1_{2\times 2},\sigma_z,(0,0),0]^{\oplus 2}\oplus \nonumber \\
&[\sigma_z,1_{2\times 2},(0,0),0]^{\oplus 2}\oplus [-1_{2\times 2},-1_{2\times 2}, 0,1]^{\oplus 2}
\end{align}
which indicates that its structure factor is 
\begin{align}
r([1_{2\times 2},1_{2\times 2}, (0,1), (0,1)])=(2,2,2).
\label{eqn_append_z2z2z2f_relation_100}
\end{align}

\textit{1.4 Solution: $[1_{2\times 2},1_{2\times 2}, (0,1), (1,0)]$}\\

We can show that 
\begin{align}
[1_{2\times 2},1_{2\times 2}, (0,1), (1,0)]&=[1_{2\times 2},1_{2\times 2}, (0,1), (0,0)]\nonumber \\
&\oplus[1_{2\times 2},1_{2\times 2}, (0,0), (1,0)]
\label{eqn_append_z2z2z2f_relation_2}
\end{align}
which indicates its structure factor is 
\begin{align}
r([1_{2\times 2},1_{2\times 2}, (0,1), (1,0)])=(2,2,0).
\label{eqn_append_z2z2z2f_relation_2_00}
\end{align}
To show (\ref{eqn_append_z2z2z2f_relation_2}), 
 is equivalent to show that 
\begin{align}
&[1_{2\times 2},1_{2\times 2}, (0,1), (1,0)]\oplus [1_{2\times 2},1_{2\times 2}, (1,0), (0,0)]\nonumber \\
&\oplus[1_{2\times 2},1_{2\times 2}, (0,0), (0,1)]=1.
\label{eqn_append_z2z2z2f_relation_3}
\end{align}
To show (\ref{eqn_append_z2z2z2f_relation_2}), we denote the bosonic edge fields of the three states by $\phi_{1,2}$, $\varphi_{1,2}$ and $\tilde \varphi_{1,2}$ respectively, which transform under  symmetry as 
\begin{align}
g_1 : \begin{pmatrix} \phi_1\\ \phi_2\\ \varphi_1\\ \varphi_2\\\tilde\varphi_1\\ \tilde\varphi_2 \end{pmatrix}\rightarrow \begin{pmatrix}\phi_1\\ \phi_2+\pi \\ \varphi_1+\pi\\ \varphi_2\\\tilde\varphi_1\\ \tilde\varphi_2 \end{pmatrix},\,
g_2 : \begin{pmatrix} \phi_1\\ \phi_2\\ \varphi_1\\ \varphi_2\\\tilde\varphi_1\\ \tilde\varphi_2\end{pmatrix}\rightarrow \begin{pmatrix} \phi_1+\pi \\ \phi_2\\\varphi_1\\ \varphi_2\\\tilde\varphi_1\\ \tilde\varphi_2+\pi\end{pmatrix}\label{eqn_append_z2z2z2f_relation_1}
\end{align}

From these bosonic edge fields, we define the six majorana fermions as (\ref{append_z2z2z2f_majorana_3_1})-(\ref{append_z2z2z2f_majorana_3_3}). We can  fully gap out the edge fields by adding the mass terms 
\begin{align}
im_{1i} \eta_R^i\xi_L^i+im_{2i} \xi_R^i\chi_L^i+im_{3i} \chi_R^i\eta_L^i.
\end{align}
The symmetry properties can be  inherited from those of bosonic edge fields, and it turns out that all the mass terms are symmetric. Therefore, the stacking system (\ref{eqn_append_z2z2z2f_relation_3}) is trivial.\\

\textit{1.5 Solution: $[1_{2\times 2},1_{2\times 2}, (0,1), (1,1)]$}\\

We can show that 
\begin{align}
[1_{2\times 2},1_{2\times 2}, (0,1), (1,1)]&=[1_{2\times 2},1_{2\times 2}, (0,0), (1,0)]\nonumber \\
&\oplus[1_{2\times 2},1_{2\times 2}, (0,1), (0,1)]
\label{eqn_append_z2z2z2f_relation_4}
\end{align}
Since from (\ref{eqn_z2z2z2f_append_relation_root3_1}) $[1_{2\times 2},1_{2\times 2}, (0,0), (1,0)]$ is the inverse of $[1_{2\times 2},1_{2\times 2}, (0,0), (0,1)]$, 
the structure factor of $[1_{2\times 2},1_{2\times 2}, (0,1), (1,1)]$ is 
\begin{align}
r([1_{2\times 2},1_{2\times 2}, (0,1), (1,1)])=(2,0,2).
\label{eqn_z2z2z2f_append_relation_8}
\end{align}
which means that  $[1_{2\times 2},1_{2\times 2}, (0,1), (1,1)]$ is equivalent to $[1_{2\times 2},1_{2\times 2}, (0,1), (0,0)]$.
To show (\ref{eqn_append_z2z2z2f_relation_4}) is equivalent to show that 
\begin{align}
&[1_{2\times 2},1_{2\times 2}, (0,1), (1,1)]\oplus [1_{2\times 2},1_{2\times 2}, (0,0), (1,0)]\nonumber \\
&\oplus[1_{2\times 2},1_{2\times 2}, (1,0), (1,0)]=1.
\label{eqn_append_z2z2z2f_relation_5}
\end{align}
To show (\ref{eqn_append_z2z2z2f_relation_5}), we denote the bosonic edge fields of the three states by $\phi_{1,2}$, $\varphi_{1,2}$ and $\tilde \varphi_{1,2}$ respectively, which transform under  symmetry as 
\begin{align}
g_1 : \begin{pmatrix} \phi_1\\ \phi_2\\ \varphi_1\\ \varphi_2\\\tilde\varphi_1\\ \tilde\varphi_2 \end{pmatrix}\rightarrow \begin{pmatrix}\phi_1\\ \phi_2+\pi \\ \varphi_1\\ \varphi_2\\\tilde\varphi_1+\pi\\ \tilde\varphi_2 \end{pmatrix},\,
g_2 : \begin{pmatrix} \phi_1\\ \phi_2\\ \varphi_1\\ \varphi_2\\\tilde\varphi_1\\ \tilde\varphi_2\end{pmatrix}\rightarrow \begin{pmatrix} \phi_1+\pi \\ \phi_2+\pi \\\varphi_1\\ \varphi_2+\pi \\\tilde\varphi_1+\pi\\ \tilde\varphi_2\end{pmatrix}\label{eqn_append_z2z2z2f_relation_1_2}
\end{align}

From these bosonic edge fields, we define the six majorana fermions as (\ref{append_z2z2z2f_majorana_3_1})-(\ref{append_z2z2z2f_majorana_3_3}). We can  fully gap out the edge fields by adding the mass terms 
\begin{align}
im_{1i} \eta_R^i\chi_L^i+im_{2i} \xi_R^i\eta_L^i+im_{3i} \chi_R^i\xi_L^i.
\end{align}
The symmetry properties can be  inherited from those of bosonic edge fields, and it turns out that all the mass terms are symmetric. Therefore, the stacking system (\ref{eqn_append_z2z2z2f_relation_5}) is trivial.\\

\textit{1.6 Solution: $[1_{2\times 2},1_{2\times 2}, (1,1), (0,1)]$}\\

This case can be achieved just by exchanging the two $\Z_2$ subgroups. Therefore, we have
\begin{align}
[1_{2\times 2},1_{2\times 2}, (1,1), (0,1)]&=[1_{2\times 2},1_{2\times 2},  (1,0),(0,0)]\nonumber \\
&\oplus[1_{2\times 2},1_{2\times 2}, (0,1), (0,1)]
\label{eqn_append_z2z2z2f_relation_4}
\end{align}
Compared to the phase  $[1_{2\times 2},1_{2\times 2}, (0,1), (1,1)]$,  the structure factor of $[1_{2\times 2},1_{2\times 2}, (1,1), (0,1)]$ is
\begin{align}
r([1_{2\times 2},1_{2\times 2}, (1,1), (0,1)])=(0,2,2).
\end{align}

\textbf{2. Solutions with $W^{g_1}=1_{2\times 2}, W^{g_2}=-1_{2\times 2}$}\\

From (\ref{phig1}), we have
\begin{align}
\delta \phi^{g_1} = \pi \begin{pmatrix} t_1^{g_1} \\ t_2^{g_1}\end{pmatrix}, 
t_{1,2}^{g_1}=0,1
\end{align}
From (\ref{phig2}) and gauge transformation, we can fix $\delta\phi^{g_2}=0$.
We can denote the phases corresponding to $W^{g_1}=1_{2\times 2}, W^{g_2}=-1_{2\times 2}$ by $[1_{2\times 2},-1_{2\times 2},t_1^{g_1},t_2^{g_1}]$.

 Similar to the case discussed in Sec.\ref{sec_z4z2f_groupstru}, we have 
\begin{align}
[1_{2\times 2},-1_{2\times 2},t_1^{g_1},t_2^{g_1}]=[1_{2\times 2},-1_{2\times 2},t_2^{g_1},t_1^{g_1}]^{-1}
\label{eqn_append_z2z2z2f_relation_7}
\end{align}
and 
\begin{align}
[1_{2\times 2},-1_{2\times 2},t_1^{g_1},t_1^{g_1}]=1.
\end{align}
Therefore, we only need to consider $[1_{2\times 2},-1_{2\times 2},0,1]$, whose structure factor, as we will show, is 
\begin{align}
r([1_{2\times 2},-1_{2\times 2},0,1])=(2,0,3).
\label{eqn_append_z2z2z2f_relation_6}
\end{align} 

We now prove one phase relation
\begin{align}
&[1_{2\times 2},-1_{2\times 2},0,1]\oplus [-1_{2\times 2},-1_{2\times 2},1,0]\nonumber \\
&\oplus [1_{2\times 2},1_{2\times 2},(1,0),(1,1)]=1.
\label{eqn_append_z2z2z2f_relation_6_1}
\end{align}

To show (\ref{eqn_append_z2z2z2f_relation_6_1}), we denote the bosonic edge fields of the three states by $\phi_{1,2}$, $\varphi_{1,2}$ and $\tilde \varphi_{1,2}$ respectively, which transform under  symmetry as 
\begin{align}
g_1 : \begin{pmatrix} \phi_1\\ \phi_2\\ \varphi_1\\ \varphi_2\\\tilde\varphi_1\\ \tilde\varphi_2 \end{pmatrix}\rightarrow \begin{pmatrix}\phi_1\\ \phi_2+\pi \\ -\varphi_1+\pi\\ -\varphi_2\\\tilde\varphi_1+\pi\\ \tilde\varphi_2 \end{pmatrix},\,
g_2 : \begin{pmatrix} \phi_1\\ \phi_2\\ \varphi_1\\ \varphi_2\\\tilde\varphi_1\\ \tilde\varphi_2\end{pmatrix}\rightarrow \begin{pmatrix} -\phi_1 \\- \phi_2\\-\varphi_1\\ -\varphi_2\\\tilde\varphi_1\\ \tilde\varphi_2\end{pmatrix}\label{eqn_append_z2z2z2f_relation_6_1_1}
\end{align}
From these bosonic edge fields, we define the six majorana fermions as (\ref{append_z2z2z2f_majorana_3_1})-(\ref{append_z2z2z2f_majorana_3_3}). We can  fully gap out the edge fields by adding the mass terms 
\begin{align}
&im_{1} \eta_R^1\xi_L^1+im_{2} \eta_R^2\chi_L^1+im_{3} \xi_R^1\eta_L^1\nonumber \\
+&im_4 \chi_R^1\eta_L^2+im_5 \xi_R^2\chi_L^2+im_6\chi_R^2\xi_L^2.
\end{align}
The symmetry properties can be  inherited from those of bosonic edge fields, and it turns out that all the mass terms are symmetric. Therefore, the stacking system (\ref{eqn_append_z2z2z2f_relation_6_1}) is trivial. From (\ref{eqn_z2z2z2f_root_relation_00}), (\ref{eqn_z2z2z2f_append_relation_root3_1}) and (\ref{eqn_z2z2z2f_append_relation_8}), we then obtain (\ref{eqn_append_z2z2z2f_relation_6}).\\

\textbf{3. Solutions with $W^{g_1}=1_{2\times 2}, W^{g_2}=-\sigma_z$}\\

From (\ref{phig1}), we have
\begin{align}
\delta \phi^{g_1} = \pi \begin{pmatrix} t_1^{g_1} \\ t_2^{g_1}\end{pmatrix}, 
t_{1,2}^{g_1, g_2}=0,1
\end{align}

From (\ref{phig2}) and gauge transformation, we have
\begin{align}
\delta \phi^{g_2} = \pi \begin{pmatrix} 0 \\ t_2^{g_2} \end{pmatrix}, t_{2}^{g_2}=0,1.
\end{align}
%where we have used the gauge transformaion to set $\delta \phi_1^{g_2}=0$.

%From (\ref{phig1g2}), we have
%\begin{align}
%t_{1}^{g_1}=0 \text{ mod }2, \quad t_{2}^{g_1}+t_2^{g_2}=0 \text{ mod } 2,
%\end{align}
%therefore
%\begin{align}
%\delta \phi^{g_1}=\delta \phi^{g_2}=\pi \begin{pmatrix} 0 \\ t\end{pmatrix}, t=0,1.
%\end{align}
We use $[1_{2\times 2}, -\sigma_z, (t_1^{g_1},t_2^{g_1}),t_2^{g_2}]$ to denote different  phases. We can generally show that the phases can be related to $[1_{2\times 2}, \sigma_z, (t_2^{g_1},t_1^{g_1}),t_2^{g_2}]$, namely we have the relation
\begin{align}
[1_{2\times 2}, -\sigma_z, (t_1,t_2),t_3]\oplus [1_{2\times 2}, \sigma_z, (t_2,t_1),t_3]=1.
\label{eqn_append_relation_13_1}
\end{align}
We denote the bosonic edge fields of the two phases by $\phi_{1,2}$ and $\tilde \phi_{1,2}$, which transform under symmetry as 
\begin{align}
&g_1:\begin{pmatrix} \phi_1 \\ \phi_2 \\\tilde \phi_1\\\tilde \phi_2\end{pmatrix} \rightarrow   \begin{pmatrix} \phi_1 +t_1\pi \\ \phi_2+t_1\pi \\\tilde \phi_1+t_2\pi\\\tilde \phi_2 +t_1\pi \end{pmatrix} \\
&g_2: \begin{pmatrix} \phi_1 \\ \phi_2 \\\tilde \phi_1\\\tilde \phi_2\end{pmatrix} \rightarrow   \begin{pmatrix} -\phi_1 \\ \phi_2+t_3\pi \\\tilde \phi_1+t_3\pi \\-\tilde \phi_2 \end{pmatrix}
\end{align}
Therefore, we can symmetrically gap out all the edge fields by the Higgs terms
\begin{align}
&\text{cos}(\phi_1+\tilde \phi_2)\nonumber \\
&\text{cos}(\phi_2+\tilde \phi_1)
\end{align}
Therefore, the stacking system (\ref{eqn_append_relation_13_1}) is trivial.\\

\textbf{4. Solutions with $W^{g_1}=-1_{2\times 2}$,  $W^{g_2}=\sigma_z$, }\\

From (\ref{phig1}) and gauge transformation, we have
\begin{align}
\delta \phi^{g_1} =0
\end{align}

From (\ref{phig2}) and (\ref{phig1g2}) and gauge transformation, we have
\begin{align}
\delta \phi^{g_2} = \pi \begin{pmatrix} t_1^{g_2} \\t_2^{g_2} \end{pmatrix}, 
t_{1,2}^{g_2}=0,1
\end{align}

Therefore, we use $[ -1_{2\times 2}, \sigma_z, t_1^{g_2},t_2^{g_2}]$ to denote different phases.
Below, we are going to show that 
\begin{align}
[ -1_{2\times 2}, \sigma_z,t_1,t_2]=&[ -1_{2\times 2}, 1_{2\times 2},(t_1,0)]\nonumber \\
\oplus &[ 1_{2\times 2}, \sigma_z,(1,1),t_2],
\label{eqn_append_z2z2z2f_relation_4_0}
\end{align}
or equivalently
\begin{align}
&[-1_{2\times 2}, \sigma_z,t_1,t_2]+[ -1_{2\times 2}, 1_{2\times 2},(0, t_1)]\nonumber \\
= &[ 1_{2\times 2}, \sigma_z,(1,1),t_2].
\label{eqn_append_z2z2z2f_relation_4_1}
\end{align}

Before proceeding, we first recall the effective edge theory of $[ 1_{2\times 2}, \sigma_z,(1,1),0]$, which is similar to the discussion in Sec.\ref{sec_z2z2z2f_root_z8}. If we denote the bosonic edge fields of $[ 1_{2\times 2}, \sigma_z,(1,1),0]$ as $\tilde \varphi_{1,2}$, we can define the four majorana fermions $\chi_{R,L}^{1,2}$ as (\ref{append_z2z2z2f_majorana_3_3}). Accordingly, they transform under symmetry as 
\begin{align}
&g_{1}: \begin{pmatrix} \chi_R^1 \\ \chi_R^2 \\ \chi_L^{1} \\ \chi_L^2 \end{pmatrix} \rightarrow  \begin{pmatrix}- \chi_R^1 \\-\chi_R^2 \\ -\chi_L^1\\ -\chi_L^2 \end{pmatrix}    \\
&g_{2}: \begin{pmatrix} \chi_R^1 \\ \chi_R^2 \\ \chi_L^{1} \\ \chi_L^2 \end{pmatrix} \rightarrow  \begin{pmatrix} \chi_R^1 \\-\chi_R^2 \\ (-)^{t_2} \chi_L^1\\ (-)^{t_2}\chi_L^2 \end{pmatrix} 
\end{align}
Therefore, by adding mass term $im\chi_R^{\tau(t_2)}\chi_L^{\tau(t_2)}$ which is symmetric, only $\eta_{R,L}^{\bar\tau(t_2)}$ are left gapless, which transform under symmetry as
\begin{align}
&g_{1}:   \begin{pmatrix} \chi_R^{\bar\tau(t_2)} \\\chi_L^{\bar\tau(t_2)} \end{pmatrix} \rightarrow  \begin{pmatrix}-\chi_R^{\bar\tau(t_2)} \\ -\chi_L^{\bar\tau(t_2)} \end{pmatrix}\label{eqn_append_z2z2z2f_symmetry_4_2_1}  \\
&g_{2}: \begin{pmatrix}  \chi_R^{\bar\tau(t_2)} \\ \chi_L^{\bar\tau(t_2)} \end{pmatrix} \rightarrow  \begin{pmatrix} -\chi_R^{\bar\tau(t_2)} \\  \chi_L^{\bar\tau(t_2)} \end{pmatrix} .\label{eqn_append_z2z2z2f_symmetry_4_2_2}
\end{align}
We note that $\tau(t_2)$ is a function on $t_2$ which takes $1$ for $t_2=0$ and 0 for $t_2=1$ and $\bar\tau(t_2)=t_2=0,1$.

%\begin{align}
%&g_{1}:   \begin{pmatrix} \chi_R^2 \\\chi_L^2 \end{pmatrix} \rightarrow  \begin{pmatrix}-\chi_R^2 \\ -\chi_L^2 \end{pmatrix}\label{eqn_append_z2z2z2f_symmetry_4_2_1}  \\
%&g_{2}: \begin{pmatrix}  \chi_R^2 \\ \chi_L^2 \end{pmatrix} \rightarrow  \begin{pmatrix} -\chi_R^2 \\  \chi_L^2 \end{pmatrix} \label{eqn_append_z2z2z2f_symmetry_4_2_2}
%\end{align}

Now we consider the left hand side of  (\ref{eqn_append_z2z2z2f_relation_4_1}). We denote the bosonic edge fields of the two phases as $\phi_{1,2}$ and $\varphi_{1,2}$ respectively, and similar to (\ref{append_z2z2z2f_majorana_3_1}) and (\ref{append_z2z2z2f_majorana_3_2}), we define the majorana fermions $\eta_{R,L}^{1,2}$ and $\xi_{R,L}^{1,2}$, which transform under symmetry as 
\begin{align}
&g_{1}: \begin{pmatrix} \eta_R^1 \\ \eta_R^2 \\ \eta_L^{1} \\ \eta_L^2 \end{pmatrix} \rightarrow  \begin{pmatrix} \eta_R^1 \\-\eta_R^2 \\ \eta_L^1\\ -\eta_L^2 \end{pmatrix}  ,\, \begin{pmatrix} \xi_R^1 \\ \xi_R^2 \\ \xi_L^{1} \\ \xi_L^2 \end{pmatrix} \rightarrow  \begin{pmatrix} \xi_R^1 \\-\xi_R^2 \\ \xi_L^1\\ -\xi_L^2 \end{pmatrix},
 \\
&g_{2}: \begin{pmatrix} \eta_R^1 \\ \eta_R^2 \\ \eta_L^{1} \\ \eta_L^2 \end{pmatrix} \rightarrow  \begin{pmatrix} (-)^{t_2} \eta_R^1 \\(-)^{1+t_2}\eta_R^2 \\(-)^{t_1} \eta_L^1\\ (-)^{t_1}\eta_L^2 \end{pmatrix} ,\,  \begin{pmatrix} \xi_R^1 \\ \xi_R^2 \\ \xi_L^{1} \\ \xi_L^2 \end{pmatrix} \rightarrow  \begin{pmatrix} (-)^{t_1}\xi_R^1 \\(-)^{t_1}\xi_R^2 \\ \xi_L^1\\ \xi_L^2 \end{pmatrix},
\end{align}
so that we can add symmetric mass terms $im_i \xi_R^i\eta_L^i$ and $im_3\eta_R^\tau(t_2)\xi_L^\tau(t_2)$ to gap out six majorana fermions $\xi_R^i$, $\xi_L^\tau(t_2)$, $\eta_L^i$ and $\eta_R^\tau(t_2)$. Now  only  two majorana fermions $\eta_R^{\bar \tau(t_2)}$ and $\xi_L^{\bar \tau(t_2)}$ are left gapless, which transform under symmetry in the same way as (\ref{eqn_append_z2z2z2f_symmetry_4_2_1}) and (\ref{eqn_append_z2z2z2f_symmetry_4_2_2}). Therefore, they must have the same symmetry anomaly, namely we prove (\ref{eqn_append_z2z2z2f_relation_4_1}).\\

\textbf{5. Solutions with $W^{g_1}=-1_{2\times 2}$, $W^{g_2}=-\sigma_z$}\\

From (\ref{phig1}) and gauge transformation, we have
\begin{align}
\delta \phi^{g_1} =0
\end{align}

From (\ref{phig2}) and (\ref{phig1g2}) and gauge transformation, we have
\begin{align}
\delta \phi^{g_2} = \pi \begin{pmatrix} t_1^{g_2} \\t_2^{g_2} \end{pmatrix}, 
t_{1,2}^{g_2}=0,1
\end{align}

Therefore, we use $[ -1_{2\times 2}, -\sigma_z, t_1^{g_2},t_2^{g_2}]$ to denote different phases.

Here we show that all the phases $[ -1_{2\times 2}, -\sigma_z, t_1,t_2]$ can be related to $[ -1_{2\times 2}, \sigma_z, t_2,t_1]$, i.e., 
\begin{align}
[ -1_{2\times 2}, -\sigma_z, t_1,t_2]\oplus [ -1_{2\times 2}, \sigma_z, t_2,t_1]=1
\label{eqn_append_z2z2z2f_relation_5_0}
\end{align}
which, according to (\ref{eqn_append_z2z2z2f_relation_4_0}), indicates that 
\begin{align}
[ -1_{2\times 2}, -\sigma_z, t_1,t_2]=&[ -1_{2\times 2}, 1_{2\times 2},(0, t_2)]\nonumber \\
\oplus &[ 1_{2\times 2}, \sigma_z,(1,1),t_1]^{-1}.
\label{eqn_append_z2z2z2f_relation_5_1}
\end{align}

Now we show (\ref{eqn_append_z2z2z2f_relation_5_0}).  We denote the bosonic edge fields of these two phases by $\phi_{1,2}$ and $\varphi_{1,2}$, which transform under symmetry as
\begin{align}
&g_{1}: \begin{pmatrix} \phi_1 \\ \phi_2\\\varphi_1\\\varphi_2  \end{pmatrix} \rightarrow  \begin{pmatrix}- \phi_1 \\ -\phi_2\\-\varphi_1\\-\varphi_2  \end{pmatrix}  \nonumber  \\
&g_{2}: \begin{pmatrix} \phi_1 \\ \phi_2\\\varphi_1\\\varphi_2  \end{pmatrix} \rightarrow  \begin{pmatrix}- \phi_1+t_1\pi \\ \phi_2+t_2\pi \\\varphi_1+t_2\pi \\-\varphi_2+t_1\pi   \end{pmatrix}   \label{eqn_append_6_symmetry_1}
\end{align}
We can symmetrically gap out the all the edge fields by Higgs terms $\text{cos}(\phi_1+\tilde \phi_2)$ and $\text{cos}(\phi_2+\tilde \phi_1)$. So we prove (\ref{eqn_append_z2z2z2f_relation_5_0}).\\

\textbf{6. Solutions with $W^{g_1}=\sigma_z$, $W^{g_2}=\sigma_z$}\\

From (\ref{phig1}) and (\ref{phig2}) and gauge transformation, we get
\begin{align}
&\delta \phi^{g_1} = \pi \begin{pmatrix} t_1^{g_1} \\t_2^{g_1} \end{pmatrix}, 
t_{1,2}^{g_1}=0,1
\\
&\delta \phi^{g_2} = \pi \begin{pmatrix} t_1^{g_2} \\0 \end{pmatrix}, 
t_{1}^{g_2}=0,1.
\end{align}

%From (\ref{phig1g2}), we have 
%$$t_1^{g_1}=t_1^{g_2} \text{ mod } 2.$$
%
%Therefore
%\begin{align}
%\delta \phi^g=\delta \phi^{g_1}=\delta \phi^{g_2} = \pi \begin{pmatrix} t\\0 \end{pmatrix}, 
%t=0,1
%\end{align}
%
We use notation $[\sigma_z, \sigma_z, (t_1^{g_1},t_2^{g_1}), t_1^{g_2}]$ to denote these two solutions.
Here we will  show that the phase $[\sigma_z, \sigma_z, (t_1,t_2), t_3]$ can be related to the phases with $[-\sigma_z, -1_{2\times 2}, (t_2,0)]$ and $[1_{2\times 2}, -\sigma_z, (0,t_1),t_3]$, i.e.,
\begin{align}
[\sigma_z, \sigma_z, (t_1,t_2), t_3]=&[-\sigma_z, -1_{2\times 2}, (t_2,0)]^{-1}\nonumber \\
\oplus &[1_{2\times 2}, -\sigma_z, (0,t_1),t_3]^{-1}.
\label{eqn_append_z2z2z2f_relation_6_0}
\end{align}
According to (\ref{eqn_append_relation_13_1}) and (\ref{eqn_append_z2z2z2f_relation_5_1}), we can obtain that
\begin{align}
[\sigma_z, \sigma_z, (t_1,t_2), t_3]=&[\sigma_z, 1_{2\times 2}, (1,1), t_2]\nonumber \\
\oplus &[ 1_{2\times 2},\sigma_z,  (t_1,0), t_3].
\end{align}

Now we are going to show (\ref{eqn_append_z2z2z2f_relation_6_0}), which is equivalent to show the stacking system 
\begin{align}
&[\sigma_z, \sigma_z, (t_1,t_2), t_3]\nonumber \\
\oplus &[-\sigma_z, -1_{2\times 2}, (t_2,0)]\nonumber \\
\oplus &[1_{2\times 2}, -\sigma_z, (0,t_1),t_3].
\label{eqn_append_z2z2z2f_relation_6_1}
\end{align}
is trivial.  We denote the bosonic edge fields of these three phases by $\phi_{1,2}$, $\varphi_{1,2}$ and $\tilde\varphi_{1,2}$ respectively. The following Higgs terms 
\begin{align}
&\text{cos}(\phi_1+\tilde \varphi_2)\nonumber \\
&\text{cos}(\phi_2+ \varphi_1)\nonumber \\
&\text{cos}(\varphi_1+\tilde \varphi_1)
\end{align}
can symmetrically gap out all these edge fields. Therefore, we prove (\ref{eqn_append_z2z2z2f_relation_6_0}).\\

\textbf{7. Solutions with $W^{g_1}=\sigma_z$, $W^{g_2}=-\sigma_z$}\\

From (\ref{phig1}) and gauge transformation, we get
\begin{align}
\delta \phi^{g_1} = \pi \begin{pmatrix} t_1^{g_1} \\0 \end{pmatrix}, 
t_{1}^{g_1}=0,1
\end{align}

From (\ref{phig2}) and gauge transformation, we have
\begin{align}
\delta \phi^{g_2} = \pi \begin{pmatrix} 0\\ t_2^{g_2}  \end{pmatrix}, 
t_{2}^{g_2}=0,1
\end{align}
We then denote the phase by $[\sigma_z, -\sigma_z, t_1^{g_1}, t_2^{g_2}]$. 
%Condition (\ref{phig1g2}) does not provide any more constraint on $\delta \phi^{g_{1,2}}$. Therefore, there are four solutions, which we denote as $[\sigma_z, -\sigma_z, t_1^{g_1}, t_2^{g_2}]$.\\
Here we we will show that 
\begin{align}
[\sigma_z, -\sigma_z, t_1, t_2]=&[ -\sigma_z, 1_{2\times 2}, (t_2,0),0]^{-1}\nonumber \\
\oplus &[  1_{2\times 2}, \sigma_z, (0, t_1),0]^{-1}.
\label{eqn_append_z2z2z2f_relation_7_0}
\end{align}
To show (\ref{eqn_append_z2z2z2f_relation_7_0}) is equivalent to show that the stacking system 
\begin{align}
[\sigma_z, -\sigma_z, t_1, t_2]\oplus &[ -\sigma_z, 1_{2\times 2}, (t_2,0),0]\nonumber \\
\oplus &[  1_{2\times 2}, \sigma_z, (0, t_1),0]
\label{eqn_append_z2z2z2f_relation_7_0}
\end{align}
is trivial. 
We denote the bosonic edge fields of these three phases by $\phi_{1,2}$, $\varphi_{1,2}$ and $\tilde\varphi_{1,2}$ respectively. The following Higgs terms 
\begin{align}
&\text{cos}(\phi_1+\tilde \varphi_2)\nonumber \\
&\text{cos}(\phi_2+ \varphi_1)\nonumber \\
&\text{cos}(\varphi_1+\tilde \varphi_1)
\end{align}
can symmetrically gap out all these edge fields. Therefore, we prove (\ref{eqn_append_z2z2z2f_relation_7_0}).\\

\textbf{8. Solutions with $W^{g_1}=W^{g_2}=-\sigma_z$}\\

From (\ref{phig1})-(\ref{phig1g2}) and and gauge transformation, we get
\begin{align}
&\delta \phi^{g_1} = \pi \begin{pmatrix} t_1^{g_1}\\t_2^{g_1}  \end{pmatrix}, 
t_{1}^{g_1}=0,1\\
&\delta \phi^{g_2} = \pi \begin{pmatrix}  t_1^{g_2} \\0 \end{pmatrix}, 
t_{2}^{g_2}=0,1
\end{align}
We then denote the phase by $[-\sigma_z, -\sigma_z, (t_1^{g_1}, t_2^{g_1}),t_1^{g_2}]$. 
We can show that 
\begin{align}
[-\sigma_z, -\sigma_z, (t_1, t_2),t_3]\oplus [\sigma_z, \sigma_z, (t_2, t_1),t_3]=1.
\label{eqn_append_z2z2z2f_relation_8_0}
\end{align}
We denote the bosonic edge fields of these two phases by $\phi_{1,2}$ and $\varphi_{1,2}$ respectively.  We can symmetrically gap out all these edge fields by the following Higgs terms
\begin{align}
&\text{cos}(\phi_1+\tilde \phi_2) \nonumber \\
&\text{cos}(\phi_2+\tilde \phi_1). \nonumber 
\end{align}
Therefore, we prove (\ref{eqn_append_z2z2z2f_relation_8_0}).

\section{$\Z_4^f$ symmetry}
\label{z4f}

From the below relations

\begin{align}
&g^2=P_f \\
&(W^g)^2=1\\
&(W^g)^TK W^g=K \\
&(1+W^g) \delta \phi^g= \pi \begin{pmatrix} 1 \\ 1 \end{pmatrix}  \label{z4f1}
\end{align}

From these, we get $W^g= \pm 1, \pm \sigma_z$. From The last relation, it is easy to see that $W^g$ can only take $1$.

%\subsection{Case one: $W^g=1$}

From (\ref{z4f1}), we get
\begin{align}
&2\delta \phi^g= \pi \begin{pmatrix} 1 \\ 1 \end{pmatrix} \\
&\delta \phi^g= \pi \begin{pmatrix} t_1 \\ t_2 \end{pmatrix}+ \frac{\pi}{2}\begin{pmatrix} 1 \\ 1 \end{pmatrix}, t_1, t_2=0,1
\end{align}

We denote the solutions as $[1, t_1, t_2]$.
It is easy to show that $[1, t_2, t_1]=[1, t_1, t_2]^{-1}$ and $[1, t, t]$ is trivial. Therefore, we only need to concentrate on $t_1<t_2$.  There is only case:  $[1, 0, 1]$.\\

\textit{1. Solution: $[1, 0, 1]$}\\

Under symmetry, 

 \begin{align}
   &g: \begin{pmatrix}   \phi_1 \\ \phi_2 \end{pmatrix} \rightarrow   \begin{pmatrix}  \phi_1 + \frac{\pi}{2}\\ \phi_2- \frac{\pi}{2} \end{pmatrix} 
%\end{align}
\end{align} 

It is easy to see that we can symmetrically gap out the edge fields by Higgs term $\text{cos}(\phi_1+\phi_2)$.

%\subsection{Case two: $W^g=-1$}
%From (\ref{z4f1}), there is no consistent $\delta \phi^g$. Therefore, $W^g=-1$ is not a consistent solution.
%
%
%\subsection{Case three: $W^g=\sigma_z$}
%From (\ref{z4f1}), there is no consistent $\delta \phi^g$. Therefore, $W^g=\sigma_z$ is not a consistent solution.
%
%
%\subsection{Case four: $W^g=-\sigma_z$}
%From (\ref{z4f1}), there is no consistent $\delta \phi^g$. Therefore, $W^g=-\sigma_z$ is not a consistent solutio.
%

Before conclude that there is no nontrivial SPT phase protected by $\Z_4^f$ symmetry, we need to trivialize the nontrivial bosonic SPT phase protected by $\Z_2$ symmetry.

The nontrivial bosonic $\Z_2 $ SPT: $K=\sigma_x$ and under symmetry $\phi_{1,2} \rightarrow \phi_{1,2}+ \pi $. To trivialize this bosonic SPT, we stack a trivial fSPT, say, $[1, 0, 1]$,  on it as $K=\sigma_x \oplus \sigma_z$. Therefore, under symmetry, 
\begin{align}
   &g: \begin{pmatrix}   \phi_1  \\ \phi_2 \\ \tilde \phi_1 \\ \tilde \phi_2 \end{pmatrix} \rightarrow   \begin{pmatrix}  \phi_1 + \pi \\ \phi_2+\pi  \\ \tilde \phi_1 +\frac{\pi}{2}\\ \tilde \phi_2 -\frac{\pi}{2}\end{pmatrix} 
%\end{align}
\end{align} 

This edge can be symmetrically gapped out by the following Higgs terms
\begin{align} 
&\text{cos}(\tilde \phi_1 +\tilde \phi_2-2 \phi_1) \\
&\text{cos}(\tilde \phi_1 -\tilde \phi_2- \phi_1) 
\end{align}

Similar trivialization of the $\Z_2$ bosonic SPT in ferminic system with $\Z_4^f$  symmetry is first discussed in Ref\cite{Chenjie1610}.

\end{document}